\RequirePackage{fix-cm}
\documentclass[natbib]{svjour3}                     
\smartqed  

\usepackage[normalem]{ulem}

\newcommand{\arcdeg}{\ensuremath{^{\circ}}}

\usepackage{graphicx}
\usepackage{epsf}
\usepackage{amssymb}
\usepackage{amsmath}
\usepackage{placeins}
\usepackage{color}

 \journalname{SSRv}
\newcommand{\farcs}{\mbox{\ensuremath{.\!\!^{\prime\prime}}}}
\newcommand{\be}{\begin{equation}}
\newcommand{\ee}{\end{equation}}
\newcommand{\beq}{\begin{eqnarray}}
\newcommand{\eeq}{\end{eqnarray}}
\newcommand\subsun[1]{{$_{\normalsize\odot}$}}

\newcommand{\kms}{~km~s$^{-1}$}

\newcommand{\Alfic}{Alfv\'enic}

\definecolor{pink}{rgb}{0.91, 0.67, 0.81}
\definecolor{violet}{rgb}{0.93, 0.51, 0.93}

\def\araa{ARA\&A}%
\def\apj{ApJ}%
\def\apjl{ApJ}%
\def\apjs{ApJS}%
\def\aap{A\&A}%
\def\jcap{J. Cosmology Astropart. Phys.}%
\def\mnras{MNRAS}%
\def\pasa{PASA}%
\def\physrep{Phys.~Rep.}%
\def\pasj{PASJ}%
\def\sovast{Soviet~Ast.}%
\def\ssr{Space~Sci.~Rev.}%
\def\nat{Nature}%
\def\memsai{Mem.~Soc.~Astron.~Italiana}%
\def\physrep{Phys.~Rep.}%
\def\procspie{Proc.~SPIE}%
\def\lsim{\;\raise0.3ex\hbox{$<$\kern-0.75em\raise-1.1ex\hbox{$\sim$}}\;}
\def\gsim{\;\raise0.3ex\hbox{$>$\kern-0.75em\raise-1.1ex\hbox{$\sim$}}\;}

\def \cms {\rm ~cm~s^{-1}}
\def\cmc{~cm$^{-3}$}

\def\cmsec {\rm ~cm~s^{-1}}
\def\ergs{\rm ~erg~s^{-1}}

\def\enf{\rm ~erg~cm^{-2}~s^{-1}}
\def\arcmin{\hbox{$^\prime$}}
\def\arcsec{\hbox{$^{\prime\prime}$}}

\def\pasj{PASJ}
\def\lsim{\;\raise0.3ex\hbox{$<$\kern-0.75em\raise-1.1ex\hbox{$\sim$}}\;}
\def\gsim{\;\raise0.3ex\hbox{$>$\kern-0.75em\raise-1.1ex\hbox{$\sim$}}\;}
\def \cmsec {\rm ~cm~s^{-1}}
\def\cmc{\rm ~cm^{-3}}
\def\kms{\rm ~km~s^{-1}}

\def\cmc{\rm ~cm^{-3}}

\def \kms {\rm ~km~s^{-1}}
\def \cmsec {\rm ~cm~s^{-1}}
\def\ergs{\rm ~erg~s^{-1}}

\def\enf{\rm ~erg~cm^{-2}~s^{-1}}
\def\arcmin{\hbox{$^\prime$}}
\def\arcsec{\hbox{$^{\prime\prime}$}}

\def\apj{ApJ}
\def\mnras{MNRAS}
\def\nat{Nat}

\def\araa{ARA\&A}                
\def\aap{A\&A}                   
\def\apjs{ApJS}                  
\def\apjl{ApJ}                   
\def\pasj{PASJ}

\def\lsim{\;\raise0.3ex\hbox{$<$\kern-0.75em\raise-1.1ex\hbox{$\sim$}}\;}
\def\gsim{\;\raise0.3ex\hbox{$>$\kern-0.75em\raise-1.1ex\hbox{$\sim$}}\;}
\def \cmsec {\rm ~cm~s^{-1}}

\definecolor{purple}{rgb}{0.63, 0.36, 0.94}

\def\kms{\rm ~km~s^{-1}}
\def\ergs{\rm ~erg~s^{-1}}

\def\cmc{\rm ~cm^{-3}}
\def\cms{\rm ~cm^{-2}}
\def\jcap{J. Cosmology Astroparticle Physics }
\def\aap{A\&A}
\def\apj{ApJ}
\def\apjl{ApJ Letters}
\def\mnras{MNRAS}
\def\physrep{Physics Reports}

\def\ssr{ Space Sci. Rev.}

\def\nat{Nature}

\def\araa{ARA\&A}%
\def\apj{ApJ}%
\def\apjl{ApJ}%
\def\apjs{ApJS}%
\def\aap{A\&A}%
\def\jcap{J. Cosmology Astropart. Phys.}%
\def\mnras{MNRAS}%
\def\pasa{PASA}%
\def\physrep{Phys.~Rep.}%
\def\pasj{PASJ}%
\def\sovast{Soviet~Ast.}%
\def\ssr{Space~Sci.~Rev.}%
\def\nat{Nature}%
\def\memsai{Mem.~Soc.~Astron.~Italiana}%
\def\physrep{Phys.~Rep.}%
\def\procspie{Proc.~SPIE}%

\begin{document}
\title{Pulsar wind nebulae with bow shocks: non-thermal radiation and cosmic ray leptons}

\titlerunning{Pulsar wind nebulae with bow shocks}        

\author{A.M.~Bykov \and E.~Amato \and A.E.~Petrov
        \and A.M.~Krassilchtchikov  \and K.P.~Levenfish}

\authorrunning{A.M.~Bykov et al.} 

\institute{A.M.~Bykov \at
           Ioffe Institute, 194021, St. Petersburg, Russia;\\ St. Petersburg Polytechnic University, Russia;\\ International Space Science Institute, Bern, Switzerland\\
           \email{byk@astro.ioffe.ru}
           \and E.~Amato \at INAF - Osservatorio Astrofisico di Arcetri, Largo E. Fermi, 5, 50125, Firenze, Italy\\
           \email{amato@arcetri.astro.it}
           \and A.E.~Petrov \at Ioffe Institute, 194021, St. Petersburg, Russia \\
       \and A.M.~Krassilchtchikov \at Ioffe Institute, 194021, St. Petersburg, Russia \\
       \and K.P.~Levenfish \at Ioffe Institute, 194021, St. Petersburg, Russia}

\date{Received: 10.04.2017 / Accepted: 24.04.2017}

\maketitle

\begin{abstract}

Pulsars with high spin-down power produce relativistic winds radiating a
non-negligible fraction of this power over the whole
electromagnetic range from radio to gamma-rays in the pulsar wind nebulae
(PWNe). The rest of the power is dissipated in the interactions of the PWNe
with the ambient interstellar medium (ISM). Some of the PWNe are moving
relative to the ambient ISM with supersonic speeds producing bow shocks.
In this case, the ultrarelativistic particles accelerated at the
termination surface of the pulsar wind may undergo reacceleration in the
converging flow system formed by the plasma outflowing from the wind
termination shock and the plasma inflowing from the bow shock. The presence
of magnetic perturbations in the flow, produced by instabilities induced by
the accelerated particles themselves, is essential for the process to work.
A generic outcome of this type of reacceleration is the creation of
particle distributions with very hard spectra, such as are indeed required
to explain the observed spectra of synchrotron radiation with photon
indices ${\rm \Gamma} \lsim$1.5. The presence of this hard spectral component is
specific to PWNe with bow shocks (\mbox{BSPWNe}). The accelerated particles, mainly
electrons and positrons, may end up containing a substantial fraction of
the shock ram pressure. In addition, for typical ISM and pulsar
parameters, the $e^+$ released by these systems in the Galaxy are numerous
enough to contribute a substantial fraction of the positrons detected as
cosmic ray (CR) particles above few tens of GeV and up to several hundred
GeV. The escape of ultrarelativistic particles from a BSPWN --- and hence,
its appearance in the far-UV and X-ray bands --- is determined
by the relative directions of the interstellar magnetic field,
the velocity of the astrosphere and the pulsar rotation axis. In this
respect we review the observed appearance and multiwavelength spectra of
three different types of BSPWNe: PSR~J0437-4715, the Guitar
and Lighthouse nebulae, and Vela-like objects. We argue that high
resolution imaging of such objects provides unique information both on
pulsar winds and on the ISM. We discuss the interpretation of
imaging observations in the context of the model outlined above
and estimate the BSPWN contribution to the positron flux observed
at the Earth.

\end{abstract}


\section{Introduction}
\label{sec:intro}

Pulsars are magnetized fast rotating neutron stars, whose rotation
energy is carried away by relativistic winds blowing out of their
magnetic poles and mainly consisting of electron-positron (e$^\pm$) pairs.
Multiband emission observed from these winds allows one to
study the complex physics of energy transport in space
\citep[][]{2002ASPC..271...71A,2006ARA&A..44...17G,2009ASSL..357..421K,
2012SSRv..173..341A,2014AN....335..234B,2015SSRv..191..391K}.
Multi-wavelength observations have revealed a vast variety of
radiative efficiencies of pulsar winds (PWs) and pulsar wind
nebulae (PWNe) produced by the winds, ranging from some tens percent
in the Crab nebula \citep[see e.g.][]{2014RPPh...77f6901B}
to much lower values in a number of other PWNe \citep[see, e.g.,][]{2008ApJ...682.1166L,KP08}.
Relativistic particles leaving the observable PWNe interact
with the ambient medium on larger scales (upto tens of parsec)
and thus provide a tool to reveal the structure of magnetic field
and some other parameters of the interstellar medium (ISM), as
discussed below in \S\ref{extended}. This is a probable case for
the fast enough, supersonically moving pulsars.

An analysis of radio pulsar velocity distribution based
on the large scale survey at 0.4 GHz by \citet{2002ApJ...568..289A}
suggested a two-component velocity distribution with characteristic
velocities of 90 and 500$\kms$ and inferred that $\sim$15\% of
the pulsars have natal velocities greater than 1000$\kms$.
The authors also concluded that under some simplified assumptions on
the supernova remnant (SNR) evolution, about 10\% of the pulsars
would leave the host remnant before they are 20~kyr old.

The catalog of 233 pulsars (both young and recycled)
compiled by \citet{2005MNRAS.360..974H} contains
derived pulsar speeds measured in one coordinate (1D) and the
transverse velocities (2D). No evidence for a bimodal velocity
distribution was found.
The authors derived the mean 1D pulsar speed of
152$\pm$10$\kms$ for the normal pulsars and
54$\pm$6$\kms$ for the recycled ones.
The corresponding mean 2D velocities in the catalog are
246$\pm$22$\kms$ and 87$\pm$13$\kms$, with the highest values
$\sim 1,600 \kms$ inferred for the pulsars B2224+64 and B2011+38.
The mean 3D natal pulsar velocity was derived to be 400$\pm$40$\kms$.

For typical pulsar and ISM parameters, most pulsars are expected
to leave their parent SNRs during the Sedov-Taylor phase of
remnant expansion. The time $T_{\rm esc}$ at which this occurs
can be estimated as $V_{\rm PSR} T_{\rm esc} = R_{\rm ST} (T_{\rm
esc}/t_{\rm ST})^{2/5}$, where $V_{\rm PSR}$ is the velocity of
pulsar proper motion, $t_{\rm ST}$ and $R_{\rm ST}$ are the time
and remnant radius at the beginning of the Sedov-Taylor phase.
With an average estimated velocity of about 400$\kms$,
an estimate of the time at which a pulsar escapes from its SNR
can be found as
\be T_{\rm esc}=\left(\frac{R_{\rm ST}}{V_{\rm PSR} T_{\rm
ST}^{2/5}}\right)^{\frac{5}{3}}\approx 45~{\rm kyr}
\left(\frac{E_{\rm SN}}{10^{51} {\rm erg}}\right)^\frac{1}{3}\
\left(\frac{n_H}{{\rm cm}^{-3}}\right)^{-\frac{1}{3}}\
\left(\frac{V_{\rm PSR}}{400~{\rm km/s}} \right)^{-\frac{5}{3}}\ ,
\label{eq:psresc} \ee

A pulsar aged a few tens of kyr still has a non-negligible fraction
of its original spin-down energy to spend. In general, one can write
$\dot \Omega \propto \Omega^n$, with $n$ -- the pulsar braking index,
$\Omega$ and $\dot \Omega$ -- the star rotation frequency and its time
derivative, respectively. For dipole spin-down one expects $n=3$,
since $\dot E=I \Omega \dot \Omega \propto B_*^2 \Omega^4$,
where $I\approx10^{45}$ g cm$^2$ is the star momentum of inertia and $B_*$ --
its surface magnetic field. While it is true that for the few cases, where
the braking index has been actually measured, its value has
always been found to be less than 3, with an average of 2.5 (see, e.g.,
\cite{2006ApJ...643..332F} and references therein), in the following we
will assume the standard dipole spin-down law. Hence, one can compute
the rotational energy the pulsar is left with at the moment $T_{\rm esc}$
when it leaves the remnant.
From the equation:
\be I\Omega \dot \Omega=\frac{4}{9}
\frac{B_*^2 R_*^6 \Omega^4}{c^3}\ , \label{eq:spindown} \ee
where $R_*\approx10$ km is the neutron star radius, we find:
\be
\Omega(t)=\frac{\Omega_0}{\left(1+t/\tau_b\right)^\frac{1}{2}}, \,
\, \, \, {\rm with}\, \, \, \, \tau_b=\frac{9 I c^3}{8B_*^2 R_*^6
\Omega_0^2}, \label{eq:omegat}
\ee
and $\Omega_0=2\pi/P_0$ -- the initial star rotation frequency
($P_0$ is the initial star spin period).
Finally, the residual rotation energy of the star at
$T_{\rm esc}$ is: \be E_{\rm res}=\frac{1}{2}I\Omega^2(T_{\rm
esc})=\frac{1}{2}\frac{I \Omega_0^2}{\left(1+t_{\rm
esc}/\tau_b\right)}. \label{eq:eres} \ee

The average value of $E_{\rm res}$ can be estimated by taking the
average values of the
pulsar surface magnetic field and initial period. From the work of
\citet{2006ApJ...643..332F} we have: $\langle B_*\rangle \approx 4.5\times
10^{12}$ G and $\langle P_0 \rangle=300$ ms. Using the estimate of
$T_{\rm esc}$ from Eq.~(\ref{eq:psresc}) one obtains $E_{\rm res}\approx
10^{47}$~erg and $\dot E_{\rm res}\approx 5 \times 10^{34}$~erg/s.

The fast average proper motion means that a non-negligible fraction of the
pulsars propagating through the ISM may form a cometary-like bow shock
structure. Optical and radio observations discovered a few bow shocks
associated with fast moving pulsars \citep[see, e.g.,][]{1988Natur.335..801K,1993Natur.362..133C,1996ApJ...464L.165F,2012ApJ...746..105N,2014ApJ...784..154B}.
High resolution X-ray observations of PWNe
revealed a rich collection of their morphologies \citep{KP08} including
a number of cometary type objects, some of which harbour bow shocks.

Bow-shock nebulae are known to originate from the interaction of
winds from moving stars with the ambient medium and appear
in association with very different types of stars: of solar type
\citep[see, e.g.,][]{1971SPhD...15..791B}, young massive stars
\citep[see, e.g.,][]{1977ApJ...218..377W}, and fast rotating magnetized
neutron stars \citep[see, e.g.,][]{2001A&A...375.1032B,2001ApJ...563..806B}.
The dynamics of astrophysical bow shocks has been modelled both analytically,
within the thin shell approximation, and by means of numerical hydrodynamics
schemes \citep[see, e.g.,][]{1993A&A...276..648B,Wilkin}.

The rich set of observed morphological and spectral appearances
of PWNe traces the differences in the intrinsic properties of
the parent pulsars (spin-down power and inclination), in their proper
velocities, and in the evolutionary stages during the interaction
of PWNe with the ISM \citep{2006ARA&A..44...17G}.

The structure of PWNe can be heavily affected by the supersonic
motion of the medium in which the nebula is expanding. This can be
either associated with the fast proper motion of the pulsar through
the normal ISM or with the reverse shock of the supernova
propagating back towards the origin of the explosion. The effect of
the external ram pressure of the fast moving ambient medium inside
the supernova remnant on the PWN structure was studied in the frame
of magnetohydrodynamical models by
\citet{2003A&A...404..939V,2004A&A...420..937V}. The authors applied
the model to three SNRs: N157B, G327.1-1.1, and W44, and concluded
that the head of the PWN is not bounded by a bow shock in the case
of N157B, G327.1-1.1, while in the case of W44 they argued for the
supersonic scenario with a bow shock PWN (BSPWN) due to the fast
motion of the pulsar. Another case of a bow shock type structure
occurs when the PW located inside an SNR is interacting with a
mildly supersonic flow of Mach number $\gsim$~1 produced in the
transition phase when the reverse shock reaches the center of the
remnant \citep{chevreyn11}. A detailed one-zone model of spectral
evolution of PWNe inside SNRs was presented by
\citet{2011MNRAS.410..381B}. The authors found that the one-zone
model provides a satisfactory description of a number of PWNe
assuming that relativistic pairs are injected in these nebulae with
a spectral distribution in the form of a broken power law, rather
flat at low energy (energy spectral index close to 1) and steepening
(spectral index larger than 2) above some break energy. They found
that the intrinsic spectral break turns out to occur at similar
energies for PWNe with rather different characteristics. A more
sophisticated modeling of BSPWN dynamics was undertaken by
\citet{2005A&A...434..189B}, who investigated it in the framework of
axisymmetric relativistic MHD, highlighting the dependence of the
dynamics on the PW magnetization.

The \Alfic\ Mach number of a bow shock of speed $v_{\rm sh}$
(at the bow apex it is about the pulsar speed) is
\begin{equation}
{\cal M}_{\rm a} = v_{\rm sh} (4\pi \rho_{\rm a})^{1/2}/B_0 \approx
460\, v_8\, n_{\rm a}^{1/2}/ \, B_{-6},
\end{equation}\label{Ma}
where  $n_{\rm a}$ is the ionized ambient gas number density
measured in $\cmc$, $B_{-6}$ is the local magnetic field in the
shock upstream measured in $\mu$G and $v_8$ is the shock velocity in
10$^8~\cmsec$. The sonic Mach number for a shock propagating in the
interstellar plasma of the standard (Solar) abundance is
\begin{equation}
{\cal M}_{\rm s} \approx 85\, v_8\, \cdot [T_4 \cdot (1 + f_{\rm
ei})]^{-1/2},
\end{equation}\label{Ms}
where $T_4$ is the plasma ion temperature measured in 10$^4$ K and
$f_{\rm ei} = T_{\rm e}/T_{\rm i}$ is the ratio between the
electron and ion temperatures.

Consider a pulsar moving supersonically with a velocity $V_{\rm
psr}$ relative to the local rest frame of the ambient medium. The
standoff distance $R_{\rm cd}$ of the PW termination shock in this frame
can be estimated as
\begin{equation}
R_{\rm cd} \approx \sqrt{\frac{\zeta_{\rm K} \dot{E}}{6\pi \rho_{\rm ism}
V_{\rm psr}^2 c}} \approx 2\times 10^{16}  \dot{E}^{1/2}_{35}
n_{\rm ism}^{-1/2} V^{-1}_{200}\,\, {\rm cm},\label{RCD}
\end{equation}
where $\dot{E}$ is the pulsar spin-down power, $\rho_{\rm ism}$
is the ambient ISM density.
In fact, $R_{\rm cd}$ is rather the location of the contact
discontinuity, than either of the PWN termination shock, or of the bow
shock. The parameter $\zeta_{\rm K}$ depends on the magnetic inclination angle
of the pulsar; $\zeta_{\rm K} =$1 if the PW is spherically symmetric
\citep[e.g.,][]{swaluw_ea_03}. However, observations of
the Crab nebula as well as {\sl particle-in-cell} and MHD
modeling of the relativistic pulsar wind indicate that the wind power is
likely highly anisotropic. Such anisotropy affects the value of
$\zeta_{\rm K}$ and, therefore, $R_{\rm cd}$, which scales $\propto \zeta^{1/2}$.
This means that if the direction of the pulsar proper motion is
close to the equatorial plane of a low-inclination rotator (i.e., the
direction of the maximal wind power) the position of the contact
discontinuity will be at an angular distance $\geq 6\arcsec
\zeta_{\rm K}^{1/2} d_{250}^{-1}$ for the most likely pulsar and ISM
parameters.
The images like the one presented in Fig.~\ref{fig:VelaXC} may
allow us to constrain the wind power anisotropy.

The model of a PWN evolving inside the remnant of the parent
supernova developed by \citet{2004A&A...420..937V} employed a hydrodynamical
approach to simulate the evolution of such a system when the pulsar's velocity
is high. The authors considered four different stages of PWN evolution:
the supersonic expansion stage, the stage of reverse shock interaction
followed by the subsonic expansion stage and, finally, the bow shock stage.
Within this model the bow shock stage occurs at roughly half the crossing time,
when the pulsar is located at about 0.68 times the radius of the SNR forward shock.
The authors suggested that the pulsar in the W44 SNR has a bow shock
because of its supersonic proper motion. However, \citet{chevreyn11} argued
that even in the case of the Vela pulsar, which is located closer than 0.68
of the SNR radius to the center of the remnant, and whose measured proper
velocity is below 70 $\kms$ \citep{2003ApJ...596.1137D}, a bow shock still can be formed.
It is likely that the SNR reverse shock has recently passed over
the Vela pulsar and a mildly supersonic flow of ${\cal M}_{\rm s} \sim$ 1.3 is naturally produced.
\citet{chevreyn11} noted that \citet{2004A&A...420..937V} had reached their conclusion
having assumed that the SNR flow can be described by the Sedov solution,
while there is a transition solution preceding the Sedov stage, which is likely to describe
the current evolutionary stage of Vela more accurately.
Within such a refined description, \citet{chevreyn11} found that for Vela-like pulsars
the flow velocity downstream of the reverse shock of speed $v_{\rm sh}$ is characterized
by a velocity 0.75$\, v_{\rm sh}$ which appears to be about 1.3 times the speed of sound.

In what follows, we discuss in some detail how particles are
accelerated in PWNe (\S~\ref{sec:PA}), with a specific attention to
acceleration in the PWNe driving bow shocks. Namely, efficient
acceleration of particles in colliding shock flows (CSFs) between the
pulsar wind termination surfaces and bow shocks is considered in
\S~\ref{CSFs} and a Monte-Carlo model is discussed in
\S~\ref{geomMC}, which is further employed to simulate acceleration, transport, and synchrotron
emission of the high energy e$^{\pm}$ pairs in \mbox{BSPWNe}. In
\S~\ref{Obs} we review far-UV and X-ray observations of emission
structures in the vicinities of some fast moving pulsars with \mbox{BSPWNe}
and apply the model of \S~\ref{geomMC} to simulate the synchrotron
images and spectra of the nebula around PSR~J0437-4715.
\S~\ref{extended} is devoted to peculiar extended X-ray structures
associated with the Geminga PWN, the ``Guitar nebula'', and the
``Lighthouse nebula''. The PWNe  are moving supersonically and have
hard photon indexes of X-ray emission in the vicinity of the bow
shocks. The fascinating extended structures observed in X-rays can
be associated with multi-TeV regime particles escaping from the bow
shock PWN astrospheres to the ISM producing the magnetized ballistic
beams of accelerated particles along the interstellar magnetic field
lines. In \S\ref{snrbspwn} we discuss the X-ray observations of
several \mbox{BSPWNe} possibly interacting with the remnants of their natal
supernovae, and discuss the Vela~X nebula in \S\ref{VelaX} as a
generic case, which also includes a broadband modeling of its
spectrum. Finally, in \S\ref{sec:posintro}, we discuss the role of
Bow Shock PWNe as plausible sources of the positron excess recently
highlighted by PAMELA and AMS-02 measurements.


\section{Particle Acceleration in Pulsar Wind Nebulae}
\label{sec:PA}

PWNe are powered by highly magnetized relativistic winds
which are terminated at some distance from the star.
At the termination surface a considerable fraction of
the pulsar spin-down power is transferred to accelerated particles
producing non-thermal radiation
\citep[see][]{1974MNRAS.167....1R,1984ApJ...283..694K,2000ApJ...539L..45C,2012SSRv..173..341A,2015SSRv..191..519S}.

The maximal energy of accelerated particles is limited according
to a given magnetic luminosity of the source ${\cal L_M}$
\citep[see, e.g.,][]{2009JCAP...11..009L}. Such energy
can be estimated by the condition on the particle acceleration time scale
in the comoving frame of the flow, which should be shorter than
the dynamical time of the flow. Within the validity limits
of relativistic MHD the acceleration time exceeds the particle gyroperiod
(this condition does not apply to the region where magnetic reconnection occurs).

Let us consider acceleration of particles of charge $Z$ in a Poynting flux
dominated source of magnetic luminosity ${\cal L_M}$. According to the analysis of
\citet[][]{1995ApJ...454...60N,1995PhRvL..75..386W,2009JCAP...11..009L}, the
luminosity required to reach the energy $E$ (in the absence of energy losses) is

\begin{equation}\label{Lmin}
{\cal L_M} \approx 3\times 10^{33} \cdot \frac{{\rm \Gamma^{2}_{\rm
flow}}}{{\rm \beta_{\rm flow}}}\left(\frac{E/Z}{10^{14} {\rm
eV}}\right)^2 \ergs,
\end{equation}
where $\beta_{\rm flow}$ is the flow speed divided by the speed of
light $c$, and ${\rm \Gamma}_{\rm flow}$ the flow Lorentz factor.
Therefore the maximal energy of a particle which is achievable in
the MHD outflow without the account for energy losses is limited by:
\begin{equation}\label{Emax14}
E_{\rm max} \approx Z \times 10^{14} \cdot \frac{{\rm
\beta^{1/2}_{\rm flow}}}{{\rm \Gamma_{\rm flow}}}\left(\frac{
\dot{E} }{3 \times 10^{33} \ergs}\right)^{1/2} {\rm eV},
\end{equation}
This approximate relation indicates that the most favourable sites to
achieve the highest energy are indeed in the trans- or
sub-relativistic flows of $\beta_{\rm flow} \lsim$1 located between
the ultrarelativistic termination shock and the non-relativistic bow
shock which we will discuss in \S\ref{sec:PACF} below.

\citet[][]{2013ApJ...771...54S} provided a detailed analysis of the
maximal energies of the particles accelerated at relativistic shocks.
They found that the relativistic perpendicular shocks propagating in
e$^{\pm}$ pair plasma (which is the likely case in PWNe) can
efficiently accelerate the pairs only if the magnetization in the
upstream plasma is $\lsim$ 10$^{-3}$ (and even lower, $\lsim$
3$\times$10$^{-5}$, for the electron-ion plasmas). The maximal Lorentz
factor of the pairs accelerated at the PW termination shock was estimated
by \citet[][]{2013ApJ...771...54S} from the constraint that the
acceleration time of the highest energy pairs be shorter than
the termination shock crossing time. These authors found that

\begin{equation}\label{Gmax1SSA}
\gamma_{\rm max}  \approx 1.9 \times 10^{7}  \cdot
\left(\frac{\dot{E}}{10^{38}\ergs} \right)^{3/4}
\left(\frac{\dot{N}}{10^{40}{\rm s}^{-1}}\right)^{-1/2} ,
\end{equation}
where $\dot{N}$ is the total particle flux entering the nebula,
which is somewhat uncertain and is likely within the range 3$\times$
10$^{38} \lsim \dot{N} \lsim $10$^{40}~~{\rm s}^{-1}$. The estimated maximal
energy is rather low even for the Crab nebula in spite of its high spin-down power.
Therefore \citet[][]{2013ApJ...771...54S} concluded that Fermi
acceleration at the termination shock of PWNe is not a likely
mechanism for producing the synchrotron emitting pairs and some other mechanisms
should be considered.

A number of pulsars with observed bow shocks, including
PSR~B2224+65, which powers the Guitar nebula,\, the old recycled
pulsar J0437-4715, \, PSR~J0633+1746 (Geminga) and others, have a
spin-down power $\dot{E}_{35} \lsim$ 0.3. Several BSPWNe harbor
regions of extended X-ray emission, where the magnetic field is
estimated to be a few tens of $\mu$G. In order to produce X-ray
synchrotron photons in such a field, pairs with a Lorentz factor
$\gamma \gsim $10$^8$ are required. But according to Eq.~(\ref{Lmin})
such high energies are compatible with the amount of spin-down power
mentioned above only if ${\rm \Gamma^{2}_{\rm flow}} \lsim$3 while
$\beta_{\rm flow} \gsim$ 0.1, i.e., the outflow is transrelativistic
in the particle acceleration zone. The Vela pulsar as well as the
pulsar IGR J11014-6103 (which powers the Lighthouse nebula) have
much higher spin-down power and the condition for acceleration of
pairs above the Lorentz factor $\gamma \gsim $10$^8$ is relaxed to
${\rm \beta_{\rm flow}} \gsim $10$^{-3}$.

In the light of these estimates, we will discuss below
the acceleration of particles
in fast flows around PW termination shocks and bow shocks.

\subsection{Particle Acceleration and Magnetic Field Amplification at Bow Shocks}
\label{sec:DSA}

Diffusive shock acceleration \citep[][]{ALS77,Kry77,Bell78a,BO78}
has been demonstrated to be a very attractive mechanism of particle
acceleration in supersonic flows of a moderate magnetization. There
is a growing body of observational evidence in favor of this
mechanism acting in SNRs, AGN jets and some other objects
\citep[see, e.g.,][]{2013APh....43...56B,2012SSRv..173..369H,2016RPPh...79d6901M}.
Particle acceleration in fast MHD shocks is strongly coupled with
magnetic field amplification (MFA) which is due to the growth
of seed fluctuations by the cosmic ray (CR) driven instabilities.
A number of such instabilities resulting in MFA can be excited
during diffusive shock acceleration \citep[see, e.g.,][]{Bell2004,ab09,BER2012,SchureEtal2012,2014ApJ...789..137B}.

The maximal energy $E_{\rm max}$ of a particle of charge $Z$ accelerated by a
strong quasi-steady bow shock of velocity $v_{\rm sh}$ and
characteristic radius $R_{\rm sh}$ is limited by the shock
curvature. It can be roughly estimated from
\begin{equation}
 D(E_{\rm max}) \sim v_{\rm sh}R_{\rm sh},
 \label{diff}
\end{equation}
where $D(E)$ is the particle diffusion coefficient in the shock
upstream. Particle propagation at a pulsar bow shock depends on the shock
obliquity, i.e., on the angle between the proper velocity of the
pulsar and the direction of the local mean interstellar magnetic
field. If the shock is quasi-parallel, it is determined
by the magnetic fluctuations amplified by the CR-driven
instabilities, as shown below in \S~\ref{QPar_bow}. On the other hand,
a transverse shock geometry may reduce the radial diffusion
coefficients and therefore drastically change the conditions
of particle confinement.

\subsection{Quasi-parallel Bow Shocks}\label{QPar_bow}

The amplitude $\delta B_{\rm w}$ of the fluctuating magnetic field
amplified by CR-driven instabilities in the upstream of a shock
can be estimated from the saturation condition of \citet[][]{Bell2004}:
\begin{equation}
\frac{\delta B_{\rm w}^2}{8 \pi} \approx \frac{v_{\rm sh}}{2c} U_{\rm CR},
\label{deltaB_Bell_regime}
\end{equation}
where $U_{\rm CR} \approx \zeta_{\rm CR}\, \rho v_{\rm sh}^2/2$ is
the energy density of the CRs accelerated at the shock. Then
the amplitude of the fluctuating magnetic field is
\begin{equation}
 \delta B_{\rm w} \approx \sqrt{2\pi\zeta_{\rm CR}\rho v_{\rm sh}^3/c}
\sim 20\zeta_{\rm CR}^{1/2}\, v_{\rm sh8}^{3/2}\, n^{1/2}~ \mu\,G,
\label{dB low}
\end{equation}
where $\zeta_{\rm CR}$ is the fraction of shock ram pressure converted into the energy of
the accelerated particles, $v_{\rm sh8}$ is the shock velocity in 10$^8$ cm s$^{-1}$ and
$n$ is the number density of the ISM.
This amplitude may reach $\delta B_{\rm w} \gsim B_0$,
where $B_0$ is the mean magnetic field in the ISM.
In any case, in the vicinity of a bow shock, $\delta B_{\rm w}$ is
likely well above the estimated amplitude of the field fluctuations
in the background ISM turbulence at the scales below 10$^{15}$~cm,
which are important for the scattering of TeV range particles.

The mean free path $\Lambda(E)$ of a relativistic proton in the
quiet ISM away from SNRs or other active space objects
can be estimated from the galactic CR propagation models as
$\Lambda(E) \sim \left(3 \mbox{--} 5\right)\times 10^{18}\,(E/ \mbox{GeV})^{a}$ cm,
where the index $a = $0.3--0.6 \citep[see, e.g.,][]{SMP2007}.
Therefore, the background magnetic turbulence in the quiet ISM
can not confine the relativistic $e^{\pm}$ pairs accelerated
in the PWNe within a region of scale size $\sim R_{\rm cd}$ as given by
Eq.~(\ref{RCD}), for a bow shock speed above 100~$\kms$.
Depending on the bow shock velocity,
strong MFA by CR-driven instabilities at the bow shock
would affect the propagation and confinement
of relativistic $e^{\pm}$ pairs up to Lorentz factors $\gsim $10$^7$.

\begin{figure}
\includegraphics[width=3.0in]{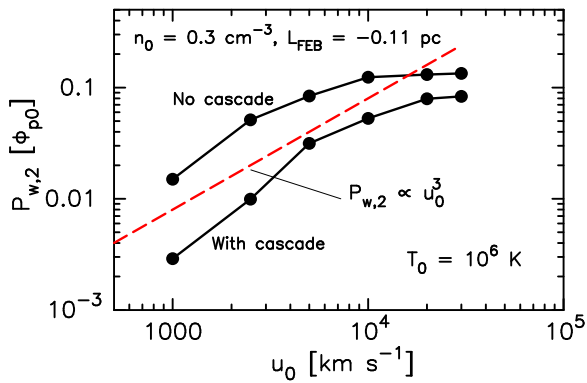}
\caption{The magnetic turbulence energy flux in the shock 
downstream $\zeta_{\rm B} \equiv \rm{P_{w,2}}$ as a function of
shock velocity simulated with a nonlinear Monte-Carlo model of
diffusive shock acceleration \citep{2014ApJ...789..137B}.
\label{fig:MFA}}
\vspace{-1.\baselineskip}
\end{figure}

Nonlinear Monte-Carlo modeling of the diffusive shock acceleration
process in the presence of MFA \citep{2014ApJ...789..137B} has
revealed that the regime consistent with (\ref{deltaB_Bell_regime})
holds for shock velocities below 5,000 $\kms$.
At higher shock velocities the energy density of the
fluctuating magnetic field scales as a fraction $\zeta_{\rm B}$ of the shock ram
pressure. The MFA efficiency is illustrated in Fig.~\ref{fig:MFA}
where the ratio $\zeta_B$ of the amplified
magnetic field energy density to the shock ram pressure
is shown as a function of the shock velocity.
In the high shock speed regime one can estimate the amplitude
of the enhanced magnetic field as
\begin{equation}
\delta B_{\rm w} \approx \sqrt{4\pi\zeta_{\rm B}\rho}\, v_{\rm sh}
\sim 50 \left[\frac{\zeta_{\rm B}}{0.01}\right]^{1/2}~v_{\rm sh8}~
n^{1/2}~ \mu G.
\label{dB high}
\end{equation}

The validity of the Bohm diffusion approach requires
strong scattering of particles by quasi-resonant fluctuations, i.e.,
$\delta B_{\rm w} \sim B_0$. In such a case the diffusion coefficient
$D(E) \approx \eta \, cE/(Ze \ \delta B_{\rm w}),$
where $\eta \gsim 1$ is a numerical factor \citep[see, e.g.,][]{reville08}.

Models of extended fast shocks
demonstrated that diffusive shock acceleration is rather efficient
in this framework with
$ \zeta_{\rm CR} >$ 0.2 \citep[see, e.g.,][]{2012SSRv..173..369H}.
The energy density of the CR-driven
magnetic fluctuations in the shock vicinity is dominated by  the
fluctuations with typical wavenumbers $k_{\ast}$ which satisfy the
condition  $k_{\ast} R_{\rm g}(E_{\rm max}) \lsim$1,
where $R_{\rm g}$ is the gyroradius of a particle.

Using Eqs.~(\ref{diff}), (\ref{RCD})
one can obtain the maximal energy of accelerated ions as
\begin{equation}
\begin{aligned}
E_{\rm max} ~ &\approx ~ 3 Ze \delta B_{\rm w} \eta^{-1} R_{\rm sh}
\frac{v_{\rm sh}}{c} ~ \approx ~  \frac{Ze}{\eta} \sqrt{\frac{3\zeta_{\rm K}\zeta_{\rm CR} \dot{E}}{c}} \left[\frac{v_{\rm
sh}}{c}\right]^{3/2} ~ \approx \\
& \approx ~ \frac{0.2 Z (\zeta_{\rm K}\zeta_{\rm
CR})^{1/2}}{\eta} \dot{E}^{1/2}_{35} v_{\rm sh8}^{3/2}\,\, {\rm
TeV}
\end{aligned}
\label{Emax low}
\end{equation}
in the regime which is described by Eq.(\ref{dB low}), while in the
regime of high shock velocity described by Eq.(\ref{dB high}) it can
be approximated by
\begin{equation}
\begin{aligned}
E_{\rm max} ~ & \approx ~ 3Ze \delta B_{\rm w} \eta^{-1} R_{\rm sh}
\frac{v_{\rm sh}}{c} ~ \approx ~ \frac{Ze}{\eta} \sqrt{\frac{6 \zeta_{\rm K}\zeta_{\rm B} \dot{E}}{c}} \frac{v_{\rm sh}}{c} ~ \approx\\
& \approx ~ \frac{4.5 Z
(\zeta_{\rm K}\zeta_{\rm B})^{1/2}}{\eta} \dot{E}^{1/2}_{35} v_{\rm
sh8}\,\, {\rm TeV}.
\end{aligned}
\label{Emax high}
\end{equation}
Under the approximations described above, the maximal energies of the
accelerated nuclei do not depend on the ambient ISM density but depend
on the bow shock velocity in the ISM rest frame.

\subsection{Propagation of Pulsar Wind-Accelerated e$^{\pm}$ Pairs at a Bow Shock}

The protons accelerated at the bow shock produce magnetic
fluctuations at wavelengths comparable to their gyroradii
in the vicinity of the bow shock. Moreover,
magnetic fluctuations of wavelengths up to a scale-length $L_{\rm
cor}$ which is much longer than the gyroradii of protons of energy
$E_{\rm max}$ can be produced via the inverse turbulent cascading and
the non-resonant CR-driven instabilities \citep[see, e.g.,][and
references therein]{bbmo13}.
These CR-driven fluctuations will
determine the diffusive transport of $e^{\pm}$ pairs
accelerated at the PW termination surface at energies
above 10~GeV in the region between the contact discontinuity and the bow
shock. The strong long-wavelength fluctuations would provide a
Bohm-type diffusion coefficient $D(E) \approx \eta \, D_B(E)$ with $\eta
\sim 1$ for e$^{\pm}$ pairs up to a Lorentz factor $\gamma_1$ determined by
the condition $R_{\rm g0}(mc^2\gamma_1) \approx L_{\rm cor}$, where
$R_{\rm g0}(E) = E/eB_0$ is the particle gyroradius in the mean magnetic
field $B_0$.

Accurate estimations of the scale-length $L_{\rm cor}$ are not available yet
because of the limited dynamical range of numerical simulations of
the CR-driven turbulence. Therefore, we consider $L_{\rm cor}$ as a free
parameter which determines the Lorentz factor $\gamma_1$ where a
transition from the Bohm-like diffusion to the small-scale scattering
regime given by Eq.~(\ref{MFP_s}) takes place.

For particles with gyroradii larger than
the scale-length of the strongly amplified magnetic turbulence
$L_{\rm cor}$, the mean free path produced by the short-scale modes
can be approximated as
\begin{equation} \label{MFP_s}
\Lambda(p) = \frac{4}{\pi}\frac{R_{\rm st}^2(p)}{L_{\rm cor}}
\propto p^2 ,
\end{equation}
where $R_{\rm st}(p) = cp/(e\ \delta B_{\rm w})$. In section
\ref{geomMC} we will apply the diffusion model to simulate
distinctive features of synchrotron images of some BSPWNe.
Apart from the effect on propagation of the $e^{\pm}$ pairs, the MHD
flow between a fast bow shock and the PW termination
shock (a colliding shock flow) may result in a specific way of
acceleration of the highest energy $e^{\pm}$ pairs. The hard energy
spectra of the pairs produced in the colliding shock flow (CSF) can be
responsible for the hard spectra of synchrotron and inverse Compton
photons (of photon indexes \mbox{${\rm \Gamma} \sim$ 1.0--1.5}) recently observed
from the heads of some of BSPWNe in the X-rays and gamma-rays, respectively.



\subsection{Quasi-Transverse Bow-Shocks}

When a shock is highly oblique in respect to the direction of the
local mean magnetic field, the transverse diffusion coefficient
plays an important role in the diffusive shock acceleration.
When the particle mean free path along the magnetic field
$\Lambda\left(p\right)$ is large compared to the gyroradius $R_{\rm g0}$,
the transverse diffusion coefficient $D_{\perp}$ is much smaller than that
along the mean field $D_{\parallel}$:
\begin{equation}
D_{\perp} \approx D_{\parallel}(p) \frac{1}{1 + \zeta^2},
\,~ D_{\parallel}(p)= \frac{v\Lambda(p)}{3},\,
\zeta = \Lambda(p)/R_{\rm g0}.
\label{diff_perp}
\end{equation}
When $\zeta \gg 1$ one obtains $D_{\perp} = [R_{\rm g0}(p)/\Lambda(p)]^2 D_{\parallel}(p)$.
Thus, the orientation of the local magnetic field can significantly affect
the transport of particles across the shock front, and, consequently,
the efficiency of particle acceleration
\citep[see, e.g.,][]{1987ApJ...313..842J,2015ApJ...809...29T}.

The different bow shock obliquities may explain the
substantial differences in the luminosities and observed spectra of
BSPWNe with otherwise similar properties, in particular, the nebulae
associated with PSR~J0437-4715 and PSR~J1741-2054.

\section{Particle Acceleration in Colliding Shock Flows}
\label{CSFs}
\label{sec:PACF}

Diffusive shock acceleration with MFA is rather an
efficient way to produce relativistic particles.
However, an even more efficient realization of the Fermi acceleration
mechanism is acceleration of particles in colliding MHD flows (colliding
shock flows, CSFs)
carrying magnetic inhomogeneities, which may scatter relativistic particles
\citep{2005mshe.work...95B,2013MNRAS.429.2755B}.
This kind of MHD flows is likely realized in a number of astrophysical objects,
in particular, when an expanding shell of a supernova is closely approaching or
interacting with a fast powerful wind of a nearby young massive star.
Converging MHD flows occur in the vicinity of a contact
discontinuity of an astropause, downstream of a fast bow shock
produced by a supersonically moving astrosphere driven by a stellar wind
(the relativistic PW in the case of a PWN).

Consider a simple analytical test particle kinetic model of particle
acceleration by the Fermi mechanism in a CSF
\citep{2005mshe.work...95B}.
Two colliding plane-parallel flows of velocities $u_{\rm 1}$ and
$u_{\rm 2}$ aligned with the positive and negative directions of axis $z$ and occupying
half-spaces $z < 0$ and $z > 0$, respectively, are separated by a discontinuity in
the plane $z = 0$.
The flows carry fluctuating magnetic fields amplified by anisotropic energetic particle
distributions (e.g., the CR current driven instability).
The particles propagate in a diffusive medium described by the fast particle diffusion
coefficients $D_{\rm i}(\gamma)$ which depend on the particle
Lorentz factor $\gamma$.

This simplified model allows us to illustrate the Fermi acceleration in a CSF
of a BSPWN and obtain some important estimates.
The collision of the flow emerging from the PW termination region with that coming
from the bow shock can be viewed within the discussed simple scheme, but
in the vicinity of the contact discontinuity the flows have
a complex multi-dimensional structure. Note that the high CR pressure in this region
may be dynamically important for the flow structure
if the MHD turbulence provides a strong coupling of the CRs with the flow.
However, the highest energy e$^{\pm}$ pairs accelerated in the PW termination region
may have so high Lorentz factors that their mean free paths become larger
than the characteristic size of the multi-dimensional bypassing
flows area in the vicinity of the contact discontinuity.
Let us denote the minimal Lorentz factor corresponding to this regime as $\gamma_0$.


Then the accelerated pair distribution function above $\gamma_0$ in between
the two shocks can be approximated by the solution obtained from the
kinetic equations in the discussed simple plane-parallel model:


\begin{equation}\label{an}
N_i^{\pm}(z,\gamma,t) = \frac{N_0(\gamma_0 ,\gamma_m)
}{\gamma}\cdot H(\gamma-\gamma_0)\cdot H(t-\tau_a(\gamma))\cdot {\rm
exp}\left(-u_{\rm i}\cdot|z|/D_{\rm i}(\gamma)\right),
\end{equation}
where $H(x)$ is the standard Heaviside step function, $\gamma_m$
is the maximal Lorentz factor of an accelerated particle, and
$N_0(\gamma_0, \gamma_m)$ is a normalization constant.
The acceleration time of the highest energy e$^{\pm}$ pair is
\begin{equation}\label{time}
\tau _a ={\frac{3}{\left( {u_1 +u_2} \right)}\int\limits_{\gamma_0
}^{\gamma} \left( {\frac{D_1(\gamma) }{u_1 }+\frac{D_2(\gamma)}{u_2
}} \right)} \frac{d\gamma}{\gamma}.
\end{equation}
We take $u_1 \approx c/3$,  $u_2 \ll c$ in accordance with the typical
flow velocities downstream of the PW termination shock and at the bow shock,
respectively, while
$D_2(\gamma) = \eta c R_g(\gamma)/3$ ($\eta \geq 1$) is
the Bohm diffusion coefficient in the half-space $z>0$.
Here $R_g(\gamma)$ is the particle gyroradius in the magnetic
field amplified by the CR-driven instabilities.
In the limit $u_2 \ll c$ Eq.~(\ref{time}) simplifies to
\begin{equation}\label{time1}
\tau _a \approx \frac{9}{c}  \int\limits_{\gamma_0}^\gamma
\frac{D_2(\gamma) }{u_2 } \frac{d\gamma}{\gamma}.
\end{equation}

As the mean magnetic field behind the termination shock of a PWN is
azimuthal, the radial transport of the pairs at high energies
across the mean field is due to the transverse diffusion. The pair
propagation at the bow shock depends on the shock obliquity. The
transverse shock geometry reduces the radial diffusion coefficients.

\subsection{The Maximal Energies of e$^{\pm}$ Pairs Accelerated in CSFs}
\label{CSFEmax}

The maximal Lorentz factors $\gamma_m$ of the
e$^{\pm}$ pairs accelerated in CSFs are mainly
limited either (i) by particle escape from the accelerator
or (ii) by the particle energy losses.

In the first case, the particle confinement condition in a CSF requires
that the acceleration time Eq.(\ref{time1}) does not exceed the time of
diffusion through the region within the bow shock.
Thus, the maximal Lorentz factor should not exceed $\gamma_{m1}$
satisfying the equation
\begin{equation}
\frac{9D(\gamma_{m1})}{c u_{\rm sh}} = \frac{R_{\rm
cd}^2}{4D(\gamma_{m1})}.
\label{GMaxCFS}
\end{equation}

Therefore, assuming the Bohm diffusion in the accelerator, the upper
limit on $\gamma_{m1}$ can be obtained from
\begin{equation}
R_g(\gamma_{m1}) \approx \frac{1}{2}\sqrt{\frac{u_{\rm sh}}{c}}
R_{\rm cd}.
\label{GMaxCFS1}
\end{equation}
However, the estimations based on the analytic one-dimensional
solutions given by Eqs.~(\ref{an})--(\ref{time}) are rather rough.
Hence in \S\ref{geomMC} we will discuss test particle Monte-Carlo
simulations of the CR acceleration and transport in BSPWNe.

In the second case, the synchrotron-Compton energy loss time of the pairs,
$\tau_s$, should be longer than the acceleration time.
This provides another constraint on the maximal Lorentz
factor $\gamma_m \leq \gamma_{m2}$ given by the equation
$\tau_s(\gamma_{m2}) = \tau_a(\gamma_{m2})$.
Using a simplified approximation for $\tau_s$
which accounts for inverse Compton losses on the cosmic
microwave background photons (as appropriate for very high energy
pairs) and synchrotron losses in the magnetic field $B$:
$$
\tau_s(\gamma) \approx 7\times 10^{19}\,  \gamma^{-1}\, \left[1 +
\left(\frac{B}{3\, \mu G}\right)^2\right]^{-1}~~s,
$$
one obtains
\begin{equation}\label{max}
\gamma_{m2}^2 \approx 4\times 10^{17}\,
\eta^{-1}\,\left(\frac{u_2}{100\kms}\right)\,\,\left(\frac{B}{3\,
\mu G}\right)\,\left[1 + \left(\frac{B}{3\, \mu
G}\right)^2\right]^{-1}\ .
\end{equation}

Then the maximal Lorentz factor $\gamma_{\rm m}$ of the pairs
accelerated in a CSF can be estimated as $\gamma_{\rm m} = {\rm
min}(\gamma_{m1},\gamma_{m2})$.

The spectral upturn at the high energies, namely, the fact that a
large fraction of energy is carried by the particles with the
longest mean free path, can result in an important non-linear
modification of the flow in the region of convergence. Such an
effect was studied in the context of CR-modified shocks by
\citet{2013MNRAS.429.2755B} within a plane-parallel time-dependent
model. In the case of interest here, i.e., for the diffusive
transport and re-acceleration of the high energy end of the spectrum
of the $e^{\pm}$ pairs which were produced at the PW termination
shock, the non-planarity is very important. Therefore, in
\S~\ref{geomMC} we discuss a Monte-Carlo model of acceleration and
propagation of high energy $e^{\pm}$ pairs in the astrosphere
produced by the relativistic PW which moves supersonically through
the ambient medium.

Acceleration in a CSF can provide very fast and efficient creation
of a nonthermal particle population in between the PW termination
region and the bow shock. A very hard energy spectrum of the
e$^{\pm}$ pairs is formed, $\propto \gamma^{-b}$ with $b \sim $1,
which may contain a substantial fraction of the kinetic power
released by the pulsar in the highest energy particles. The
synchrotron X-ray emission from the extended region between the PW
termination shock and the bow shock where acceleration of pairs
occurs would have a hard photon index, 1$\leq {\rm \Gamma} \leq$1.5. The
associated inverse Compton emission may reach the TeV regime and
extend to large distances downstream of the bow shock. Such an
extended structure of hard emission may be treated as a distinctive
feature of BSPWNe.


\section{The Monte-Carlo Particle Acceleration and Transport Model}
\label{geomMC}

The modeling of the structure of PWNe interacting with supersonic
flows either inside or outside of supernova remnants with an
adequate account for the magnetic field structure is a complicated
problem
\citep[see, e.g.,][]{2001ApJ...563..806B,2001A&A...375.1032B,2004A&A...420..937V,2014AN....335..234B}.
The study of particle
transport in these systems faces the complexities related to a
non-trivial behaviour of magnetized flows in the vicinity of the
contact discontinuity. Fortunately, for electrons and positrons
emitting synchrotron radiation in the UV and X-ray range this
problem is alleviated due to their large mean free paths. Along
with the deceleration of the colliding flows and dissipation of
the transported magnetic inhomogeneities it significantly simplifies
the problem. The transport of high energy e$^{\pm}$ pairs is not very
sensitive to the complicated structure of the multidimensional flow
in the vicinity of the contact discontinuity. However, the flow advects
the low energy pairs along the nebula trail. We discuss here
a simplified Monte-Carlo model of the high energy pair transport in BSPWNe.

\begin{figure}
\includegraphics[width=3.0in]{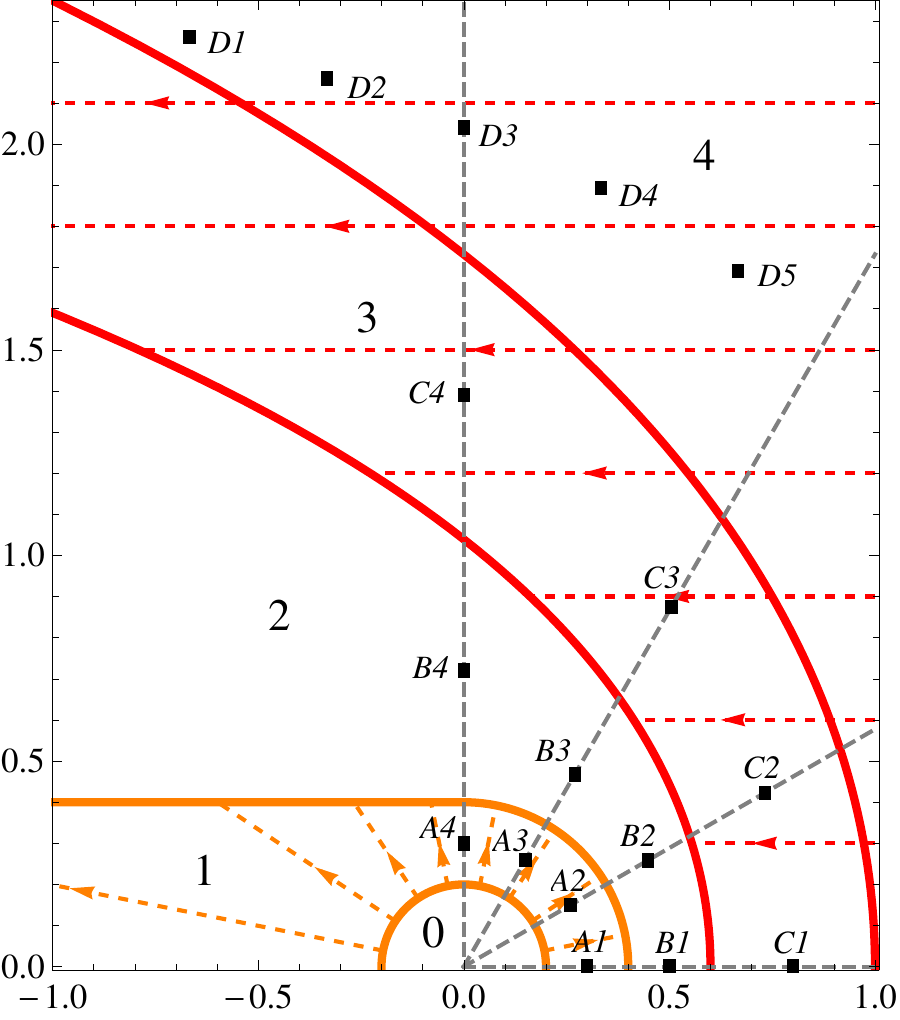}
\caption{The geometry of the model: half of the axial section of the system
by the plane perpendicular to the line of sight, crossing the pulsar (plane $PoS$).
The numbers 0 -- 4 label the regions with different diffusion modes:
0 is the cold PW, 1 (between the orange curves) is the shocked PW, 2 is
the zone near the contact discontinuity, 3 (between the red curves)
is the postshock flow of the ISM matter, 4 is the unperturbed ISM.
The relative sizes of the regions are not to the scale and depend on
the particular properties of a BSPWN. The locations (bins) chosen in
the following sections to
illustrate simulated particle energy distributions are marked by
labelled black squares ($A1-D5$). The bins are located at the midpoints of
the dashed gray segments of the lines connecting the pulsar and the
surfaces, which separate regions 0 -- 4 
(see \S\ref{geomMC:geometry}).
The gray lines are directed at 0$^{\circ}$,
30$^{\circ}$, 60$^{\circ}$, and 90$^{\circ}$ with respect to the
system symmetry axis.}
\label{fig:Geom1} 
\vspace{-1.\baselineskip}
\end{figure}

\subsection{The Model Approach}

Simulation of the relativistic PW particle acceleration and
transport through a PWN interacting with a supersonic flow is
performed via a stationary Monte-Carlo approach.
The e$^{\pm}$ pairs injected into the system at the PW termination
surface are treated as test particles and propagated through the PWN
with a fixed set of diffusion parameters, flow velocities, and
geometry. An important feature of the approach is the account of the
main geometrical properties of the modelled system.

\subsection{The Model Geometry}
\label{geomMC:geometry}

The modelled system is assumed to be axisymmetic and treated as a
composition of axisymmetric regions with spatially homogeneous but
energy-dependent particle diffusion coefficients.
A sketch of the modelled system geometry is presented in Fig.~\ref{fig:Geom1}.
The spherical coordinates $r$, $\theta$, and $\phi$ are centered
at the pulsar position and the zenith direction is aligned with
the pulsar velocity vector. Also the cylindrical
coordinates ($\rho$, $\alpha$, $x$) orientated along the
symmetry axis of the system ($Ox$) are used.

The innermost spherical region with radius $r_{ts}$ corresponds to
the cold pulsar wind. Through its boundary -- the termination shock --
the particles are injected into the system. The region labeled
``1'' corresponds to the shocked pulsar wind (the PWN). Its outer
boundary has the following shape:
\begin{equation}
r\left(\theta\right) = \left\{
\begin{aligned}
& a_{in}\mbox{, if } \theta \leq \pi/2\\
& a_{in} / \sin\theta \mbox{, if } \theta > \pi/2. \label{TS_form}
\end{aligned}
\right.
\end{equation}
The scaling factor $a_{in}$ is a model parameter.
The shocked PW flow velocity is directed radially and its value is
$u_w = \frac{c}{3}r_{ts}^2/r^2$ at distance $r$ from the pulsar. The
region labeled ``3'' corresponds to the postshock flow of the ambient ISM.
Its boundaries are described by the equation for the bow shock shape
derived by \citet{Wilkin} in the thin shell approximation:

\begin{equation}
r\left(\theta, \sigma\right) = \sigma R_{cd} \csc{\theta} \sqrt{ 3
\left(1 - \theta~ \mathrm{\cot} \theta\right)} \label{BS_form}\ .
\end{equation}
Here $R_{\rm cd}$ is the standoff distance defined by Eq.~(\ref{RCD}).
The positions of inner and outer boundaries of region ``3"
are determined by the model parameter $\sigma$.
The apexes of these surfaces are denoted as $a_{\rm out}$ and $a_{\rm sys}$,
respectively. The velocity of the flow
through the bow shock in the pulsar rest frame is fixed and
directed oppositely to the pulsar velocity in both regions ``3" and ``4".
In region ``2", located near the contact discontinuity, advection is neglected.
Region ``4" corresponds to the unperturbed ambient ISM.
The system is bounded by a cylinder whose axis coincides
with the system symmetry axis and whose height is $h_{\rm sys} = x_{\rm FEB} +
a_{\rm sys}$, where $x = - x_{\rm FEB}$ is the coordinate of the
tail-oriented base of the cylinder. The cylinder radius $\rho_{\rm sys}$
matches the radius of section of region ``3" outer boundary by the $x = -
x_{\rm FEB}$ plane. At the cylinder, free escape boundary conditions
are imposed.

\subsection{Particle Propagation}
\label{geomMC:propagation}

In each of the regions ``1''-``4'' a spatially uniform mean value
of magnetic induction $B$ is specified. The parameters of
particle diffusion are chosen as follows.
In the PWN we adopt the mean free path $\lambda_{1}$
proportional to the particle gyroradius $R_g$:
\begin{equation}
\lambda_1\left(B, \eta, \gamma\right) = \eta R_g\left(B,\gamma\right) = \eta
\frac{m_e c^2 \gamma}{eB}\ . \label{lambda_1}
\end{equation}
For the ambient
medium the mean free path is taken in the form:
\begin{equation}
\lambda_{\rm am}\left(\gamma\right) = 3\times 10^{18} \left(\frac{E }{ 1 \mbox{
GeV}}\right) ^{1/3} \mbox{ cm}\ . \label{am_lambda}
\end{equation}
In regions "2"
and "3" the mean free path is chosen in the form:
\begin{equation}
\lambda_{\rm bow}\left(B,\eta,\gamma\right) = \left\{
\begin{aligned}
& \eta R_g\left(B,\gamma\right) \mbox{, if } \gamma \leq \gamma_{1}\\
& \eta R_g\left(B,\gamma\right) \gamma / \gamma_1 \mbox{, if } \gamma_{1} < \gamma \leq \gamma_{2}\\
& \lambda_{\rm am} \mbox{, if } \gamma > \gamma_2
\end{aligned}
\right. \ .\label{lambda_br}
\end{equation}

In Eqs.~(\ref{lambda_1})--(\ref{lambda_br}) $e$ and $m_e$ are the
charge and mass of an electron or a positron, and $E = m_e c^2 \gamma$ is
its energy. The values of $\eta \geq 1$ and $\gamma_1$ are
free parameters of the scattering model, while $\gamma_2$ is derived
from the continuity of the mean free path value.

At the beginning of a simulation the initial spectrum of particles,
with a given shape, is generated. Then, in the course of Monte-Carlo
modeling of particle acceleration and transport, the generated
particles are injected into the system one by one. In the developed
numerical code the injection is implemented as follows. Two random
numbers are generated for every given particle to define its initial
position on the termination shock. Two more are generated to specify
the initial velocity direction.
The particle then propagates for a distance equal to its mean free path
in region ``1'' and after that is scattered for the first time.

The scatterings of the particles are isotropic in the local plasma
rest frame. At each scattering the value of particle momentum in the
local plasma rest frame is calculated. Then two random numbers which
define two scattering angles are generated and the particle momentum
is rotated according to the paper of \citet{Ostrowski1991}. After
that the new particle momentum in the pulsar rest frame is
calculated. The value of $\delta\Omega_{\rm max}$, introduced by
\citet{Ostrowski1991} is taken equal to $\pi$ (due to that, $\eta$
actually should not be less than $2$).

After injection, the particle is propagated through the system.
The particle moves along a straight line during the
time interval $dt_{\rm mfp} = \Lambda / v$ (where $\Lambda$ is its mean free path
at the current location, and $v \approx c$ is its velocity),
and then is scattered. This cycle repeats until the particle finally
crosses the free escape boundary, and, therefore, leaves the system.
Then the next particle is injected and its propagation starts.
When the particle crosses the border between two regions
with different scattering parameters, the crossing point
is determined and the particle is moved from this point
in the same direction for a distance equal to
$\lambda_{\rm new}(1-\Delta x / \lambda_{\rm old})$, where $\lambda_{\rm old}$
is the mean free path in the region it has left, $\lambda_{\rm new}$
is the mean free path in the region it has entered and $\Delta x$
is the path of the particle in the region it has left after the last scattering.
If the particle enters the cold PW region, it is reflected
from its boundary in the case when its gyroradius in the
magnetic field downstream of the PW is smaller than $r_{ts}$
and crosses it without scattering otherwise.

The phase space of the system is divided into bins -- small volumes
of phase space, which are used to 'detect' the particle spatial
and momentum distribution. The coordinate space of the system is
permeated by a cylindrical 'grid' of bins.
The cylinder containing the
system is splitted into $N_x$ equal cylinders along the $Ox$ axis.
Each of these smaller cylinders is divided into $N_{\rho}$ coaxial
cylindrical layers, whose outer radii grow linearly as $\rho_n =
n\rho_{\rm sys}/N_{\rho}$, where $n = 1,2, \dots, N_{\rho}$.
Each cylindrical layer is divided into $N_{\alpha}$ equal segments by
semi-planes $\alpha = 2\pi k / N_{\alpha}$, $k = 1, \dots, N_{\alpha}$.
To simulate the particle energy distribution, recording of the
absolute values of
particle momenta at each detector is required.
In order to do this, we introduce a momentum range
which is somewhat wider than the range of particle momenta in the initial
(injected) particle distribution. This 'detector' range is
logarithmically binned; hence $N_m$ momentum bins are introduced.
In case of considerable anisotropy of particle energy distribution function
detection of particle angular distribution is also of some interest. Thus, the ranges of
$\mu=\cos\theta$ and $\phi$ values are splitted into $N_{\mu}$ and $N_{\phi}$ bins, respectively.
The
particle is detected by the bin with a given set of numbers ($n_x$,
$n_{\rho}$, $n_{\alpha}$, $n_m$, $n_{\mu}$, $n_{\phi}$) if its energy corresponds to the $n_m$-th
momentum bin, direction of velocity to $n_{\mu}$-th and $n_{\phi}$-th angular bins,
and it crosses the left boundary of the spatial bin
($n_x$, $n_{\rho}$, $n_{\alpha}$) corresponding to
$x = -x_{\rm FEB} + (n_x - 1) h_{\rm sys} / N_x$.
To obtain the particle distribution function from the
'detection' data, we employ the formulae analogous to those presented in
\S 3.1.4 of \citep{Vladimirov2009d}. In order
to compensate for the growth of the bin size with the increasing
distance from the axis, the values measured by the detectors are divided
by the ratio of the current bin volume to the volume of the bin at $n = $1.

To take into account the effect of energy losses due to synchrotron
radiation, after each straight line piece of trajectory, the
particle momentum value is re-evaluated as $p_2 = p_1 / (1 + p_1 S dt)$.
Here $p_1$ is the momentum in the local plasma rest frame at the
beginning of the current piece of trajectory, $p_2$ is the new momentum
value in the same frame, $dt$ is the propagation time, and $S$ is
the synchrotron loss coefficient: $${dp\over dt} = - S p^2 = - \frac{4}{9}\frac{e^4
B^2}{m^4c^6} p^2.$$ Since for all the explored sets of model parameters
the particles do not suffer strong energy losses over a diffusion length,
the particle energy is effectively decreased due to these losses
only after the particle has crossed all the 'detectors' along a mean free path.
If a particle crosses the boundary between regions with different
values of $B$, the decrease of its energy is calculated consequently
for pieces of trajectory in each region.

\subsection{The Model Spectra}
\label{geomMC:spectra}

The Monte-Carlo procedure described in \S\ref{geomMC:propagation}
allows us to simulate particle spectra at certain locations inside
the BSPWN system as well as spectra of synchrotron emission radiated
by the accelerated particles. The intensity of synchrotron emission
at various photon energies can be calculated for the lines of sight
crossing these locations, thus allowing us to simulate emission maps
in the sky plane.
The locations chosen to present the particle spectrum modeling results
are specified as follows.
We choose a certain axial section plane of the axisymmetric BSPWN system.
A right-handed Cartesian coordinate system $(xyz)$ centered at the
point $P$, the pulsar position, is introduced.
The $x$ axis matches the symmetry axis and the plane $Pxz$ matches
the fixed axial section.
We choose four groups of points in the $Pxz$ plane. Namely, we take
four rays starting at $P$, lying in $z \geq 0$ and oriented
at angles 0, 30, 60 and 90 degrees with respect to the symmetry axis.
Along each ray we select three locations, each in the center of the
interval delimited by the intersections of the ray with the surfaces
separating regions with different diffusion properties.
Thus, we select 3$\times$4 = 12 points in regions ``1''--``3''.
We include five additional points in region ``4''.
Five rays parallel to the $Pz$ axis and lying in the
$Pxz$ plane ($z \geq 0$) are chosen at $x$ coordinates
$x_j = - x_{\rm FEB} + j \left(a_{sys} + x_{\rm FEB}\right) / 6,~~ j = 1..5$.
The locations are taken at the centers of the intervals between
the points of intersection of these rays with the outer boundary
of region ``3'' and with the outer boundary of the whole system.

The orientation of the considered system with respect to the
direction to the observer is defined as follows. We introduce an
axis $Y$, directed from the observer towards the pulsar (point $P_0$)
and the plane of the sky (PoS), orthogonal to
this axis and containing $P_0$. The orientation of the source is
determined by three angles $\Phi$, $\Theta$, and $\Psi$. $\Phi$
is the angle between the $Py$ axis and the line $l_{Y}$ of
intersection of the planes $Pzy$ and PoS (chosen in the positive direction:
counterclockwise when looking at the $Pzy$ plane against
the $Px$ axis direction). $\Theta$ is the angle between $Y$ and $Px$ axes.
$\Psi$ is the angle between $l_{Y}$ and the North direction in the PoS.
Thus, one may introduce a coordinate system $(XYZ)$ centered at $P_0$, $X$ and
$Z$ axes in the PoS, where the $Z$ axis is aligned with the direction to the
North. Then the angles $\Phi$, $\Theta$, and $\Psi$ would be the three
Euler angles defining the orientation of the $XYZ$ system with respect
to the $xyz$ system. In the following and in Fig.~\ref{fig:Geom1} we use $\Theta = 90^{\circ}$.


\section{Observations and Modeling
of X-ray and Far-UV Emission from BSPWNe }\label{Obs}

Multi-wavelength observations in the radio, optical and X-ray bands
have revealed the presence of bow-shock type structures in a number of PWNe
\citep[see, e.g.,][]{2004MSAIS...5..195P,2006ARA&A..44...17G}.

An example of an apparent bow shock PWN (BSPWN) is the Mouse
nebula powered by the radio pulsar J1747-2958, which is moving
with transverse velocity (306 $\pm$ 43)$\times ({d / 5~{\rm kpc}}) \kms$
\citep{2004ApJ...616..383G,2009ApJ...706.1316H}.
The spin-down power of J1747-2958 is about 2.5$\times 10^{36}\ergs$,
while the X-ray luminosity of the nebula
in the 0.5-8 keV range is about 5$\times 10^{34}\ergs$,
and the observed PWN structure shows a narrow 45$\arcsec$-long X-ray tail
(the length is $\sim$~1.1~pc at the 5 kpc distance).
The best-fit power
law photon index of the X-ray spectrum ${\rm \Gamma}$ = $2.0 \pm 0.3$ was
derived from {\sl Chandra} observations by \citet{2004ApJ...616..383G},
who also noticed some steepening of the spectrum from the halo
to the tail of the PWN.

High resolution observations of PWNe in H$_{\alpha}$ \citep[see,
e.g.,][]{1988Natur.335..801K,1993Natur.362..133C} have provided
images of cometary-like bow shock structures ahead of fast moving pulsars.
A recent survey of such bow shocks \citep[][]{2014ApJ...784..154B}
listed 9 pulsars with resolved H$_{\alpha}$ emission at the apex
and a number of pulsars with upper limits set on their H$_{\alpha}$ flux.

\subsection{Synchrotron Spectra and Images of the Far-UV Bow Shock
of PSR J0437-4715}\label{J0437}

One of the most interesting and best studied objects in the
compilation of \citet[][]{2014ApJ...784..154B} is the bow shock of
the nearest millisecond pulsar J0437-4715. The pulsar is located at
a reliably estimated distance of about 160~pc in a binary system with
measured parallax and transverse velocity $\sim$ 104$\kms$ derived
from the motion of the pulsar companion and the bow shock apex
measured over 17 yrs. The estimated spin-down power of J0437-4715
is $\dot{E}~\sim$~6$\times $10$^{33}\ergs$.

Extended X-ray emission from the $\sim 5\arcsec$ vicinity of J0437-4715
was detected with {\sl Chandra} by \citet{2016ApJ...831..129R}.
This emission indicates the presence of a faint ($L \sim$ 3$\times $10$^{28} \ergs$)
PWN produced by J0437-4715. The X-ray photon index of the PWN is
${\rm \Gamma}=1.8\pm0.4$.

Far-ultraviolet (FUV) imaging of the vicinity of J0437-4715 performed
by \citet{2016ApJ...831..129R} with the {\sl Hubble Space Telescope}
has revealed a bow-like structure which stands 10$\arcsec$ ahead of the pulsar
and is coincident with the apex part of the H$_{\alpha}$ bow shock
earlier observed by \citet[][]{2014ApJ...784..154B}.
The observed 1250--2000 \AA\ luminosity of the FUV bow
is about 5$\times 10^{28} \ergs$ -- an order of magnitude higher
than the H$_{\alpha}$ luminosity of the same structure.
The FUV observations also revealed a likely flux enhancement
in the extended (about 3$\arcsec$) region at the limb of the bow,
which is not seen in the wide-band optical/near-UV images.

Both H$_{\alpha}$ and FUV emission at the bow of J0437-4715
can be associated with line and continuum radiation of
shocked interstellar plasma.
In particular, even a simple one-dimensional model of gas flowing
through a fast enough interstellar shock \citep[see, e.g.,][]{BykovEtal2013}
can be used to explain the fluxes of broadband FUV and H$_{\alpha}$ emission
observed by \citet{2016ApJ...831..129R}. Within such a model
the 1250--2000~\AA\ band emission is generated in the hot downstream
of the shock and dominated by several individual lines such as C~IV 1549~\AA,
O~IV 1403~\AA, Si~IV 1397~\AA, C~II 1335~\AA, and He~II 1640~\AA\
rather than by continuum UV radiation.

Alternatively, synchrotron radiation of
accelerated electrons and positrons in the magnetic field amplified
at the bow shock could also provide the observed optical and FUV fluxes.
As 
illustrated
in Fig.~\ref{fits_UV_X}, even the observed morphology
of the FUV and X-ray emission around J0437-4715 can be understood
within such an interpretation.

Modeling of broadband synchrotron emission from PWNe with a single
power law particle injection spectrum fails to reproduce the observed spectral
energy distributions
\citep[e.g.,][]{1984ApJ...278..630R,1996MNRAS.278..525A,2000A&A...359.1107A}.
Hence, a spectral break is inferred from the data of radio and X-ray observations
of young PWNe in supernova remnants \citep[see, e.g.,][]{2009ApJ...703.2051G, 2011MNRAS.410..381B}.
Namely, \citet{2011MNRAS.410..381B} found that injection of a particle spectrum
in the form of a broken power law with $3\times 10^4\lsim \gamma\lsim 10^6$ and
high energy injection index $\geq$ 2.14  can yield a satisfactory
description of the available observational data for a number of PWNe.
The maximal Lorentz factors of e$^{\pm}$ pairs estimated from the voltage
of the total magnetospheric potential of a pulsar
are $\sim 10^{8}$ for the spin-down power $\dot{E} \sim$ 6$\times $10$^{33}\ergs$.

\begin{figure}
\includegraphics[width=5.0in]{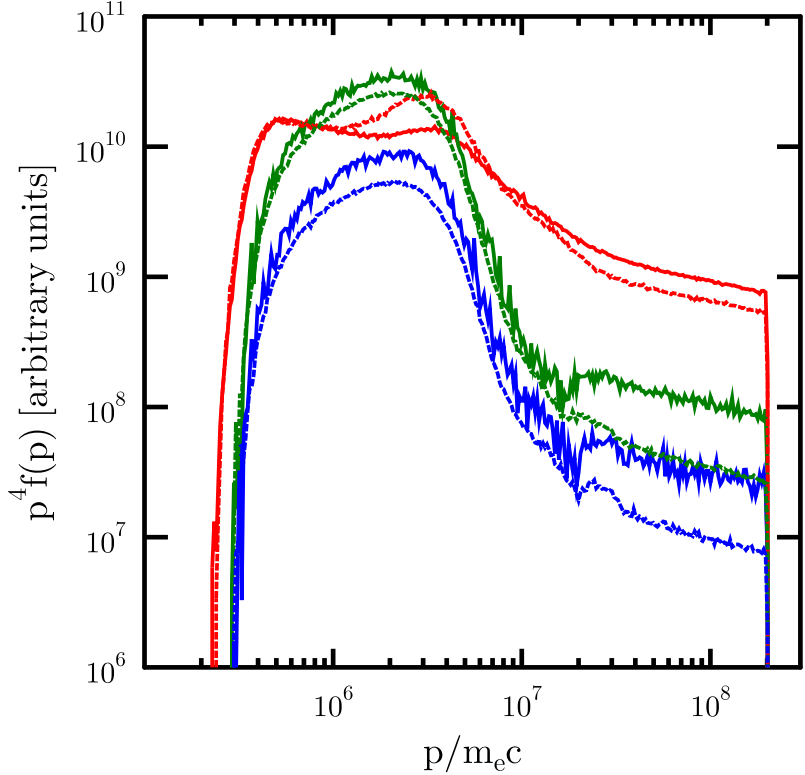}
\caption{Non-smoothed Monte-Carlo simulated local spectra of e$^{\pm}$ pairs in
different zones of a BSPWN (see \S\ref{geomMC}) for parameters
similar to those of the source associated with PSR~J0437-4715.
The red curves correspond to the PW termination shock downstream:
the dashed one refers to the $A2$ bin, the solid one refers to
the $A4$ bin viewed along the ray directed to the observer.
The blue curves correspond to the bow shock region (bin $C1$ -- the
solid curve, bin $C3$ -- the dashed curve). The green curves
correspond to the region between the two shocks (bin $B1$ -- the solid
curve, bin $B3$ -- the dashed curve).
The locations of all the bins are shown in Fig.~\ref{fig:Geom1}.
Worse statistics for bins $B1$ and $C1$ is due to the
smaller squares of 'detectors' near the system
symmetry axis (see \S\ref{geomMC:propagation}).
}
\label{Spectra J0437P}
\end{figure}

\begin{figure}[h!]
\includegraphics[width=\linewidth]{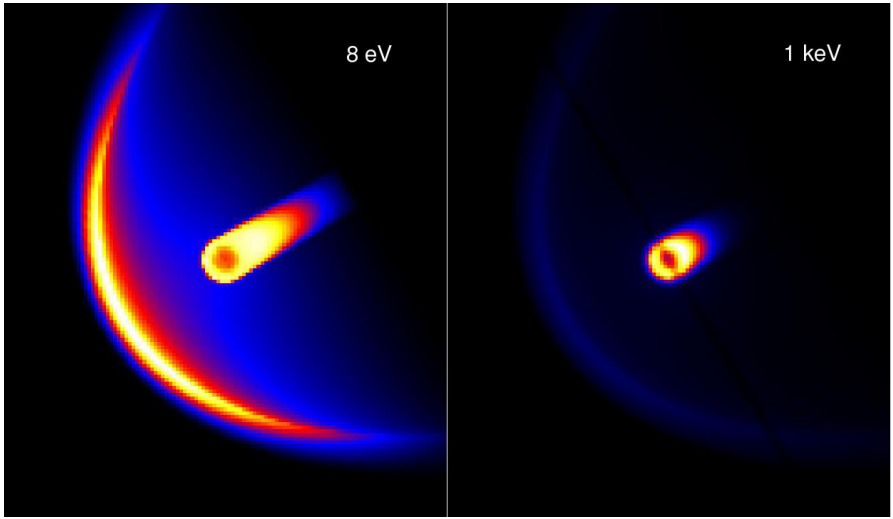}
\hfill \hspace{-1cm}
\caption{
Simulated synchrotron images of an object similar to J0437-4715,
for $E = $8~eV ($\approx$ 1550 \AA, left) and 1 keV (right). The absolute intensity
is given in normalized units to show the contrast change through
the image.
}
\label{fits_UV_X}
\end{figure}

To model the observed optical, FUV and X-ray emission of the J0437-4715, 
a power law spectrum of e$^{\pm}$ pairs is injected at the
PW termination surface with the index 2.2--2.3 (for the Lorentz
factors $\gamma_0 \leq \gamma \leq \gamma_{\rm m}$). This corresponds to
the common models of particle acceleration at relativistic shocks
\citep[see, e.g.,][]{Achterberg2001,2004APh....22..323E,2005PhRvL..94k1102K,
PLM2009,2012SSRv..173..309B}. While it is not clear that the standard Fermi
acceleration might be effective at the PW termination shock, unless highly
efficient dissipation of the PW magnetic field occurs before (or at) the termination
shock surface, it is true, nevertheless, that this spectral slope is what is inferred from
X-ray observations for the majority of PWNe. The shape of the particle spectrum below
$\gamma_0$ does not need to be specified in the present modeling.
The maximal Lorentz factor of the e$^{\pm}$
pairs accelerated at the PW termination surface is likely
scaled with the pulsar spin-down power as $\gamma_{\rm m} \propto \dot{E}^{0.5}$
(as the size of the termination shock)  and thus may reach (3--5)$\times 10^7$
for J0437-4715, i.e., it is just below the value achievable
in the magnetospheric potential gap.
Pair (re-)acceleration in the colliding shock flow (CSF) of
the PW termination shock and the bow shock would result
in substantial spectral hardening up to the maximal energy given by Eq.~(\ref{GMaxCFS1}).
For PSR~J0437-4715, with $u_{\rm sh} \sim 100 \kms$ and $R_{\rm cd}
\sim 2\times 10^{16}$ cm, from the simplified estimation of Eq.~(\ref{GMaxCFS1}) one can
obtain $\gamma_{m1} \approx 3 \times 10^6$. Indeed, the modeled spectra of e$^{\pm}$
pairs re-accelerated in the CSF demonstrate a break at
$\gamma \sim \left(2\mbox{--}3\right) \times 10^6$
in all the three considered spatial zones (Fig.~\ref{Spectra J0437P}).

In Fig.~\ref{Spectra J0437P} the accelerated e$^{\pm}$ pair
spectra in different zones of a BSPWN for parameters similar to those
of the source associated with PSR~J0437-4715
are shown.
They have been simulated within a Monte-Carlo model, where
a power law spectrum $\propto \gamma^{-s}$ with $s$ = 2.3 upto
$\gamma_{\rm max}$ = 2$\times$10$^8$ (which is $\lsim \sqrt{3\dot{E}/2c}$)
is injected in the immediate downstream of the PW termination surface.
As described in \S\ref{geomMC}, the injected pairs propagate through
the colliding flow where the efficient re-acceleration reshapes the initial
spectrum.
%
The spectra are presented for a few bins, whose locations are shown in Fig.~\ref{fig:Geom1}.
Namely, the red curves correspond to the PW termination shock downstream:
the dashed one refers to the $A2$ bin, while the solid one refers to
the bin located at the same distance from the pulsar as the $A4$ bin, but lying on the
ray directed from the pulsar to observer.
The blue curves correspond to the bow shock region (bin $C1$ -- the
solid curve, bin $C3$ -- the dashed curve). The green curves
correspond to the region between the two shocks (bin $B1$ -- the solid
curve, bin $B3$ -- the dashed curve).
Worse statistics for bins $B1$ and $C1$ is related to the
smaller squares of 'detectors' near the system
symmetry axis (see \S\ref{geomMC:propagation}).

To confront the modeling with the observational data, one needs to
construct images and spectra of synchrotron emission produced by
the modelled distributions of accelerated e$^{\pm}$.
The intensity of the synchrotron radiation averaged over the magnetic field orientations
was obtained using the approximations given by \citet{2007A&A...465..695Z}.
The modifications of the synchrotron emission spectra related to the Doppler beaming were taken into account.

Simulated synchrotron images of an object similar to J0437-4715
at $E = $ 8 eV and 1 keV are shown in Fig.~\ref{fits_UV_X}.
On the left panel of Fig.~\ref{fits_UV_X} a bright bow-like structure
can be seen, which corresponds to the inner part of region ``3'' in
Fig.~\ref{fig:Geom1}. A very faint counterpart of the bow is hardly
seen in this zone on the right panel, where the 1 keV emission is shown.
Some details of the simulated source morphology can be affected by varying the
parameters of modeling, such as the pulsar axes inclination angles and others, over the allowed parameter space.
However, the presence of the bright bow-like structure in the FUV regime and its faint appearance in the X-rays
remains.

\begin{figure}
\includegraphics[width=5.0in]{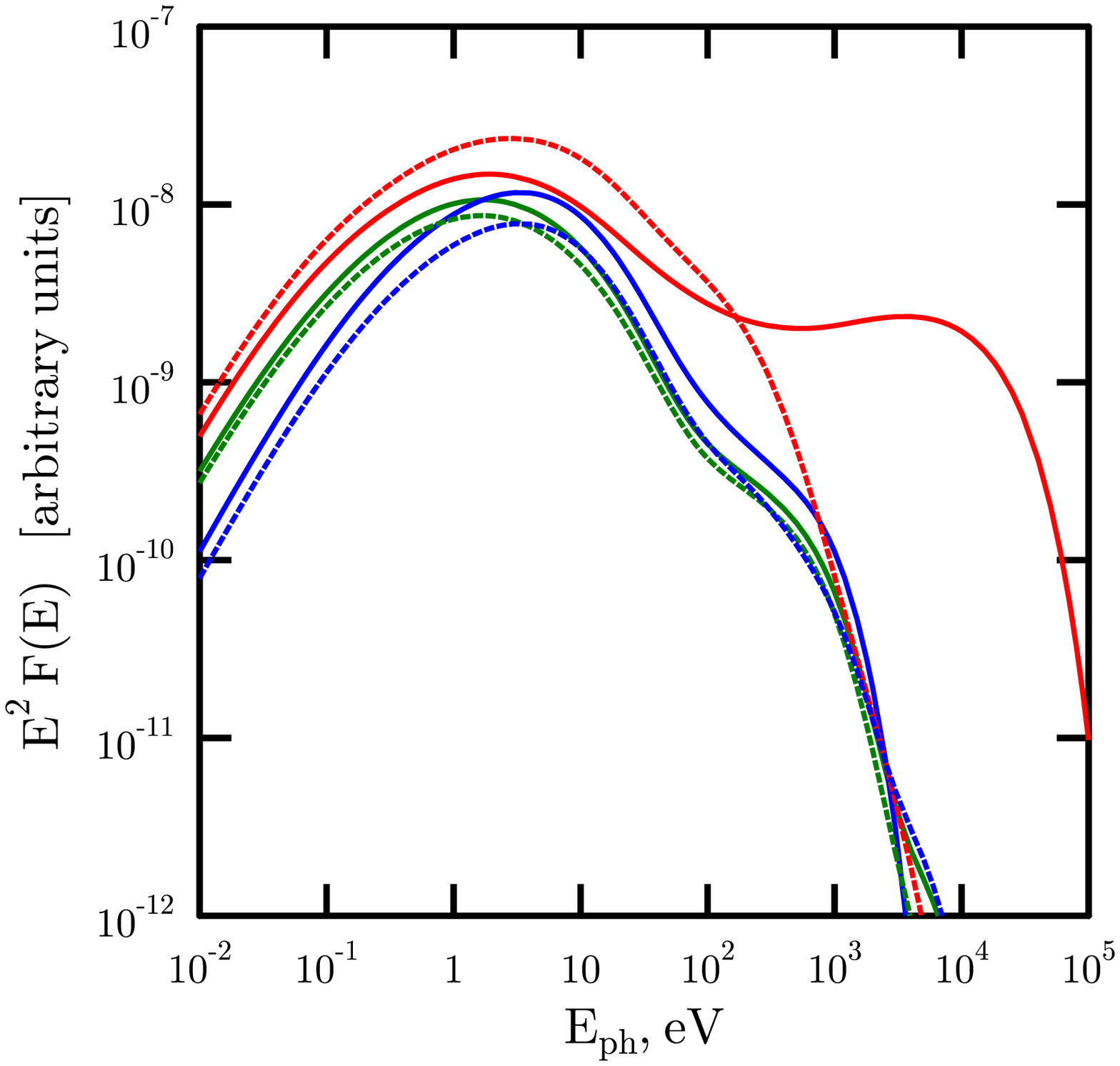}
\caption{Simulated synchrotron spectra of a source similar to
J0437-4715. The orientation of the source is given by the angles
$\Phi = 90^{\circ}$, $\Theta = 90^{\circ}$, and $\Psi =
149.5^{\circ}$ (see \S\ref{geomMC:spectra}), i.e., the pulsar proper
velocity vector lies in the plane of sky perpendicular to the line
of sight crossing the pulsar (plane $PoS$) and the position angle
$\theta_{pos} = 270^{\circ} - \Psi = 120.5^{\circ}$. The curves show
the spectral energy distribution ($\nu F_{\nu}$) calculated by
integration of the spectral emissivity along the lines of sight,
which cross the bins, indicated in the caption to Fig.~\ref{Spectra
J0437P}. The colors and styles of the curves are the same as in
Fig.~\ref{Spectra J0437P}.
}
\label{Spectra J0437R}
\end{figure}

The spectra of synchrotron emission produced by the re-accelerated
pairs in the modelled BSPWN are shown in Fig.~\ref{Spectra J0437R}.
The simulated synchrotron spectra are shown for a source similar to
J0437-4715. The orientation of the source is given by angles $\Phi =
90^{\circ}$, $\Theta = 90^{\circ}$ and $\Psi = 149.5^{\circ}$ (see
\S\ref{geomMC:spectra}), i.e. the pulsar proper velocity vector lies
in the plane of sky perpendicular to the line of sight crossing the
pulsar (plane $PoS$) and the position angle $\theta_{pos} =
270^{\circ} - \Psi = 120.5^{\circ}$. The curves show the spectral
energy distribution ($\nu F_{\nu}$) calculated by integration of the
spectral emissivity along the lines of sight crossing the bins which
were chosen to show the local particle spectra. The colors and
styles of the curves are the same as in Fig.~\ref{Spectra J0437P}.

The ratio of the modelled FUV luminosity in the bow-like structure
to the modelled X-ray luminosity in the vicinity of the TS is about 2.
The modelled luminosity of the J0437-4715 bow in the H$_{\alpha}$
band is lower than the measured value, so most of the observed
H$_{\alpha}$ emission in the WIYN filter W012
\citep{2014ApJ...784..154B} cannot be attribited to the synchrotron
radiation. In general, the simulated synchrotron spectra are
consistent with the data of multiband observations of J0437-4715
mentioned above.

The extended (about 3$\arcsec$) FUV flux enhancement (dubbed "blob" )
which is apparent at the limb of the bow, but not seen in the
wide-band optical/near-UV images of \citet{2016ApJ...831..129R}, can be
explained in the frame of the synchrotron model. The optical-UV continuum
spectrum is steep at the bow as it is clearly seen in  Fig.~\ref{Spectra J0437R}.
This means that even relatively modest fluctuations of the
magnetic field may produce a sharp structure at higher photon energies, which
is not prominent at lower energies. The intermittent clumpy
images are a characteristic feature of the systems with steep
synchrotron spectra, as it was shown by \citet[][]{2008ApJ...689L.133B}.

Another case of a bow shock driven by a supersonically moving
pulsar is the nebula of the gamma-ray pulsar J1741-2054, whose
estimated velocity is $\sim 150\kms$ (assuming the uncertain
distance of about 400 pc). The H$_{\alpha}$ bow shock standing
at $1\farcs$5  from the pulsar position was reported
by \citet{2010ApJ...724..908R}. The pulsar spin-down power
is about 9.5$\times 10^{33}\ergs$. Deep {\sl Chandra}
observations of PSR J1741-2054 performed by \citet{2015ApJ...802...68A}
have revealed a PWN with an extended ($\sim 2 \arcmin$) trail, whose
power law photon index is ${\rm \Gamma}$ = 1.67$\pm$0.06 and
the total 0.5--10~keV luminosity is about 2.4$\times$ 10$^{30}\ergs$.
The trail shows no for synchrotron cooling. Note that the total
X-ray luminosity of PSR~J1741-2054 is two orders of magnitude higher
than that of PSR~J0437-4715, though their estimated spin-down powers
and proper velocities are comparable.
The measured X-ray photon index
of J1741-2054 is consistent with the model discussed above,
which predicts harder synchrotron spectra in the optical-UV regime followed
by asymptotical softening towards the X-ray band.
The shape of the simulated energy distribution of the high-energy X-ray emitting particles
reproduces the shape of the spectra of e$^{\pm}$ pairs accelerated
at a relativistic termination shock. For the power law spectra
$\propto \gamma^{-s},$ where $s \sim 2.2$ it gives the photon
indices close to 1.6.
As the spin-down powers and velocities of the two PSRs are similar,
the break energy between the optical-UV component with
the hard spectrum produced in the CSF between the
termination and bow shocks and that produced at the termination
shock of PSR~J1741-2054 is expected to be at a few eV.
Therefore, the FUV synchrotron luminosity of the bow shock of
PSR~J1741-2054 is expected to be about a few times 10$^{28} \ergs$.

\section{Extended X-ray Structures Associated with Bow Shock Pulsars}
\label{extended}

High resolution observations with the {\sl Chandra} X-ray observatory
\citep{2014RPPh...77f6902T} have revealed a great diversity of
X-ray appearances of PWNe \citep{2006ARA&A..44...17G,KP08}. For the
subset of PWNe with bow shocks (BSPWNe), born due to fast supersonic
motion of pulsars, the reasons for diversity are even more numerous
than for the rest of the class, as in addition to the pulsar and PW
properties (such as pulsar obliquity and wind magnetization), which determine the
wind anisotropy \citep[e.g.,][]{2016MNRAS.462.2762B},  the angle between the
star proper velocity and its rotation axis as well as the direction
of the local interstellar magnetic field are also important, and greatly
extend the parameter space.
Some particular cases and configurations will be discussed below.

Peculiar extended X-ray structures associated
with PWNe --- the ``Guitar nebula'' and the ``Lighthouse nebula'' ---
were observed around two apparently fast supersonically moving pulsars PSR 2224+65
\citep[see, e.g.,][]{1997ApJ...484L.137R,2007A&A...467.1209H,2010MNRAS.408.1216J,2012ApJ...747...74H}
and IGR J11014-6103
\citep[see, e.g.,][]{2011A&A...533A..74P,2014A&A...562A.122P,2016A&A...591A..91P}.
Also, {\sl Chandra} observations revealed a highly extended X-ray
feature likely associated with the nearby $\gamma$-ray pulsar PSR~J0357+3205
\citep[see, e.g.,][]{2013ApJ...765L..19D}, whose space velocity is estimated
to be about 390$\kms$ (assuming the $\sim$~500~pc distance)
and is well aligned with the X-ray feature.

The observed structures of the Guitar and Lighthouse nebulae,
as well as the appearance of extended X-ray structures around
the Geminga pulsar may be understood within the two jets
scenario combined with re-acceleration of the hard-spectrum
e$^{\pm}$ component in colliding shock flows (CSFs).

\subsection{The Geminga PWN}\label{Gem}

A bright nearby gamma-ray pulsar, PSR~J0633+1746 (``Geminga'') ,
is located at the estimated distance of 250$^{+230}_{-80}$pc
\citep[][]{2012ApJ...755...39V}.
Its age is about 340~kyr and its spin-down power is about 3.3$\times
10^{34}\ergs$. The estimated transverse velocity of Geminga
is about 211$\kms$, which is supersonic for the most widespread
ISM phases, and thus it should make a BSPWN.
Pulsed X-ray emission up to 20 keV was recently
detected from PSR~J0633+1746 with the {\sl Nuclear Spectroscopic Telescope Array}
(NuSTAR)  by \citet{2014ApJ...793...88M}. These authors
found that two models can fit the observed spectrum.
One of the models consists of a power law component with
photon index about 1.7 and two blackbody components
of temperatures $kT \sim$ 44~eV and 195~eV. An equally
good model consists of a blackbody component of $kT \sim$ 42
eV and a broken power law with photon indexes ${\rm \Gamma} \sim$ 2.0 and
1.4 with a break at 3.4 keV. Much harder photon indexes are needed to
model the X-ray spectrum of the Geminga PWN.

With a deep {\sl X-ray Multi-Mirror Mission-Newton (XMM-Newton)}
observation of the PWN, \citet{2003Sci...301.1345C} revealed an extended
structure with two $\sim~2'$ elongated, nearly parallel, X-ray tails
trailing behind the fast moving pulsar PSR~J0633+1746.
However, no H$_{\alpha}$ emission from the nebula was found
in a 5-hour exposure of the ESO VLT/FORS1 spectrograph.
In addition to the two lateral $\sim$~0.2~pc long tails,
an axial tail of $\sim$~0.05~pc length resolved into a few individual segments
was found in deep {\sl Chandra} observations by \citet{2017ApJ...835...66P}.
The X-ray spectra of the lateral tails, which can be modelled with power law
photon indexes ${\rm \Gamma} \approx$ 1, are much harder than that of the axial tail,
where ${\rm \Gamma} \approx$~1.6. The lateral tails are connected to the pulsar,
\citet{2017ApJ...835...66P} found some indications of apparent
motions of their footpoints.

The Milagro collaboration \citep{2009ApJ...700L.127A} reported
a detection of a several degrees size TeV emission region in the
vicinity of the Geminga PWN. Recently the {\sl High Altitude Water
Cherenkov Observatory (HAWC)} confirmed the TeV source to be about 2
degrees large and found that its TeV spectral index ${\rm \Gamma}
\sim$2 \citep{2017arXiv170202992A}. Contribution of TeV range
positrons from the Geminga pulsar is considered as a potential
source of the excess over the standard predictions of secondary
production in the interstellar matter \citep[see,
e.g.,][]{2017arXiv170208436H}, which will be discussed in detail in
section \ref{sec:posintro}.

The observed lack of H$_{\alpha}$ emission from the PWN may indicate
that PSR~J0633+1746 is moving through a rarefied ambient medium:
an analysis of the geometry of the PWN tails allowed \citet{2017ApJ...835...66P}
to conclude that the ambient number density around the nebula is below 0.01$\cmc$.
For the bow shock moving at about 210$\kms$ such an extremely low density would indicate
that the surrounding ISM is hot and thus the Mach number of the shock is moderate.
The shock velocity is somewhat higher than that of PSR~J0437-4715 and, more
importantly, the estimated spin-down power of PSR~J0633+1746 is
about an order of magnitude higher than that of PSR~J0437-4715.
Hence, the hard spectrum of e$^{\pm}$ pairs re-accelerated in the colliding
shock flow between the bow shock and the PSR wind termination shock of Geminga
can be extended to sufficiently high energy to produce synchrotron
emission in the X-ray band (according to the maximal energies
of the e$^{\pm}$ pairs estimated in \S\ref{CSFEmax}).
The X-ray emitting accelerated pairs may flow away through the
two magnetic jets and form the apparent lateral tails bent behind
due to the pulsar proper motion. Within such a scenario the axial tail may
be filled with the pairs accelerated at the PW termination
shock and therefore their emission spectrum would have a steeper
spectrum, possibly with a photon index
close to the observed ${\rm \Gamma} \approx$ 1.6.

An important constraint for the colliding flow model
is the spatial distribution of spectral hardness
of the observed X-ray emission. The spectra of the ``ring" region \citep{2017ApJ...835...66P}
and of the axial tail are softer than those of the lateral tails.
The observed photon indexes of the lateral tails (${\rm \Gamma} \sim$ 1)
are substantially harder than those from the other regions of the PWN
(${\rm \Gamma} >$ 1.5) as well as that of the pulsar. Assuming
a synchrotron origin of the hard emission, one may
locate the sources of radiating electrons in the region
where the relativistic outflow behind the termination shock is
colliding with the bow shock downstream flow. In this region the
multi-TeV electrons emitting X-rays can have mean free paths long
enough to be accelerated via the Fermi mechanism between the
converging flows. This would result in the spectrum of particles
$\propto \gamma^{-s}$ with $s \gsim$ 1.

If the pulsar velocity is close to the normal to the ``wind plane",
the model of particle acceleration in the colliding flows of the
anisotropic wind and of the bow shock downstream could explain the
hard X-ray spectral indexes in the lateral tails of the Geminga PWN:
here the accelerated electrons are advected through the region of
the contact discontinuity between the shocked PW and the
interstellar plasma. Numerous simulations of planetary and
heliospheric bow shocks predict the enhanced plasma density and
magnetic field magnitude in the vicinity of the extended contact
discontinuity. The accelerated X-ray radiating electrons would
produce the ``two tails'' configuration as a projection of the
rotation figure with a thin radiating shell. The ratio of X-ray
surface brightness in the middle of the nebula, at a distance $z$
from the pulsar, to that in the tails would be then $\propto
[\delta(z)/R_{\rm cd}(z)]^{1/2},$ where $\delta(z)$ is the radiating
shell thickness.

In the frame of the discussed model, the axial tail of the Geminga PWN
may be attributed to the synchrotron X-rays from the magnetic tail
where the magnetic field reconnection regions and plasma sheets
are similar to planetary magnetotails \citep[e.g.,][]{eastwood2016}
An important difference, however, is that, contrary to the planetary
and heliospheric magnetotails, reconnecting magnetic flows in PWNe are
expected to be composed of moderately magnetized relativistic plasma.
Recent
two-dimensional particle-in-cell (PIC) simulations
of relativistic reconnection in such objects \citep{sironi_ea_2016}
have revealed a phenomenon of formation of quasi-spherical plasmoids,
which may show up as blobs apparent in the X-ray images of the
the axial tail of Geminga. To be consistent with the observed spectra
of the blobs, the simulated spectra of the plasmoids derived by
\citet{sironi_ea_2016} must have moderate magnetization parameter
3 $\leq \sigma \leq$ 10.


\subsection{Extended X-ray Structures in the Guitar Nebula}\label{Guitar}

The ``Guitar nebula" is associated with the bow shock from
PSR B2224+65 \citep[see, e.g.,][]{1993Natur.362..133C,1997ApJ...484L.137R,2004ApJ...600L..51C}.
The pulsar is likely one of the fastest pulsars observed with a
radio measured proper motion. \citet{1993MNRAS.261..113H} estimated
its transverse velocity to be about 986 $\kms$ at the distance of 1.1~kpc,
derived from the dispersion measure. The multi-wavelength study
of the nebula by \citet{2004ApJ...600L..51C} resulted in the
distance estimation between 1 and 2 kpc suggesting even faster
velocity of PSR B2224+65. The pulsar has a modest spin-down power $\sim $10$^{33}\ergs$.
Its observed X-ray spectrum is well described by a
photon index ${\rm \Gamma} = $1.58$^{+0.43}_{-0.33}$
\citep[see][]{2007A&A...467.1209H}.

The X-ray observations of PSR B2224+65 revealed a jet-like feature
extended to $\sim 2.4\arcmin$ (i.e. of about a parsec length) away
to the north-west from the pulsar and offset by about 118$\arcdeg$
from the pulsar direction of motion \citep[see, e.g.,][and the
references therein]{2007A&A...467.1209H,2010MNRAS.408.1216J,2012ApJ...747...74H}.
The observed spectrum of the extended X-ray filament is much harder
than of that of the pulsar. \citet{2010MNRAS.408.1216J} derived a
photon index of ${\rm \Gamma} =$~1.00$^{+0.53}_{-0.47}$ in the
extended X-ray filament, while a softer spectrum, with ${\rm
\Gamma} = 1.69^{+0.39}_{-0.34}$, was observed from the head segment of the
PWN. \citet{2012ApJ...747...74H} discussed the possible progenitor
star of PSR B2224+65, which is running away from the Cygnus OB9 association.
The authors also suggested that the inverse Compton scattering of e$^{\pm}$ pairs
of $\gamma \sim 10^4$ could produce the extended X-ray feature. The
total magnetospheric potential of PSR B2224+65 can provide a maximum
particle energy $\sqrt{3\dot{E}/2c} \lsim$7.5$\times 10^{13}$ eV, which
is likely below the energy of the X-ray emitting pairs in the
observed extended filament, assuming synchrotron emission
in the typical ISM magnetic field $\sim$ 3 $\mu G$.
We will discuss further the acceleration of the e$^{\pm}$ pairs
in the PWN associated with PSR~B2224+65.

A scenario to explain the complex phenomenology of the extended
X-ray filament was proposed by \citet{2008A&A...490L...3B}. The
author suggested that the highest energy electrons of Lorentz factor
$\sim 10^8$ are accelerated at the termination shock and then
escape from the bow shock into the ISM. They emit
synchrotron X-rays in a magnetic field of strength $\sim 45~\mu G$
directed along the filament. Such magnetic field is higher than
typical ISM values and must have been amplified in some way (e.g.,
by a streaming instability). The idea behind this scenario is to attribute the
observed X-ray emission to collective radiation of accelerated
pairs diffusing into the ISM, rather than interpreting the
filament as a particle beam. It seems indeed difficult
to keep a parsec size approximately linear
structure of the beam without a strong bending, despite the high
ram pressure due to the fast motion of the pulsar. For example, the
X-ray emitting tails of the Geminga pulsar are highly bent, while the
estimated proper velocity of PSR B2224+65 is much larger than that
of Geminga.

For viability of this scenario a hard spectrum of
accelerated pairs escaping the PWN with Lorentz factors $\sim
10^8$ is needed. Indeed, the X-ray luminosity of the filament is rather high,
corresponding to about a percent of the pulsar spin-down power
(if the distance to PSR B2224+65 is about 1~kpc, or even higher for
a 2~kpc distance). The photon index of the X-ray filament,
estimated from {\sl Chandra} observations, is also hard: ${\rm \Gamma} =
1.00^{+0.53}_{-0.47}$.

Diffusive shock acceleration at the highly relativistic termination
shock would provide a pair spectrum of index 2.2-2.3 and corresponding
photon index of synchrotron radiation softer than 1.6, which is
consistent with that obtained by \citet{2010MNRAS.408.1216J} for the
head segment.
With
the moderate spin-down power derived for PSR B2224+65, the
maximal Lorentz factor of e$^{\pm}$ estimated from Eq.~(\ref{Lmin}) is $\sim
10^8$. This maximum energy can only be achieved in a
transrelativistic outflow with ${\rm \Gamma}_{{\rm flow}}^{2}/{\rm
\beta_{\rm flow}} \sim 1$. This excludes as the acceleration sites
both the highly relativistic wind before the termination shock and
the region around the non-relativistic bow shock. On the other hand,
acceleration of pairs up to $\gamma \sim 10^8$ may occur in the
transrelativistic colliding flow behind the termination shock and
the bow shock.

A model of the long linear X-ray filament in PSR~B2224+65 also has to
explain why it is seen just in the north-west direction from
the pulsar. In this context it is important that a similar X-ray
filament was reported for the pulsar IGR~J11014-6103 by
\citet{2011A&A...533A..74P,2014A&A...562A.122P,2016A&A...591A..91P}.
Recently, \citet{2017MNRAS.464.3297Y} presented hydrodynamic
simulations illustrating that the assumption of a density profile
of the ISM with inhomogeneities in the form of a series of density discontinuities
ahead of the fast moving PSR~B2224+65 can reproduce the form of the
bow shock and the multiple bubbles seen in H$_{\alpha}$
observations of the Guitar nebula. Such a model may also help to
understand the peculiar appearance of the H$_{\alpha}$ bow shock of
the recycled millisecond pulsar J2124-3358.

\subsection{Extended X-ray Structures in the Lighthouse Nebula}\label{LH}

\begin{figure}
\includegraphics[width=0.99\linewidth]{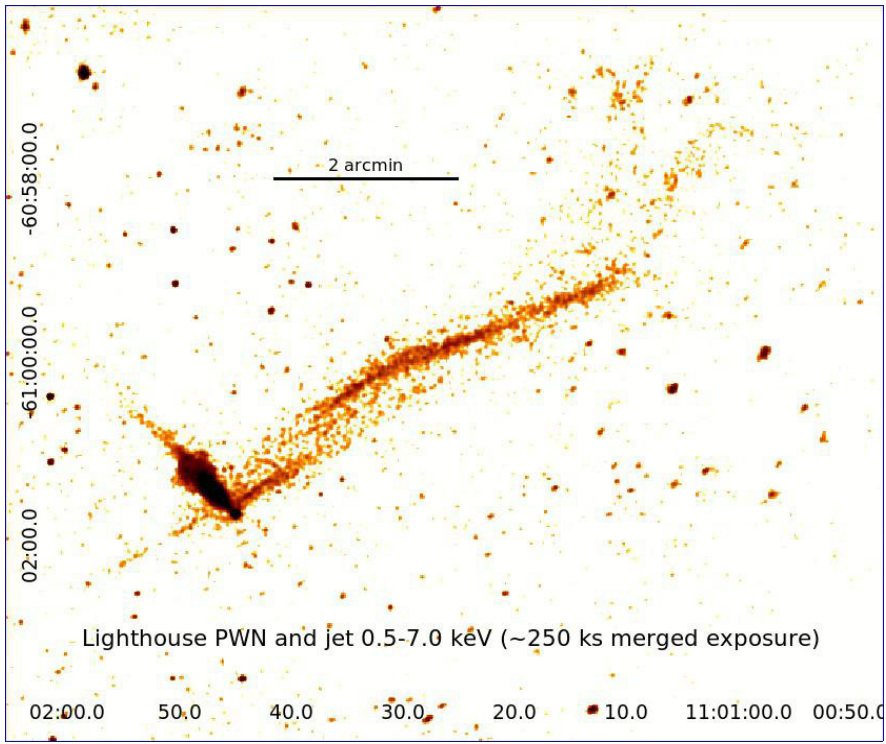}
\caption{A 0.5-7 keV {\sl Chandra} image of IGR~J11014-6103 (the Lighthouse
nebula) likely powered by PSR~J1101-6101. The image shows the
BSPWN as well as large-scale jet-like features, all launched by
IGR~J11014-6103 \citep[see][]{2016A&A...591A..91P}. \label{fig:LH}}
\vspace{-1.\baselineskip}
\end{figure}

IGR~J11014-6103 is a source discovered with the {\sl IBIS/ISGRI}
camera aboard the gamma-ray telescope {\sl INTEGRAL}
\citep{2010ApJS..186....1B}. With follow up {\sl Chandra} observations,
\citet{2011A&A...533A..74P} revealed, in addition to the point-like source,
an extended X-ray structure and a helical type tail of $\sim 4\arcmin$
extension. The tail is nearly transverse to the system proper motion (see
Fig.~\ref{fig:LH}) and somewhat resembles a similar structure in PSR~B2224+65.
The authors associated IGR~J11014-6103 with the PWN
(the "Lighthouse nebula") which is produced by the high-velocity ($\sim$1,000
$\kms$) pulsar of estimated spin-down power $\dot{E} \sim 10^{37} \ergs$.
The pulsar might originate from the supernova remnant MSH~11-61A
at an estimated distance of 7$\pm$1 kpc. At this distance
the physical extension of the tail would be about 10~pc. The spectral map
of the extended filament shows some patchy patterns with spectrum
softer than that of PSR~B2224+65. If the tail extension is indeed
well above 1~pc, the synchrotron burn-off effects may play a role.
Deep {\sl Chandra} observations indicated large deviations from a simple
helical model at small and large distances, and the possible
presence of an apparent brightness dip at about 50$\arcsec$ distance from
the pulsar \citep{2016A&A...591A..91P}.
\citet{2016A&A...591A..91P} concluded that both a ballistic jet scenario
and the scenario of \citet{2008A&A...490L...3B}, which considers
synchrotron radiation by high energy pairs propagating along
pre-existing interstellar magnetic field lines, can explain only some of the
observed features.

Apart from the Guitar and Lighthouse nebulae, several other
elongated structures associated with pulsars are known. A very
extended ($\gsim 9\arcmin$) X-ray filament in the trail of the radio-quiet
middle-aged gamma-ray pulsar J0357+3205 was revealed by
\citet{2013ApJ...765L..19D}. The space velocity of the pulsar was estimated
as $\sim 390 \kms$ if it is at a distance of about 500 pc,
though no H$_{\alpha}$ emission was detected.  
\citet{2016ApJ...819...40M} found two tail-type nebulae associated with the
old pulsar PSR J2055+2539.  The pulsar has a modest spin-down power,
$\dot{E} \sim 5\times10^{33} \ergs$, a characteristic age of about 1.2 Myr,
and the estimated distance is $\sim$ 600~pc. The bright tail has an angular
size of 12$\arcmin \times$20$\arcsec$ corresponding to  $\sim (2.1
\times  0.05)$ pc, while the faint one is of 250$\arcsec \times
30\arcsec$ size, i.e. $(0.7 \times 0.09)$ pc. The X-ray luminosity of the
bright tail in the 0.3-10 keV range is  $\sim 2\times 10^{-3}
\dot{E}$ (assuming a 600~pc distance). No information on the proper motion
of PSR~J2055+2539 is available yet, but \citet{2016ApJ...819...40M} suggested
that it is likely to produce a bow-shock.

\subsection{Magnetic Jets in BSPWNe}

The structure of an astrosphere produced by a relativistic pulsar
wind confined by a counter propagating supersonic flow has much in common with that
of the heliosphere. Recent progress in the heliosphere observations
with {\sl Voyager} 1 and 2, {\sl Cassini}, and {\sl
Interstellar Boundary Explorer} satellites, have initiated a new
model of the structure of the heliosphere developed by
\citet{2015ApJ...800L..28O} and \citet{2015ApJ...808L..44D}.
The model suggests
that a two-jet structure might provide a better description of the system than
the standard comet-like shape. The oppositely directed
jets are eventually deflected into the tail region by the motion of
the Sun through the ISM. In the models of
\citet{2015ApJ...808L..44D} and \citet{2017MNRAS.464.1065G} the heliosphere
is axisymmetric and the structure of the subsonic flows downstream of
the termination shock, in the heliosheath and heliopause, is governed
by the solar magnetic field. Tension of the solar magnetic field
produces a drop in the total pressure between the termination shock
and the heliopause. The pressure drop accelerates the plasma flows
downstream of the termination shock into the north and south
directions to form the collimated jets.

The fact that the magnetic pressure of the azimuthal field may
dominate the structure of the subsonic flow in the inner heliosheath
was revealed in the early models discussed by
\citet[][]{1972NASSP.308..609A}. Later, \citet{1992ApJ...397..187B}
proposed that the pinching effect of the toroidal magnetic field in
the subsonic flow downstream of the termination shock can be
responsible for the observed elongation of the Crab nebula.
Similar axisymmetric structures extended in the polar direction were
advocated by \citet{1994ApJ...421..225C} for the magnetic shaping of
planetary nebulae. Finally, a number of studies \citep[see, e.g.,][]
{2002MNRAS.329L..34L,2004MNRAS.349..779K,2004A&A...421.1063D,2014MNRAS.438..278P,2016JPlPh..82f6301O}
demonstrated that magnetic collimation of the plasma downstream of a
PW termination shock can be a plausible mechanism behind formation of jets in PWNe.

The double jet structure proposed for the heliosphere has not been
studied in the regime relevant for supersonically moving PWNe.
However, there are reasons to speculate that the tension of the azimuthal magnetic field
between the PWN termination shock and the bow shock may drive formation
of a double jet structure in this system as well.
There is no reason to believe that the orientation of the interstellar
magnetic field with respect to the pulsar rotation axis is
the same for all BSPWNe. In the case when the plane defined
by the directions of the interstellar field and the proper velocity
of the pulsar is nearly perpendicular to the pulsar rotation axis,
one may expect that the jets would be confined to the region between
the apex of the bow shock and the downstream of the termination
shock. This constraint may be fulfilled for a narrow subset of
bow shock PWNe. For instance, in the Vela~PWN (which is likely
a BSPWN), shown in the left panel of Fig.~\ref{fig:VelaXC} the
two apparent X-ray jets are connected to the pulsar position. This
may be not too surprising given that the bow shock formed in the
Vela~PWN due to the mildly supersonic flow behind the supernova
reverse shock \citep{chevreyn11} is likely quasi-parallel. The
available PWN jet models are axially symmetric relative to the pulsar
rotation axis, so the scenario described above needs to be confirmed
with more complex simulations.

The magnetic jets would channel away some fraction of the high
energy particles accelerated by the colliding shock flow in the
equatorial wind plane. Magnetically dominated force-free structures
are observed in the solar atmosphere and in terrestrial laboratories
in the form of magnetic flux ropes \citep[see, e.g.,][]
{1990GMS....58.....R,2007rmf..book.....B,2011NatPh...7..539D,2016PhPl...23g2901V}.
Magnetic flux ropes typically have helical field structure with the
maximum of the (untwisted) field at the rope axis. The magnetic
field lines in the jets may reconnect with the interstellar magnetic
field lines providing a way for the magnetized ultrarelativistic
particles of energies $\gsim$ 10~TeV to escape into the ISM. If the
particles are (re)accelerated in CSF, they would have hard spectra
of indexes below 2 and therefore contain most of the energy in the
highest energy end of their distribution. This is an important point
in regard of explaining the observed high efficiencies of the X-ray
emission from the extended structures in the Guitar nebula
\citep[see, e.g.,][]{2010MNRAS.408.1216J} and in the Lighthouse
nebula \citep[see, e.g,][]{2016A&A...591A..91P}. The X-ray
efficiency may reach a percent of the pulsar spin-down power, a fact
that requires {\sl (i) hard spectra of accelerated pairs} and {\sl
(ii) an amplified magnetic field} in the extended X-ray emitting
filaments. We shall discuss below, in \S\ref{MFAbeam}, a possible
mechanism of magnetic field amplification by a CR beam propagating
along the local magnetic field.

Reconnection of magnetic field lines in the jet with the oppositely
directed interstellar field would happen for only one of the two
jets associated with the bow shock PWN. This may explain the
apparent asymmetry of the X-ray emitting filaments in both the
Guitar and Lighthouse nebulae. The X-ray appearance would depend on
the orientation of the pulsar rotation axis (providing the jet
directions) as well as the direction of the pulsar proper motion. In
both the Guitar and the Lighthouse nebulae the pulsar proper motion
is directed nearly transverse to the interstellar magnetic field.

\subsection{Magnetic Field Amplification in Extended Ballistic Beams}
\label{MFAbeam}

Relativistic particles leaving the accelerator
may produce streaming instabilities which would amplify fluctuations
of the interstellar magnetic fields to high magnitudes
\citep[e.g.,][]{1974ARA&A..12...71W,1980ARA&A..18..289C,1987PhR...154....1B,1990acr..book.....B,LB2000}.

A fast, non-resonant, almost purely-growing instability driven by
CR current was discovered by \citet{Bell2004,Bell2005}. The growth
rate for the Bell's mode propagating along the mean field is
\begin{equation}\label{DispU1}
{\rm \Gamma_{\rm B}} = v_{\rm a}\sqrt{k_{\rm c}|k|-k^{2}}
\end{equation}
where
\begin{equation}\label{k1}
k_{\rm c} = \frac{4 \pi J_{\rm CR}}{c B},~~ k_{\rm c}R_g > 1.
\end{equation}
Here $J_{\rm CR}$ is the electric current carried by the
ultrarelativistic particles in the beam escaping the accelerator.
Note that in the case of CRs leaving a BSPWN the CR beam is composed
of ultrarelativistic e$^{\pm}$ pairs and protons. The highest
energy CRs can be accelerated in colliding shock flows (see
\S\ref{sec:PACF}) which energize the highest energy end of the CR
spectrum. While the pairs are likely injected into CSF from the PW,
the relativistic protons can be injected due to the diffusive shock
acceleration at the bow shock. The exact CR composition in the PWNe
is somewhat uncertain: both injectures are subject of the intensive
studies. However, the CR current $J_{\rm CR}$ exists in the beam
leaving the accelerator if the net charge is nonzero in the CR beam.

The saturation level $B_{\rm b}$ of the short scale magnetic field
in the beam amplified by Bell's instability can be estimated from
the condition $k_{\rm c}R_g \sim$ 1 \citep{Bell2004}. Then using
Eq.~(\ref{k1}) one can get
\begin{equation}\label{BB}
B_{\rm b} \approx  \frac{\sqrt{\eta_{\rm CR} \dot{E}/c}}{R_{\rm b}},
\end{equation}
where $\eta_{\rm CR}$ is the fraction of pulsar spin-down power
converted into the accelerated high energy particles forming the beam
and $R_{\rm b}$ is the radius of the beam which is $R_g(\gamma_{\rm max}) \lsim
R_{\rm b} < R_{\rm cd}$.

The scale length $l_{\rm b}$ of the beam $\sim c/V_{\rm psr}\times
R_g$, unless the length is limited by the particle energy losses
which is the case for a small $V_{\rm psr}$. The leading edge of the
beam is located at a distance $\delta_{\rm bmf} \sim c {\rm
\Gamma_{\rm B}}^{-1}$ ahead of the region filled with the amplified
magnetic field. The maximal growth rate is ${\rm \Gamma_{\rm
B}^{max}} \sim v_{\rm a} k_{\rm c}/2$, and therefore
\begin{equation}\label{deltab}
\delta_{\rm bmf} \approx \frac{2c}{v_{\rm a}k_{\rm c}} = \frac{c^2
\sqrt{\rho}}{\sqrt{\pi} J_{\rm CR}}.
\end{equation}
The condition $\delta_{\rm bmf} < l_{\rm b}$ must be fulfilled to
fill the beam with the magnetic field amplified by the CR driven
short scale Bell's instability. Bell's instability
amplifies highly only the magnetic field fluctuations of short scales
$kR_g~>~1$, which are inefficient for scattering the highest
energy pairs. The known long-wavelength instabilities \citep[see,
e.g.,][]{bbmo13} are too slow to be able to fill the beam with
long wavelength turbulence. Therefore the high energy pair mean free
paths are determined by the background interstellar magnetic
fluctuations. The synchrotron-Compton losses of $\gsim$ 10 TeV pairs
in the amplified magnetic field can limit the X-ray filament length
not to exceed greatly $\sim$ 10 pc.

The mean free path $\Lambda(p)$ of a 10 TeV energy particle along
the interstellar magnetic field is a few parsecs
\citep[e.g.,][]{SMP2007}, while its gyroradius $R_g$ is about
10$^{16}$ cm. Therefore the particles are highly magnetized and
their diffusion coefficient transverse to the mean magnetic field
estimated from Eq.~(\ref{diff_perp}) is $D_{\perp} \approx vR_g/3
\times (R_g/\Lambda)$. The transverse diffusion coefficient is small
enough to keep the transverse size of the beam $l_{\rm b \perp}$ to
be about a few $R_g$.

The main X-ray filament in the Lighthouse nebula extends for about
11~pc at the estimated distance of about 7 kpc \citep[see,
e.g.,][]{2016A&A...591A..91P}, while the scale size of the X-ray
filament in the Guitar nebula is of about a parsec length \citep[see,
e.g.,][]{2010MNRAS.408.1216J}. The X-ray filament length $l_{\rm b}$
in this simplified model scales as $\sqrt{\dot{E}}/V_{\rm psr}$.
In the case of the Lighthouse nebula the synchrotron-Compton
losses of the X-ray emitting pairs are likely to limit the filament
scale size, while in the Guitar nebula, the filament size
is roughly consistent with the discussed scaling.

The essential feature of the model described above is
formation of a hard spectrum of particles thanks to acceleration in the
CSF between the PW termination shock and the bow shock:
these particles may power the extended X-ray filaments.
The mechanism efficiently converts a sizeable fraction of the wind power
into CR luminosity dominated by the highest energy pairs. The
accelerated pairs can escape the system through the jets which can
be formed in the magnetized region between the shocks. These jets,
bent by the ram pressure due to the supersonic motion of the pulsar, can
be an explanation of the hard tails observed in the Geminga PWN.
If a fast pulsar moves transverse to the interstellar magnetic field,
then the reconnection of the ISM magnetic field with the field in a
pulsar jet can launch a beam of ultrarelativistic pairs along the
interstellar field, which may explain the extended X-ray features
observed around PSR~B2224+65 and IGR~J11014-6103.


In Fig. 7 (right panel) a simulated synchrotron image of the
magnetized ballistic beam is presented. A simplified kinetic
modeling is performed for a set of parameters likely relevant for
the elongated structure of the Guitar PWN. In the discussed scenario
only the particles at the highest energies achieved in the colliding
shock flow can escape the accelerator \citep[see Fig.~2
in][]{2013MNRAS.429.2755B} to be injected into the beam.
The angular distribution of 
these particles is anisotropic. While the particle pitch-angle
distribution is nearly isotropic inside the acceleration region,
only particles with pitch-angles $\theta \leq \theta_c < \pi / 2$
are injected into the beam. The magnetic field amplification by the
CR-current driven instability provided $B_{aism} = $ 30 $\mu$G along
the beam. An apparent bright region in the central part of the
observed structure is related to an inhomogeneity of the field in
the magnetic rope. In the presented simulation a circularly
polarized field variation with the wavelength 1.3 pc and amplitude
$\delta B_{aism} = $ 15 $\mu$G propagating along the beam was
imposed onto the background field. The beam was supposed to be
directed towards the observer at an angle of 45$^{\circ}$ relative
to the line of sight. The beam had a cylindrical shape and its
radius was $3\times 10^{17}$ cm. The distance to the source was
$d_{\rm beam} = 1.7$ kpc. The particle pitch angle evolution along
the beam conserves the adiabatic invariant $p_{\perp}^2 / B$, since
the mean free path of the X-ray emitting pairs is larger than the
beam length.

\begin{figure}
\includegraphics[width=0.99\linewidth]{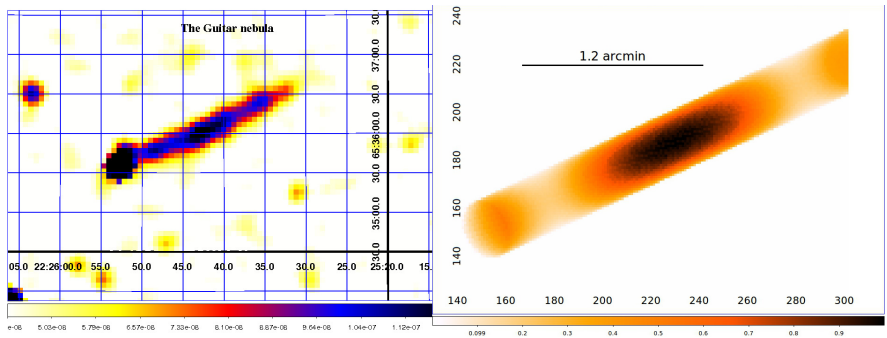}
\caption{Left panel: a 0.5-7~keV {\sl Chandra} image of the Guitar
nebula constructed from the merged archive observations of B2224+65
(PI: D.Q.Wang). Right panel: a simulated synchrotron image of the
nebula. The simulated X-ray synchrotron structure is produced by a beam
of particles accelerating in the colliding shock flows of the
BSPWN, then escaping from the acceleration site and propagating along
the ISM magnetic field lines. The ISM magnetic field is
supposed to be amplified by the CR current-driven instability (see
section \ref{MFAbeam} for details).}  \label{fig:Guitar}
\vspace{-1.\baselineskip}
\end{figure}


\section{BSPWNe Possibly Associated with SNRs}\label{snrbspwn}

Some of the fast moving young pulsars with PWNe
reviewed by \citet{KP08} are located within young supernova
remnants (SNRs). Radio observations of the young active pulsar PSR~B1853+01
by \citet{1996ApJ...464L.165F} revealed that the pulsar, of the
spin-down power about 4.3$\times 10^{35} \ergs$, is located at the
apex of an extended feature which has a cometary morphology. The
pulsar has a derived transverse velocity of about 375~$\kms$.
\citet{2002ApJ...579..404P} observed in X-rays 
the pulsar PSR~B1853+01 and its associated PWN, embedded into
the supernova remnant W44. The authors found
the photon index of the extended X-ray nebula to be about
2.2, much steeper than that of PSR~B1853+01 which is $\sim$1.4.

A similar cometary shape PWN around the putative
pulsar CXOU J061705.3+222127 was observed in the young supernova
remnant IC~443 \citep[see, e.g.,][]{2001A&A...376..248B,2015ApJ...808...84S}.
Deep X-ray observations reported by \citet{2015ApJ...808...84S} revealed
a ring-like structure surrounding the pulsar and a jet-like feature.
The authors found no evidence for a bow shock or contact
discontinuity neither spectrally nor morphologically, since the
nearly circular ring is not distorted by motion through the ambient
medium. They placed an upper limit on the proper motion of the pulsar of
about 310$\kms$ at a distance of 1.5~kpc. Earlier
\citet[][]{2001A&A...376..248B} revealed
a significant spectral softening along the extended
cometary shape X-ray structure, with the power law photon index of
about 1.6 in the head and about 2.2 in the tail of the nebula. This is consistent with
the expectations based on the simple one zone model of PWNe \citep[see e.g.][]{2000ApJ...539L..45C}.
However, the softening is not typical for many of the bow shock PWNe which we discuss here.

The PWN energized by PSR~B1951+32 is located in the field
of the supernova remnant CTB 80 (G69.0+2.7). This is another
interesting case of a BSPWN. PSR~B1951+32 is an energetic
pulsar at a 2~kpc distance with a spin-down power $\dot{E} = 3.7 \times 10^{36} \ergs$,
an estimated age $\sim 10^5$ yrs, and a moderate proper velocity of
$240\pm 40 \kms$ \citep{2004ApJ...610L..33M,2005ApJ...628..931L}.
Radio and optical H$_{\alpha}$ observations revealed a bow shock and
\citet{2004ApJ...610L..33M} estimated that the contact discontinuity
between the shocked PW and the shocked ambient ISM is at a
distance about 0.05~pc from the pulsar.

Spatially resolved X-ray spectra of the main components of the
PWN were studied with {\sl Chandra} by \citet{2005ApJ...628..931L}.
They found that the photon index of the emission from the bow shock region is
$1.6^{+0.1}_{-0.2}$ and it is harder than that of the entire
nebula, $1.73 \pm 0.03$. Moreover, \citet{2009AIPC.1126..259M}
reported a significant detection of hard X-ray emission from CTB~80/PSR~B1951+32
up to 70 keV  with the {\sl IBIS/ISGRI} camera aboard the {\sl INTEGRAL} mission.

\citet{2005ApJ...628..931L} pointed out that the observed hard
emission spectrum may be due either to the acceleration of PW
particles at the termination shock or to a contribution from
a second $e^{\pm}$ component. They proposed that
the new component of emitting $e^{\pm}$ may originate at the bow
shock, which was suggested by \citet{1988ApJ...331L.121H} to explain
the optical filaments in CTB 80. PSR~B1951+32 has spin-down power
and proper velocity not very different from those of the Vela pulsar
and we will argue in \S\ref{VelaX} that the hard component may
naturally arise due to the acceleration of relativistic $e^{\pm}$ pairs in
a CSF between the termination and bow shocks.

An extended (arcminute size) faint X-ray nebula to the
south of PSR J0633+0632 was detected by
\citet{2011ApJS..194...17R} with {\sl Chandra}. The pulsar,
which is a gamma-ray active {\sl Fermi} source
\citep[see][]{2013ApJS..208...17A}, was resolved in the X-ray
observation, while its optical counterpart is not yet found
\citep{2016MNRAS.461.4317M}. The X-ray spectrum
of the large scale extended PWN has a very hard photon index of
\mbox{0.74--1.29} for a fixed $N_H = 0.15 \times 10^{22} \cms$
(0.9$^{+0.5}_{-0.4}$ for a varied column density). Such a hard
spectrum is very difficult to explain as synchrotron emission of
the relativistic $e^{\pm}$ accelerated only at the PW
termination shock. The radio-quiet pulsar has an estimated
spin-down power $\dot{E} \sim 10^{35} \ergs$ and the pulsar position
suggests the Rosette nebula as its possible birth place \citep[see
for a discussion][]{2015PASA...32...38D}. In this case the distance
to the pulsar would be $\sim 1.5$ kpc and then its proper velocity is
$\sim 1,000 \kms$.

Now we will discuss a representative case of a PWN interacting with its
SNR. The Vela PWN is a bright nearby object, well studied
in multi-wavelength observations. The Vela pulsar is located
within its maternal SNR and thus the PWN is interacting with the SNR.

\begin{figure}
\includegraphics[width=4.0in]{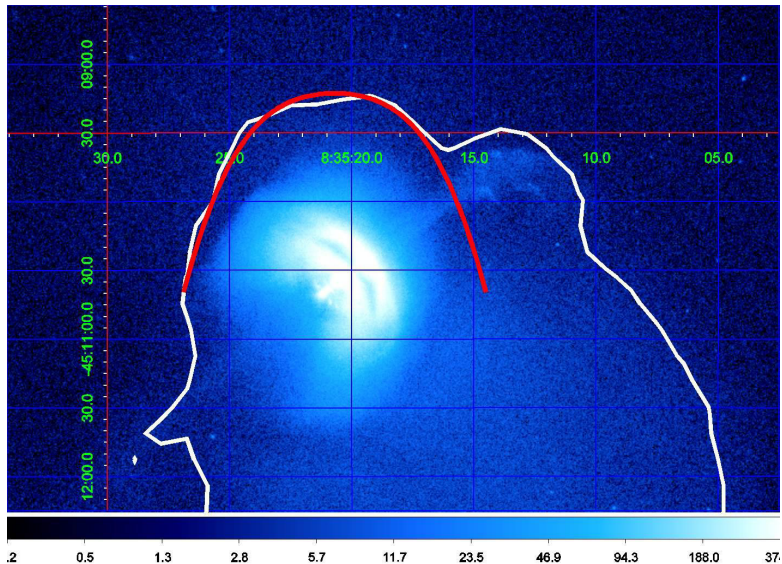}
\caption{The X-ray image of the Vela PWN in the 0.5-7 keV energy
band. The image is composed of about 250 ks of merged {\sl Chandra
ACIS} observations at different epochs. The white contour shows the
boundary of the 0.5-7 keV diffuse emission. The red curve is an
overimposed shape of the bow shock of moderate Mach number
1.3, as suggested by \citet{chevreyn11}. The image illustrates the
apparent structure of the asymmetric bow shock nebula surrounding
the double ring structure of the PWN around PSR B0833--45
\citep{2001ApJ...556..380H,2001ApJ...552L.129P}
and the jet-like structure discussed in detail by \citet{2013ApJ...763...72D}.}
\label{fig:VelaXC} \vspace{-1.\baselineskip}
\end{figure}

\begin{figure}
\includegraphics[width=4.0in]{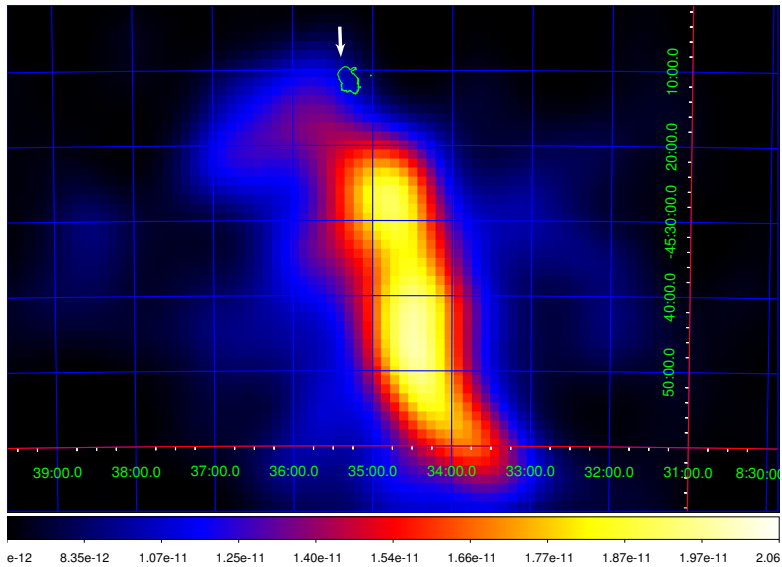}
\caption{The very high energy gamma-ray (0.75 to 70 TeV) image of
Vela X region as observed by the H.E.S.S. telescope
\citep[HESS J0835-455, ][]{2006A&A...448L..43A,2012A&A...548A..38A}.
The small green
contours at the north correspond to the X-ray {\sl Chandra ACIS} image in 0.5-7 keV energy
band of the arcminute size Vela PWN shown in Fig.~\ref{fig:VelaXC}.
The white arrow
shows the expected direction of the velocity of the large scale flow
behind the reverse shock of the Vela SNR. \label{fig:VelaXH}}
\vspace{-1.\baselineskip}
\end{figure}

\begin{figure}
\includegraphics[width=5.0in]{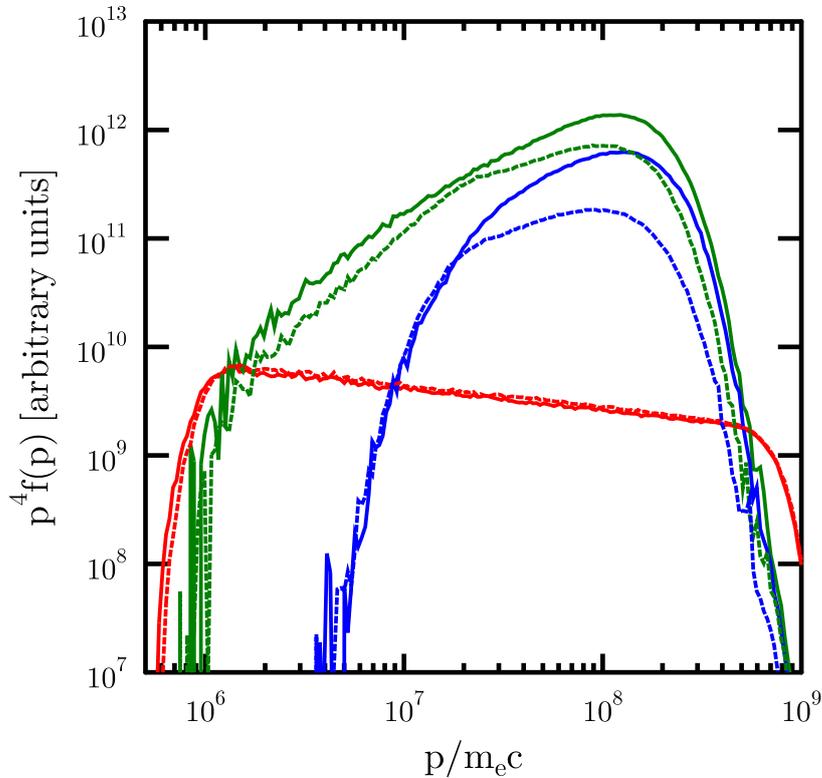}
\caption{The simulated local spectra of e$^{\pm}$ pairs in different
zones of a BSPWN (see Fig.~\ref{fig:Geom1})
for parameters similar to those estimated for the Vela PWN.
The red curves correspond to the PW termination shock downstream:
the solid one refers to the $A2$ bin, the dashed one refers to
the $A4$ bin.
The blue curves correspond to the bow shock region (bin $C2$ -- the
solid curve, bin $C4$ -- the dashed curve). The green curves
correspond to the region between the two shocks (bin $B2$ -- the solid
curve, bin $B4$ -- the dashed curve).
The locations of all the bins are shown in Fig.~\ref{fig:Geom1}.
}
\label{Spectra_VelaP}
\end{figure}

\begin{figure}
\includegraphics[width=5.0in]{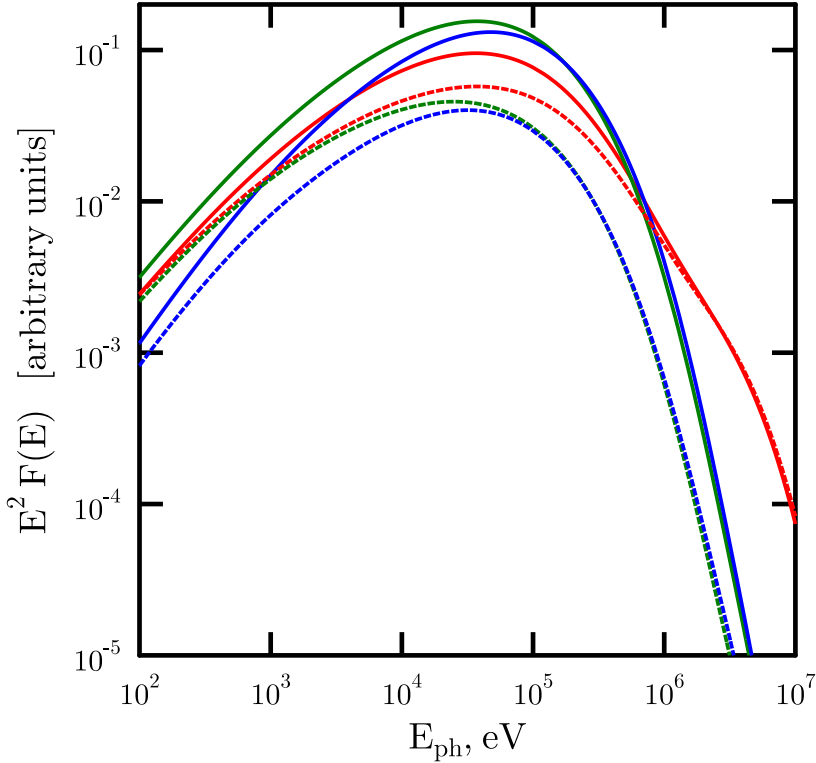}
\caption{The simulated synchrotron spectra of a source similar to
the Vela PWN, derived from the spectra of e$^{\pm}$ pairs
illustrated in Fig.~\ref{Spectra_VelaP}. The orientation of the
source is given by the angles $\Phi = 90^{\circ}$, $\Theta =
90^{\circ}$, and $\Psi = 270^{\circ}$ (see \S\ref{geomMC:spectra}),
i.e., the pulsar velocity in the local ambient matter rest frame
lies in the plane of sky perpendicular to the line of sight crossing
the pulsar (plane $PoS$) and is directed to the north. The curves
show the spectral energy distribution ($\nu F_{\nu}$) calculated by
integration of the spectral emissivity along the lines of sight,
which cross the bins, indicated in the caption to
Fig.~\ref{Spectra_VelaP}. The colors and styles of the curves are
the same as in Fig.~\ref{Spectra_VelaP}. } \label{Spectra VelaR}
\end{figure}

\begin{figure}
\includegraphics[width=5.0in]{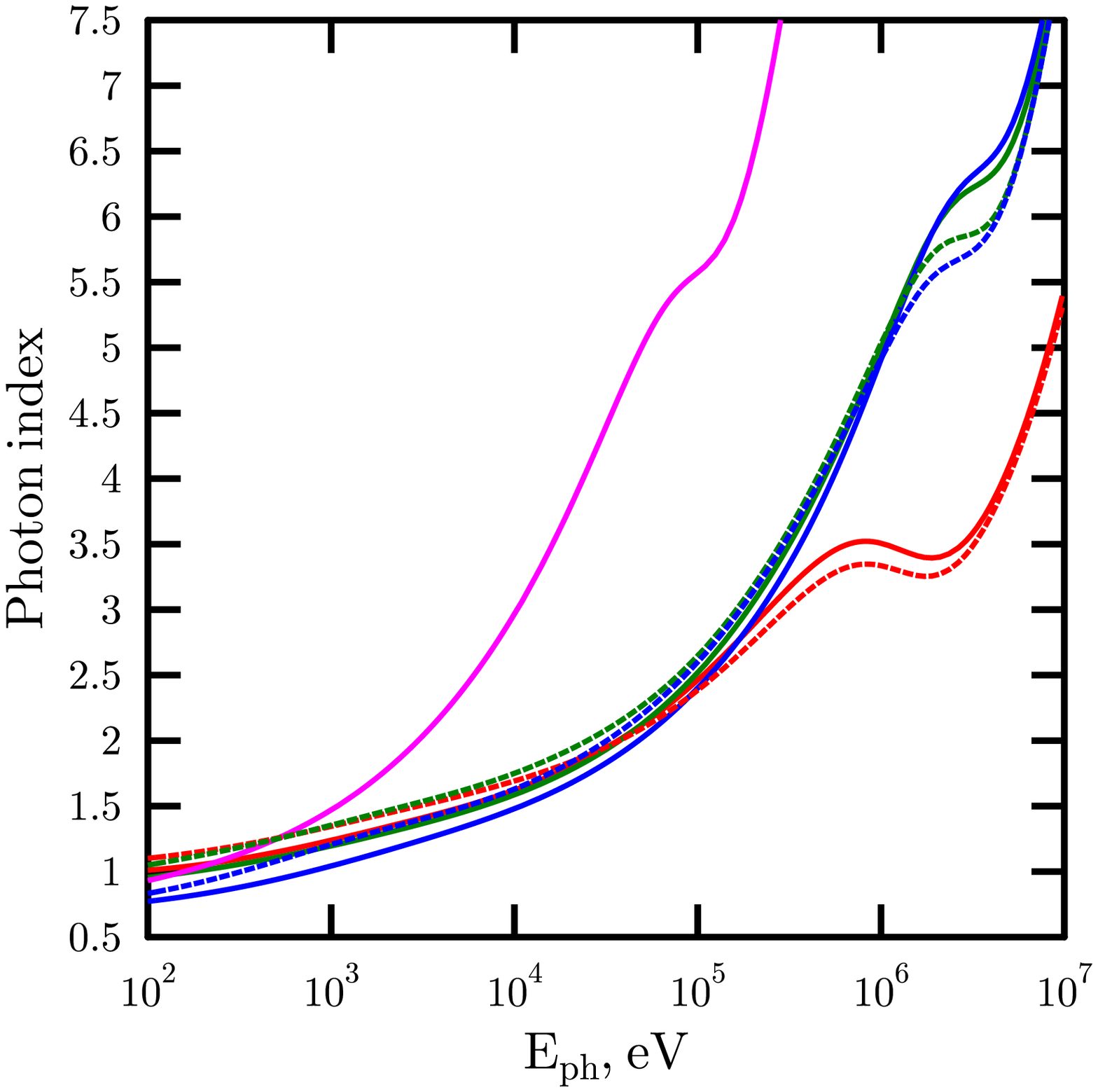}
\caption{The power law photon spectral indexes obtained by fitting of
the simulated synchrotron spectra of a source similar to the Vela PWN in the
wide energy range. The magenta line corresponds to a line of sight
crossing the SNR outside the BSPWN. This line of sight crosses the
bin $D3$ (see Fig.~\ref{fig:Geom1}).
Other curves correspond to the lines of sight defined in the caption to
Fig.\ref{Spectra VelaR}.
The colors and styles of those curves are the same as
in Fig.~\ref{Spectra_VelaP}.
}
\label{fig:Vela_index} \vspace{-1.\baselineskip}
\end{figure}


\section{The Vela PWN and Vela-like Objects}\label{VelaX}

The bright radio and gamma-ray pulsar PSR B0833--45 (the Vela
pulsar) has a period about 89.3 $ms$, characteristic spin-down age
of 11,000 yr, and a spin-down power of $\sim 7\times10^{36} \ergs$.
It is located at a distance of 287$^{+19}_{-17}$ pc, derived from
the parallax measured through the very long baseline interferometry
observation by \citet{2003ApJ...596.1137D}. The authors derived the
proper transverse space velocity of pulsar PSR B0833--45 to be 61
$\pm 2 \kms$. This velocity is likely subsonic being the pulsar
located inside the Vela SNR, where \citet{1999A&A...342..839B}
derived from {\sl ROSAT} observations the presence of a
multi-temperature inhomogeneous plasma with a temperature range from
about 0.1 to $\sim$ 1 keV and a typical number density 0.1-0.5
$\cmc$.

The nearby Vela supernova remnant appears in multi-wavelength images
as a nearly circular outer shell of about 8$\arcdeg$ in diameter.
The bright Vela X region, of about $2\arcdeg \times 3\arcdeg$ scale
size, is located near the center of the remnant to the south from
the pulsar PSR B0833--45. This region, which has a high degree of
radio polarization and a radio spectral index (0.08$\pm$0.10) much
flatter than the main SNR shell, is likely powered by the pulsar
\citep[see e.g.][]{1980A&A....90..269W,1997ApJ...475..224F}. VLA
radio observations of the Vela X region by
\citet[][]{1997ApJ...475..224F} revealed a network of fine, extended
overlapping linear filaments over the region. The authors also
estimated an equipartition magnetic field $\sim 20 \mu G$ in the
region.

Radio observations of the vicinity of the Vela pulsar by
\citet{2003MNRAS.343..116D} with {\sl Australia Telescope Compact
Array} revealed a highly polarized structure of a few arcminutes
extent with two emission lobes to the south and north of the Vela
pulsar. While the radio structure surrounds the pulsar, no excess at
the position of the X-ray bright nebula was detected. The observed
polarization vectors show a symmetry with respect to the spin axis
of the Vela pulsar. \citet{2003MNRAS.343..116D} suggested that the
complex emission from the degree scale size Vela X region is the
result of the integrated evolutionary history of the Vela pulsar,
which makes it different from the few arcminutes vicinity of the
pulsar, where emission is dominated by the currently injected
particle population (see also the high frequency radio observations
of Vela X by \citet{2004ApJ...613..977H}). The observed polarized
radio structure was modeled by \citet{chevreyn11} as an initially
thick toroidal structure produced by the PW which
is interacting with a mildly supersonic inward flow from the
supernova remnant. The flow can be produced in a transient phase of
the SNR evolution when the reverse shock reaches the center of the
remnant.

The high resolution {\sl Chandra} X-ray observations have revealed a
fascinating synchrotron nebula around the Vela pulsar with a bright
structured torus and two jets
\citep{2001ApJ...556..380H,2001ApJ...552L.129P,2001ApJ...554L.189P,2003ApJ...591.1157P}.
The total 1-8 keV luminosity of the bright nebula within 42$\arcsec$
from the pulsar is about 6$\times 10^{32} \ergs$ \citep[see
e.g.][]{2003ApJ...591.1157P} (assuming a distance of 300 pc) which
is a very small fraction ($\sim 10^{-4}$) of the pulsar spin-down
power. The photon indexes of the power law fits vary across the
nebula being about $1.3\pm 0.1$ for the outer jet of X-ray
luminosity about 3$\times 10^{30} \ergs$, close to 1.1-1.2 for the
inner jet and between 1.2 and 1.5 within the bright torii
\citep[e.g.][]{2003ApJ...591.1157P}. A two arcminute scale size
northern jet demonstrated helical type dynamical behavior
\citep{2013ApJ...763...72D} and a number of fine scale structures in
the nebula were identified by \citet{2013MmSAI..84..588L} in deep
{\sl Chandra} images of the Vela PWN. The few arcminutes size nebula
with polarized radio lobes and bright synchrotron X-ray structures
is most likely a result of the recent activity of the Vela pulsar
over a few decades time scale after the reverse shock passed through
the nebula \citep[see e.g.][]{chevreyn11}. A few degrees size
extended Vela X region in this scenario is produced by the
integrated past activity of the Vela pulsar with account of the
relativistic particle escape, their transport and magnetic field
structure
\citep[see e.g.][]{2006A&A...448L..43A,2008ApJ...689L.125D,2011ApJ...743L...7H,2013ApJ...774..110G}.

Hard X-ray and gamma-ray observations are very useful to study the
radiation of the  highest energy particles escaped from the Vela
PWN. {\sl Suzaku/XIS} observations just outside the few degrees size
radio emitting Vela X region were reported by
\citet{2011PASJ...63S.827K} who found a  large scale hard X-ray
emission of constant power law photon index $\approx$ 2.4 extending
throughout the field of view. While the photon index remains
constant, the observed flux is decreasing with increasing distance
from the Vela pulsar. Therefore, the authors suggested that the
hard X-ray emission is associated with the Vela pulsar wind nebula
and that distinct populations of relativistic e$^{\pm}$ pairs are
needed to explain both the X-ray PWN and Vela X region.

A wide field hard X-ray image of the Vela X region was obtained by
\citet{2011ApJ...743L..18M} with the {\sl IBIS/ISGRI} camera aboard
the {\sl INTEGRAL} satellite. The image revealed a significant
extended emission of $\sim$ a degree scale size both in the northern
and southern sides of the region around the Vela pulsar.   The
spectrum of the Vela nebula in the 18-400 keV energy range was
measured by {\sl IBIS/ISGRI} and {\sl SPI} on board of the {\sl
INTEGRAL} satellite. Given the point  spread function  of the coded
mask instrument {\sl IBIS/ISGRI} it was possible to obtain the
spectrum within $6\arcmin$ from the Vela pulsar. The fit to the
combined  {\sl INTEGRAL} and {\sl Suzaku XIS} data yielded a broken
power law with the break energy 27$\pm 3$ keV and the photon indexes
1.642$^{+0.005}_{-0.006}$ and 2.07$\pm 0.05$ below and above the
break respectively.  The plasma temperature in the hot Vela
supernova remnant environment was derived as  $kT = 0.214^{+0.003}_{
-0.005}$ keV. The authors pointed out that the observed change of
the slope at $\sim$ 25 keV is compatible with the standard value of
0.5 due to the cooling break in models with a continuous particle
injection and radiative losses \citep[][]{2000ApJ...539L..45C}. The
slope before the break is consistent with the value of photon index
1.6 which is expected in the model of diffusive particle
acceleration at relativistic shock. However, this slope is steeper
than the photon index 1.3-1.4 observed with {\sl Chandra}
 within the bright $\sim$ arcminute size torus-jet nebula around the Vela pulsar.
 This very hard photon spectrum is indeed reminiscent of those
observed at keV photon energies from a number of bow shock PWNe
\citep{KP08}.

The {\sl AGILE} satellite studied the region and the particle
population there in the energy range from 100 MeV to 3 GeV
\citep{2010Sci...327..663P}. The source AGL J0834-4539 was detected
at the significance level about 5$\sigma$ and the spectrum sampled
in three energy bands (0.1-0.5 GeV, 0.5-1 GeV, 1-3 GeV) yields a
power law fit with a photon index  $1.67 \pm 0.25$.

Gamma-ray emission from the Vela X region was detected by {\sl the
Fermi Large Area Telescope},  and the spectrum in the 200 MeV to 20
GeV range was measured in the first year of observations
\citep{2010ApJ...713..146A}. With a detailed morphological and
spectral analysis of the Vela X region  based on four years of
observations in the 0.3 - 100 GeV range, \citet{2013ApJ...774..110G}
 detected the emission from the northern wing (from the pulsar) of
the Vela X region.

The total gamma-ray luminosity of the Vela X region detected by {\sl
Fermi LAT} above 0.2 GeV is 2.4$\times 10^{33} \ergs$ which is
0.03\% of the pulsar spin-down power \citep{2013ApJ...774..110G}.
These authors studied separately the spectra of the northern wing
(to the Vela pulsar position) of the {\sl Fermi LAT} image which can
be fitted with a single power law of photon index $2.25 \pm 0.07 \pm 0.20$.
Fitting the southern wing required instead two power laws, with a
break at an energy of $2.1 \pm 0.5 \pm 0.6$ GeV: the photon index
before the break was found to be $1.81 \pm 0.10 \pm 0.24$ while
after the break it would steepen to $2.90 \pm 0.25 \pm 0.07$.
\citet{2013ApJ...774..110G} pointed out that while the energy break
at 2.1 GeV is significant, the spectral difference between the
northern and southern wings is only marginal. {\sl Fermi LAT} data
analysis by \citet{2010ApJ...713..146A} favored a scenario with two
distinct electron populations to understand the multiwavelength
spectrum of Vela X. Fast diffusion of particles accelerated in the
vicinity of the pulsar was discussed by \citet{2011ApJ...743L...7H}
and \citet{2013ApJ...774..110G} to explain the origin of the steep
{\sl Fermi LAT} spectrum. Moreover, \citet{2011ApJ...743L...7H}
suggested that if the accelerated particles being released early in
the evolution of the nebula then travel with a large diffusion
coefficient to avoid severe adiabatic losses,  this could help to
understand the rising positron fraction in the local CR spectrum, an
issue which we shall discuss in some detail below.

Very high energy gamma-ray emission  from 0.75 to 70 TeV was
detected with the groundbased Cherenkov telescope {\sl H.E.S.S.}
from the Vela  X region
\citep{2006A&A...448L..43A,2012A&A...548A..38A}. The source HESS
J0835-455 is extended to the south from the pulsar PSR B0833--45 and
the morphology of the source was characterized as an ellipse with a
Gaussian width along the major axis of $0.52\arcdeg \pm
0.02\arcdeg$, while $0.12\arcdeg \pm 0.02\arcdeg$ is the width along
the minor axis. The source shape is different from that obtained by
\citet{2013ApJ...774..110G} with the four years of {\sl Fermi LAT}
data, where the northern wing of the emission was detected. However
the source is positionally coincident with the source AGL J0834-4539
imaged by the {\sl AGILE} telescope in the energy range from 0.1 to
3 GeV. The spectral model of the entire  Vela X region (about $1.2
\arcdeg$ circle around the central position)
in range from 0.75 to 70 TeV
is the power law with a
photon index $1.32 \pm 0.06_{\rm stat} \pm 0.11_{\rm sys}$ and exponential cut off
at $14.0 \pm 1.6_{\rm stat} \pm 2.6_{\rm sys}$ TeV, while the index is even
somewhat harder for the external ring from 0.8 to 1.2 degrees
\citep{2012A&A...548A..38A}.

Recently \citet{2015arXiv150100278M} discussed optical observations
of the Vela X region using {\sl Digital Sky Survey} and SAAO 1.0 m
telescope observations in an attempt to constrain the
multi-component models of the high energy lepton injection suggested
by multiwavelength studies of Vela X.

 A PWN that is rather similar to Vela \citep{chevreyn11} is that
powered by the pulsar J2229+6114 of spin-down power $\sim 2.2\times
10^{37}\ergs$ and associated with SNR G106.6+2.9 (also known as the
``Boomerang"). The comet-shaped structure of G106.6+2.9 appears as a
highly polarized incomplete radio shell with a flat spectrum and an
X-ray source of photon index $\sim 1.5$
\citep{2001ApJ...547..323H,2001ApJ...552L.125H}.  The authors
suggested that the X-ray source may be a wind-blown bubble or
bow-shock nebula produced by a young pulsar. A harder X-ray photon
index ${\rm \Gamma} =1.3\pm 0.1$ and an X-ray luminosity of $8.2\times
10^{32} \ergs$ were estimated for the PWN (assumed at a distance of
3 kpc as estimated from X-ray absorption) by \citet{KP08}.

A study by \citet{2001ApJ...560..236K} of the kinematics of the
neutral hydrogen and molecular material associated with the SNR has
suggested a distance of $\sim$ 0.8 kpc to the SNR and a size of
about 14$\times$6 pc. The J2229+6114 PWN would have a diameter of
0.8 pc. This estimate of the distance is supported by the highly
polarized emission. On the other hand the distance derived from
X-ray absorption (about 3 kpc) and the pulsar dispersion measure
($\sim$ 7 kpc) are much larger, making the distance  still very
uncertain \citep{chevreyn11}.

\subsection{A Model of X-ray Emission from the Vela BSPWN}\label{VelaXmod}

An analysis of the {\sl ROSAT HRI} X-ray image by
\citet{1998MmSAI..69..927M} suggested that the compact $\sim$ 2
arcmin "kidney bean" shaped X-ray nebula is due to  the bow shock
produced by
 the fast moving Vela pulsar,  with an inferred velocity of 260$\kms$.
 However, as we referred above, the derived proper transverse space velocity of pulsar PSR B0833--45
 is only 61$\pm 2 \kms$ \citep{2003ApJ...596.1137D}, while the estimated  plasma temperature
 inside the Vela SNR is kT $\gsim 0.2$ keV (though inhomogeneous).
 Therefore the proper velocity of PSR B0833--45 is too slow to launch a
 bow shock in such a hot medium. On the other hand the flow behind
 the SNR reverse shock \citep[see e.g.][]{1999ApJS..120..299T} may
 produce the bow shock. \citet{chevreyn11} noted that there is a transition period before the
Sedov solution is reached and this would allow to produce the mildly
supersonic flow for the centrally located pulsar PSR B0833--45. They
suggested that the reverse shock has recently passed over the Vela
pulsar and the plasma flow is mildly supersonic with
 sonic Mach number ${\cal M}_{\rm s} \sim$ 1.3 at the pulsar position.

As we discussed in \S\ref{CSFs}, the colliding shock flow is expected
to occur between the termination surface of the highly relativistic
PW and the pulsar bow shock. The magnetized PW is likely highly
anisotropic \citep[see
e.g.][]{2003MNRAS.344L..93K,2004A&A...421.1063D,2009ASSL..357..421K,2014MNRAS.438..278P}.
Relativistic magnetohydrodynamic simulations, supplied with some
recipe for the transport and spectral evolution of the accelerated
particles,  have been extensively used to assess the relation
between the wind properties and nebular emission. This has been done
through synthetic maps of PWNe, representing their synchrotron
emissivity, spectral index and polarization \citep[see
e.g.][]{2006A&A...453..621D,2008A&A...485..337V,2016JPlPh..82f6301O,2016MNRAS.460.4135P}.

Recently  \citet{2016MNRAS.462.2762B} performed  axisymmetric
relativistic magnetohydrodynamic simulations to test the dependence
of the PWN morphology appearance on the PW magnetization and
obliquity of the pulsar. The simulations suggested that in order to
reproduce qualitatively the double rings appearance of the Vela PWN,
the pulsar obliquity angle should be about $\pi/4$ and the  wind
magnetization should be rather high, with $\sigma_0 \sim 3$ before
annihilation of the magnetic field in the striped region, and
$\sigma \sim 0.12$ after that \citep[see
e.g.][]{2011ApJ...741...39S}. The authors simulated the PWN
synchrotron emission following the evolution of the PW and assuming
that the photon emissivity depends on the Doppler factor of the
local flow and magnetic field transverse to the line of sight. The
origin and spectral evolution of the accelerated particles were not
directly modeled in these RMHD simulations: particles were just
assumed to be there, due to acceleration either at the termination
shock or further downstream in the nebula. The relatively large
residual magnetization of the flow in the simulation by
\citet{2016MNRAS.462.2762B} even after the field annihilation
region makes diffusive shock acceleration of the pairs at the
ultrarelativistic termination shock rather problematic
\citep{2013ApJ...771...54S}, and the maximal energies estimated
according to the Eq.(\ref{Gmax1SSA}) are not enough to provide the
extended hard X-ray emission observed by
\citet{2011PASJ...63S.827K,2011ApJ...743L..18M}.

Since we aim to discuss here mostly the effect of the bow shock on
the Vela PWN high energy emission spectrum, let us assume that a
power law distribution of accelerated pairs $\propto Q_0
\gamma^{-s}$, with $s= 2.2$ up to $\gamma_{\rm max} = 10^9$ (which
is below the value estimated from the magnetospheric potential $\sim \sqrt{3\dot{E}/2c}$),
is injected at the pulsar wind
termination surface. Then we model the particle propagation and
re-acceleration in the bow shock PWN through the Monte-Carlo model
which described in \S\ref{geomMC}.

Several studies have addressed the propagation of accelerated pairs
across a PWN and the consequent variations of spectral index.
\citet{2012ApJ...752...83T} concluded that a purely diffusive model
can  well reproduce the spectral index maps as well as the nebular
size for PWNe without an apparent bow shock signature, like the Crab
nebula, G21.5-0.9 and 3C 58. This is because in these PWNe
one-dimensional advection of the pairs dominates only close to the
pulsar, while on larger scales the streamlines become so twisted
that the motion can be described as turbulent diffusion.
\citet{2016MNRAS.460.4135P} recently modeled the transport of
high-energy accelerated pairs in PWNe using a three-dimensional
magnetohydrodynamic simulation of the nebular dynamics and a
test-particle propagation model for the accelerated pairs. The
authors found that strong  fluctuations of the flow velocity
resulted in energy independent diffusion in PWNe with
P$\grave{e}$clet number of the order of unit. The bow shock flows
were not modeled in these papers.

As extensively discussed above, in the case of bow shock PWNe, the
MHD flows are important further away from the pulsar and the
colliding flows may result in pair re-acceleration causing a
spectral hardening. This is clearly seen in Fig.~\ref{Spectra_VelaP}
where the local spectra in the different parts of the bow shock PWN
are presented. The simulation run is for a magnetic field strength
of 200 $\mu G$ immediately downstream of the PW termination shock
(section 1 in Fig.\ref{fig:Geom1}), 120 $\mu G$ at the bow shock
(section 3 in Fig.\ref{fig:Geom1}), 20 $\mu G$ in the region between
the two shocks (section 2 in Fig.\ref{fig:Geom1}). The flow velocity
is taken to be at 500 $\kms$.

In Fig.~\ref{Spectra_VelaP} we illustrate the simulated local spectra
of e$^{\pm}$  pairs in different zones of a BSPWN (see \S\ref{geomMC})
for parameters similar to  those estimated for the Vela PWN. Spectra are shown for a
few bins of the Monte-Carlo model, whose locations are given in Fig.~\ref{fig:Geom1}.
The red curves correspond to the PW termination shock downstream:
the solid one refers to the $A2$ bin, the dashed one --- to
the $A4$ bin.
The blue curves correspond to the bow shock region (bin $C2$ -- the
solid curve, bin $C4$ -- the dashed curve), while the green ones ---
to the region between the two shocks (bin $B2$ -- the solid
curve, bin $B4$ -- the dashed curve).
Because of the efficient $e^{\pm}$ pairs
re-acceleration in the colliding flows the pairs spectra in the
latter two regions are much harder than what was injected at the
wind termination surface.


The radiation spectra in the multi-zone model of a Vela-like PWN are
shown in Fig.~\ref{Spectra VelaR}. The orientation of the source is
given by the angles $\Phi = 90^{\circ}$, $\Theta = 90^{\circ}$, and
$\Psi = 270^{\circ}$ (see \S\ref{geomMC:spectra}), i.e., the pulsar
velocity in the local ambient matter rest frame lies in the plane of
sky perpendicular to the line of sight crossing the pulsar (plane
$PoS$) and is directed to the north. The curves show the spectral
energy distribution ($\nu F_{\nu}$) calculated by integration of the
spectral emissivity  along the lines of sight crossing the same
locations. The colors and styles of the curves are the same as in
Fig.~\ref{Spectra_VelaP}

 It should be noted here that the radiation spectra presented
in Fig.~\ref{Spectra VelaR} are simulated for the outer regions of
the PWN where the flow velocity typically does not exceed 0.3~c. The
effect of particle acceleration in the colliding plasma flows
between the downstream of the relativistic termination shock and the
bow shock results in a very hard spectrum of the diffuse
X-ray emission in the vicinity of the bow shock. The bright X-ray
arcs, rings, and knots dominating the emission from the inner part
of the nebula in Fig.~\ref{fig:VelaXC} are likely highly enhanced by
the Doppler boosting in the relativistic outflow in the vicinity of
the termination shock \citep[see,
e.g.,][]{2004MNRAS.349..779K,2006A&A...453..621D,2016MNRAS.462.2762B}.
Hence, to derive the flux of the e$^{\pm}$ pairs injected into the
CSF acceleration from the observed X-ray image shown in
Fig.~\ref{fig:VelaXC} one has to use a quantitative model of the
Doppler boosted synchrotron X-ray radiation from the
transrelativistic outflow at the wind termination surface. Having
this in mind, we show in Fig.~\ref{Spectra_VelaP} the spectra of
e$^{\pm}$ pairs simulated for a given arbitrarily normalized
spectrum of the pairs, which were injected into the downstream of
the termination shock. The absolute flux of the injected pairs (the
red curve in Fig.~\ref{Spectra_VelaP}) determines the fluxes of the
pairs accelerated at the CSF (shown as blue and green curves in
Fig.~\ref{Spectra_VelaP}). The spectra were simulated within the
test particle approximation, which assumes that the pressure of the
accelerated particles is below the magnetic pressure at the bow
shock.

The spectral shape above a few keV is sensitive to the position of the apex and
the magnetic field strength at the bow shock.
The spectra presented in the Fig.~\ref{Spectra VelaR} illustrating the modeling of a Vela-like PWN
correspond to a rather large value of the bow shock apex.

The acceleration time to reach in CSF the maximal $e^{\pm}$ energies
shown in Fig.~\ref{Spectra_VelaP} (blue and green curves) according
to Eq.~(\ref{time1}) is about 20--30 yrs for the bow shock upstream
flow velocity $\sim 500 \kms$ in a Vela-like PWN.

In Fig.~\ref{fig:Vela_index} the power law photon spectral indexes
obtained by fitting
 the simulated synchrotron spectra in a wide energy range are presented. In
 the {\sl Chandra ACIS} energy range, at both the bow shock and in the
 region between the termination and bow shocks, we find photon indexes ${\rm \Gamma} <
 1.5$ consistent with what observed in the arcminute-size
 X-ray nebula \citep{2001ApJ...556..380H,2001ApJ...552L.129P}.
These indexes are substantially harder than what one would expect
for accelerated $e^{\pm}$ pairs injected at the wind termination
surface. The magenta line in Fig.~\ref{fig:Vela_index} corresponds
to a position outside the PWN, before the bow shock (the northern
region), but inside the Vela SNR where large scale X-ray emission of
power law photon index $\approx$ 2.4 was found in the 2-10 keV {\sl
Suzaku/XIS} observations reported by \citet{2011PASJ...63S.827K}.

The red curve in Fig.~\ref{Spectra VelaR} shows the spectrum of the
synchrotron radiation on  the lines of sight which are crossing the
PW termination shock downstream. The spectrum  was simulated
assuming a regular magnetic field in the emitting volume. It is
likely, however,  that the magnetic fields in PWNe are fluctuating.
This would result in a  particularly strong variability in the steep
gamma-ray tail of the spectrum, where modest fluctuations of the
magnetic field may lead to strong flares of spectral flux
\citep{2012MNRAS.421L..67B}. {\sl eASTROGAM}, a space mission
dedicated to the gamma-ray observations with sensitivity $\sim
10^{-12} \enf$ in the energy range 0.2 -- 100 MeV, extending up to
GeV energies \citep[see, e.g.,][]{2016SPIE.9905E..2NT}, can be used
to detect or constrain the flux variations in MeV energy regime from
the Vela PWN.

It is worth to note that the Vela SNR, being a nearby young SNR
hosting a PWN is  expected to be a major contributor to the observed
leptonic cosmic ray component \citep{2011ApJ...743L...7H}. The
spectrum of $e^{\pm}$ injected by the Vela SNR consists of both
particles accelerated at the SNR forward shock and pairs accelerated
in the Vela PWN which escape from the nebula and its bow shock. The
spectral index of the $e^{\pm}$ escaped the BSPWN is likely
harder than that produced at the SNR forward shock and extends up to
$\sim$ 50 TeV. Therefore the spectrum of $e^{\pm}$ injected from the
Vela SNR would have a spectral break at some energy below 50 TeV
with a softer spectrum above the break energy. A break at much lower
energy might actually be required not to overpredict the flux of CR
leptons, as constrained from AMS-02 measurements
\citep{2014PhRvL.113l1102A} and the latest analysis of Fermi/LAT
\citep{2017arXiv170300460D}. We shall discuss more extensively the
role of PWNe as the CR positron sources in the Galaxy in
\S\ref{sec:posintro}.

The ultrarelativistic $e^{\pm}$ pair spectra shown in
Fig.~\ref{Spectra_VelaP} were produced in the colliding flow behind
the bow shock. If the Bohm diffusion model is a good approximation for the
transport of pairs in region ``3" of the Monte-Carlo model (see \S\ref{geomMC:geometry}),
then, from Eq.(\ref{time1}) the time
needed to accelerate the pairs to $\gamma \gsim 3 \times 10^8$ in a
magnetic field of $\gsim 10 \mu$G is below 30 yrs. Therefore, within
this model the high energy pair spectrum of the Vela PWN is indeed
multi-component because of the colliding flow produced by the bow
shock. This study supports the interpretation of the Vela PWN as the
result of the interaction of the PW with a mildly supersonic MHD
inflow \citep{chevreyn11}. The flow  bypasses the pulsar with speed
$\gsim 500 \kms$, as might happen in a transient phase while the
reverse shock approaches the center of the progenitor supernova
remnant.

The multi-component model outlined above may help to understand the
observed X-ray morphology and spectra on the few arcminutes size
nebula. We discussed in the \S\ref{extended} the problem of the
escape of $e^{\pm}$ pairs as well as protons accelerated at a bow
shock PWN. The structure of the large scale magnetic field around
the PWN can determine the escape of some of the accelerated
particles. In the case of the Guitar nebula and the Lighthouse
nebula, as we have argued earlier, the interstellar magnetic field
is likely nearly transverse to the pulsar proper motion.

The local magnetic field  in the degree scale size region around the
Vela PWN is determined by the structure of the MHD flow inside the
Vela SNR. If the magnetic field is nearly radial and aligned with
the flow velocity behind the SNR reverse shock, then one may
understand  what observed in the very high energy gamma-rays by
H.E.S.S. and shown in Fig.~\ref{fig:VelaXH}. Indeed, pairs with
very hard spectra, with particle spectrum index $\sim$ 1.1--1.5,
accelerated  up to $\gamma \gsim
10^8$ at the colliding flow in the region between the wind
termination surface and the bow shock (see the blue curve in
Fig.~\ref{Spectra_VelaP}) can escape the source. The escaped
particles would propagate along the mean field to a few parsec scale
distances without appreciable losses (assuming the large scale
magnetic field to be below $\sim 50\mu$G). They would produce the
TeV range emission by inverse Compton scattering of the background
photons. Note that earlier X-ray studies  by
\citet{2010ApJ...719L.116B} of PWNe of different ages, up to $\sim$100
kyr, but all characterised by low surface brightness, suggested that
the accelerated electrons up to $\sim$ 80 TeV can escape from the
PWNe without relevant energy losses. Many authors \citep[see,
e.g.,][]{2008ApJ...689L.125D,2006A&A...448L..43A,2010ApJ...713..146A,2010Sci...327..663P,2013ApJ...774..110G,2012A&A...548A..38A},
who
 studied different aspects of the extended Vela X region discussed above, have
 concluded that multiple populations of accelerated particles are needed
 to explain the object. In the case of bow shock PWNe, acceleration of pairs in the colliding flow structure can
 naturally produce an extra particle population, in addition to acceleration at the termination shock.

Some Vela-like objects show hard indexes of the X-ray emission. The
pulsar wind nebula G106.6+2.9, with a thick toroidal structure, may
have much in common with the Vela PWN, and the scenario discussed
above can explain the hard photon index ${\rm \Gamma} = 1.3 \pm0.1$
observed in X-rays \citep{KP08}. Also the PWN of the young pulsar PSR B1823-13
showing properties similar to the Vela pulsar has a rather hard
photon index, ${\rm \Gamma} =1.3\pm 0.4$ \citep[see e.g.][]{KP08},
while the progenitor SNR has not been detected yet
\citep{2003ApJ...588..441G}. The radio pulsar B1046-58 is associated
with an asymmetric PWN, which may imply a supersonic proper motion
\citep{vela_like06}. The X-ray emission from this source also shows
a hard photon index ${\rm \Gamma} = 1\pm 0.2$ \citep{KP08}.  The
Vela-like object, PWN G284.0-1.8, associated with pulsar J1016-5857,
is characterized by an X-ray spectrum which can be  fitted by a
power law with index ${\rm \Gamma} =1.32\pm 0.25$
\citep{2004ApJ...616.1118C}. The PWN around the fast-spinning
energetic
pulsar PSR J0855-4644 
likely produced an elongated axisymmetric X-ray nebula  with a
longitudinal extent of 10$\arcsec$, with a very hard spectrum of
photon index ${\rm \Gamma} \sim 1.1$ \citep{2016arXiv161006803M}.
Moreover, \citet{2016arXiv161101863H} showed that at least a part of
the shell emission observed by H.E.S.S. from the supernova remnant
RX J0852.0-4622 (also dubbed Vela Junior) is likely due to the
presence of a PWN around PSR J0855-4644 located near the
south-eastern rim of the SNR. The origin and age of PSR J0855-4644
are not yet established \citep[see e.g.][]{2013A&A...551A...7A}: it
can be interacting with the ejecta of RX J0852.0-4622 SNR,  likely
located at a distance of about 1 kpc and may be associated with the
Vel OB1 association (0.8 kpc) \citep[see
e.g.][]{2015ApJ...798...82A}. Among the PWNe sources which
demonstrated a prominent GeV-TeV emission listed in
\citet{2013arXiv1305.2552K,2013ApJ...773...77A} only very few have
photon indexes of the TeV emission below 2.0. All of these very
interesting objects need further multi-wavelength studies to see if
there is indeed a subclass of Vela-like PWNe.


\section{Positrons from BSPWNe}\label{sec:posintro}

As already mentioned in previous sections, BSPWNe are likely to play
an important role in explaining one of the most intriguing anomalies
recently discovered in the field of cosmic ray physics, the
so-called {\it positron excess}. In the following we briefly review
what this excess is, why it is both mysterious and potentially very
important, and try to assess what role BSPWNe might play in
explaining it.

\subsection{The {\it Positron Excess}}\label{sec:excess}

\begin{figure}
\includegraphics[width=0.75\textwidth]{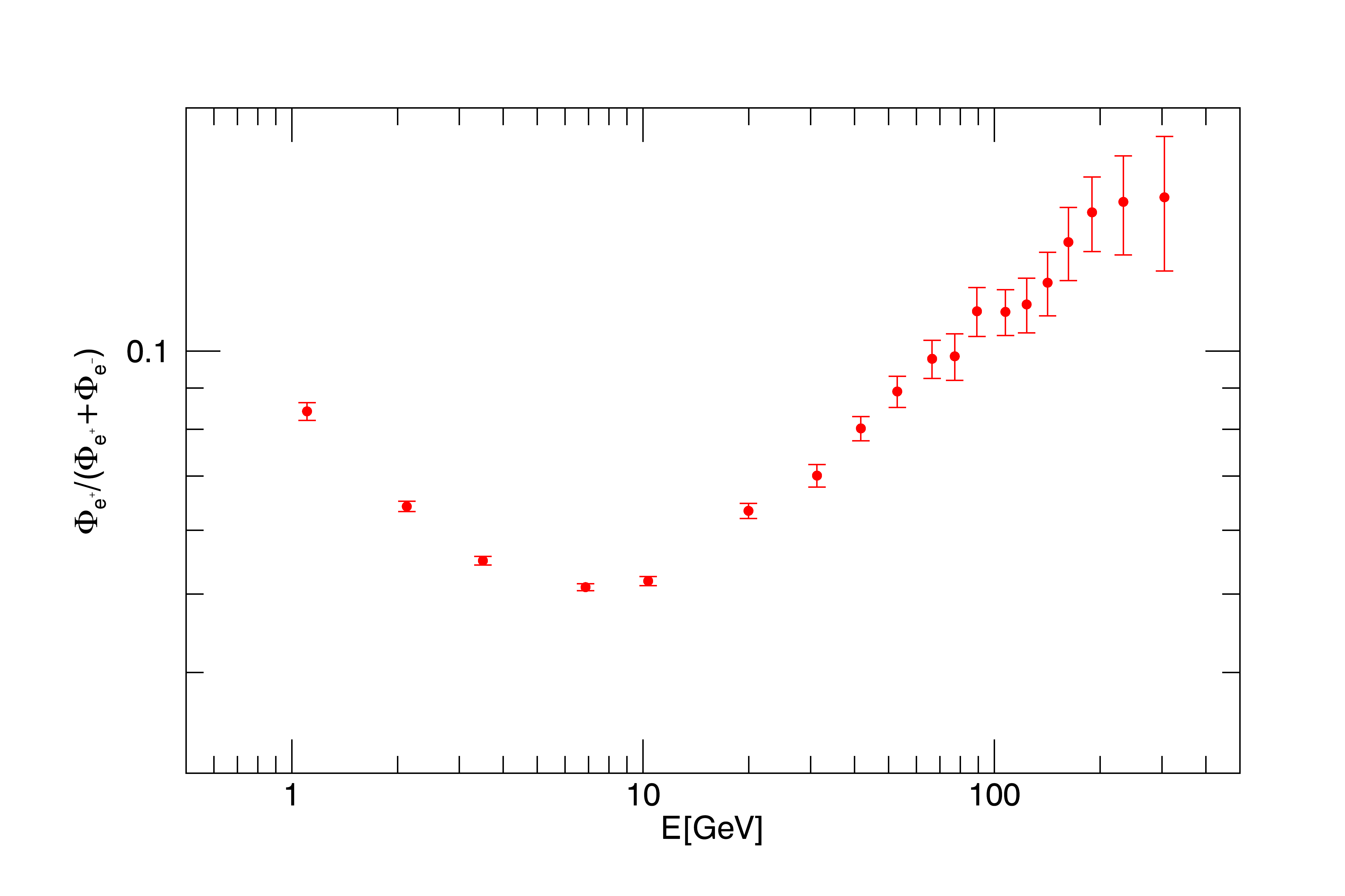}
\caption{The positron fraction measured by AMS-02 \citep{2013PhRvL.110n1102A}
as a function of energy. The figure has been created using the data
published by \cite{2013PhRvL.110n1102A} as supplemental material.}
\label{fig:amsfrac}
\end{figure}

Galactic Cosmic Rays (CRs)
are known to be mainly made of protons (90\% by number), with a
minor fraction of heavier nuclei (about 10\% of Helium in the GeV
range and minor percentages of more massive species), a tiny
fraction, about 1\%, of electrons, and an even smaller fraction of
antimatter, positrons and antiprotons. In recent years a few
experiments have been devoted to precise measurements of the
composition and spectrum of CRs, finding intriguing results on both
their hadronic and leptonic components. The most relevant results
for the discussion in this section concern the latter and have been
first obtained with sufficiently high significance by PAMELA
\citep{2009Natur.458..607A} and more recently confirmed with even higher
statistics by AMS02 \citep{2013PhRvL.110n1102A}. What these two experiments
found is that the ratio between the positron flux ($\Phi_{e^+}$) and
the total electron ($\Phi_{e^-}$) plus positron flux in CRs,
$\chi=\Phi_{e^+}/(\Phi_{e^+}+\Phi_{e^-})$, increases with energy for
energies above $\sim$ 6 GeV. The results published by AMS-02
\citep{2013PhRvL.110n1102A} are reported in Fig.~\ref{fig:amsfrac}.

A growing positron fraction is in plain contrast with the standard
scenario of CR origin and propagation and indeed, its unequivocal
detection by PAMELA (following earlier, but statistically less
significant claims by CAPRICE \citep{2001AdSpR..27..669B}, HEAT \citep{2004PhRvL..93x1102B} and
AMS-01\citep{2007PhLB..646..145A}), raised considerable interest in the scientific
community, one of the main reasons being the fact that the excess
positrons might be revealing dark matter related processes.

Before discussing how this excess could be related to the class of
sources being discussed in this article, let us briefly review why
it appears so puzzling and interesting.

In the standard model for the origin and propagation of Galactic
CRs, these particles are accelerated in the blast waves of Supernova
explosions and propagate diffusively throughout the Galaxy
interacting with the interstellar gas. The SN blast wave picks up
the particles to be accelerated from the Interstellar Medium (ISM),
and hence only ordinary matter enters the acceleration process.
Electrons are accelerated together with protons and heavier nuclei,
but positrons (and antiprotons) present in CRs have long been
considered to be of purely secondary origin, namely deriving from
the nuclear collisions of CR protons with the ISM during their journey
through the Galaxy. The most relevant chain of reactions for positron
production involves p-p scattering
leading to the generation of charge pions. These subsequently decay
into muons that finally produce positrons and neutrinos: \be
p+p\rightarrow p+n+\pi^++....\ , \, \, \, \, \, \, \, \, \,
\pi^+\rightarrow\mu^++\nu_\mu, \, \, \, \, \, \, \, \, \,
\mu^+\rightarrow e^++\nu_e+\bar \nu_\mu. \label{eq:secured} \ee A
similar reaction involving negatively charged pions lead to electron
production. In the standard scenario, all these secondary leptons
are much less numerous than primary electrons. In addition, their
ratio to primaries is expected to monotonically decrease with
increasing energy. The reason for this expectation is more easily
understood in the case of the ratio between secondaries and
primaries that do not suffer energy losses, while the actual
demonstration in the case of leptons that do suffer appreciable energy losses
in the relevant energy range is more cumbersome.

Let us therefore consider, just for illustrative purposes, the ratio
between secondary and primary nuclei that do not suffer relevant
energy losses during propagation. In relation to CR confinement, let
us describe the Galaxy as a cylinder of radius $R_d \approx 15$ kpc
(the star and gas disk radius) and height $H\approx 3$ kpc (as the
most recent propagation models suggest for the extent of the
magnetized halo of the Galaxy, see {\it e.g.}
\citet{2011ApJ...729..106T}). Let us further assume that CRs diffuse
through the volume of the Galaxy with a spatially uniform diffusion
coefficient $D(E)\propto E^\delta$, where $E$ is the CR energy and
$\delta >0$. Finally, particles are considered free to leave the
Galaxy through the vertical bounds at $\pm H$. This is the so-called
{\it Leaky Box Scenario} for CR propagation.

In the absence of energy losses, the equilibrium spectrum of primary
CRs, $N_{\rm prim}(E)$, will then simply result from balance between
injection and escape from the Galaxy \be N_{\rm prim}(E)=Q_{\rm
prim}(E)\tau_{\rm esc}\propto Q_{\rm prim}(E)
\frac{H^2}{D(E)}\propto E^{-(\gamma_{\rm inj}+\delta)}\ ,
\label{eq:primspec} \ee where we have taken the number of particles
injected by the sources per unit time and energy interval to scale
with energy as $Q_{\rm prim}(E)\propto E^{-\gamma_{\rm inj}}$.

Following the same reasoning we can also derive the equilibrium
spectrum of secondaries. Now the injection relies on the spallation
reactions of primary nuclei and is then described by \be Q_{\rm
sec}(E)\propto N_{\rm prim}(E) \sigma_{\rm sp} c n_H \propto
E^{-(\gamma_{\rm inj}+\delta)}\ , \label{eq:secinj} \ee where
$\sigma_{\rm sp}$ is the spallation cross-section (assumed to be
independent of energy, which is a good approximation in the GeV-TeV
energy range), $c$ is the speed of light and $n_H$ is the density of
target material.

As a result, the equilibrium spectrum of secondaries will be: \be
N_{\rm sec}(E)=Q_{\rm sec}(E)\tau_{\rm esc}\propto E^{-(\gamma_{\rm
inj}+2\delta)} \label{eq:secspec} \ee and the secondary-to-primary
flux ratio will scale as: \be \frac{\Phi_{\rm sec}(E)}{\Phi_{\rm
prim}(E)}\propto E^{-\delta}. \label{eq:secrat} \ee

While absence of energy losses is a good approximation for protons
and nuclei of all energies between a few GeV and the knee, the same
is not true for leptons, which lose energy through synchrotron and
Inverse Compton radiation while propagating through the Galaxy. If
we take the standard value of magnetic and radiation energy density
in the Galaxy, $U_B\approx U_{\rm rad}\approx 1$ eV cm$^{-3}$, the
radiative loss time, $\tau_{\rm loss}(E)\approx 2.5\times 10^8$ yr
$(E/{\rm GeV})^{-1}$, is shorter than the propagation time
$\tau_{\rm esc}(E)\approx 2 \times 10^8$ yr $(E/{\rm GeV})^\delta$
already at $E>$ few GeV, for $0.2<\delta<0.8$, as inferred from
recent studies of CR propagation \citep{2011ApJ...729..106T}.
In this case, in the energy
range of our interest, the equilibrium spectrum of electrons, which
we approximate as all primaries, is described by: \be N_{\rm
e^-}(E)\propto E^{-\gamma_{\rm inj,el}}\ E^{-(1+\delta)/2}\ ,
\label{eq:elspec} \ee where the first term on the {\it rhs} comes
from injection and the second from propagation. The form of the
latter is determined by the combination of diffusive propagation and
radiative energy losses that limit the number of contributing
sources with respect to the case of nuclei, imposing constraints in
terms of distance in space and time.

Analogously, for positrons, that are all secondaries, we find: \be
N_{\rm e^+}(E)\propto E^{-(\gamma_{\rm inj,p}+\delta)}\
E^{-(1+\delta)/2}\ , \label{eq:posspec} \ee where what appears in
the first term of the {\it rhs} is the equilibrium proton spectrum
($\propto E^{-\gamma_p}$). The result for the positron fraction is
then: \be
\chi=\frac{\Phi_{e^+}}{\Phi_{e^+}+\Phi_{e^-}}\approx\frac{\Phi_{e^+}}{\Phi_{e^-}}\propto
E^{-(\gamma_{\rm inj,p}-\gamma_{\rm inj,el}+\delta)}\ ,
\label{eq:chiexp} \ee and in the standard scenario, with
$\gamma_{\rm inj,p}=\gamma_{\rm inj,el}$, $\chi\propto E^{-\delta}$
is expected to monotonically decrease with energy. This is why the
observed increase was found to be puzzling and stimulated a plethora
of attempts at explaining it.

\begin{figure}
\includegraphics[width=0.75\textwidth]{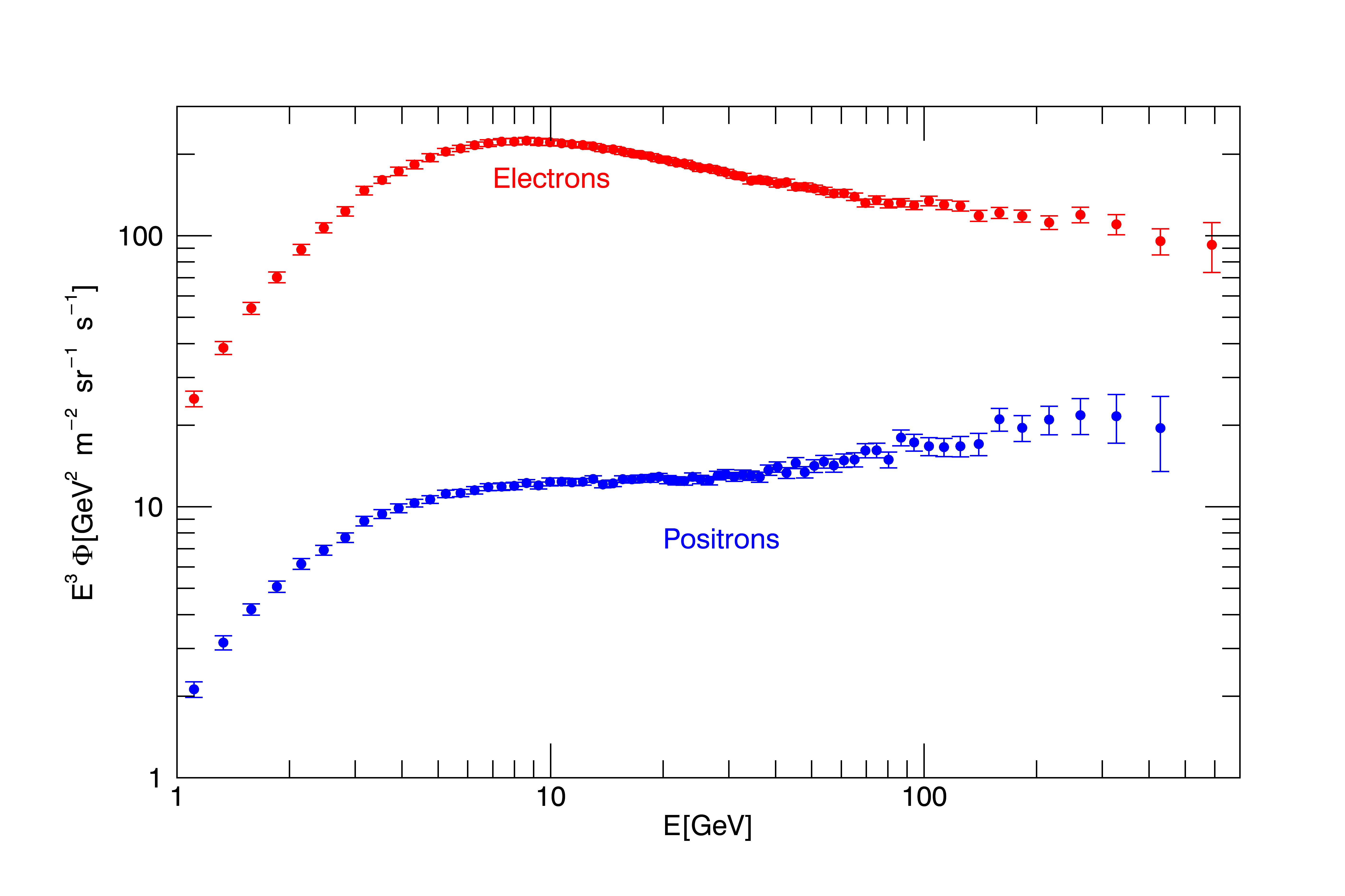}
\caption{The spectra of electrons and positrons  measured by AMS-02
\citep{2014PhRvL.113l1102A} as a function of energy. The figure has
been built from the data published by \cite{2014PhRvL.113l1102A} as
supplemental material.} \label{fig:amselpos}
\end{figure}

Before discussing the possible explanations and how BSPWNe enter the
scene, an important remark is in order. The prediction of a
decreasing positron fraction is a strong one only in a scenario that
assumes a spatially uniform distribution of sources. This is indeed
the assumption underlying Eqs.~\ref{eq:elspec} and \ref{eq:posspec},
but we know that this assumption does not apply to SNRs, the
putative sources of primary electrons: these are discrete sources in
space and time and their distribution in the Galaxy is not uniform
at all. The effects of the discreteness of the sources and their
rapidly decreasing density with Galactocentric distance were
analysed for CR nuclei by \cite{2012JCAP...01..010B}, and in that
case they were found to be poor: the resulting spectrum of all
different species still respected on average the predictions of the
simplified {\it leaky box model} (LBM hereafter). With electrons,
however, due to radiative losses, some difference might appear,
since especially at the highest energies the number of sources that
might contribute to the flux at Earth is likely to be very small,
and strongly dependent on the recent SN explosion history in the
solar neighborhood. A possibility would then be that it is actually
the electron flux that steepens with energy with respect to the LBM
prediction, rather than a flattening occurring in the positron flux.
This possibility has been ruled out by more recent analysis of the
PAMELA \citep{2011PhRvL.106t1101A} and AMS-02 data
\citep{2014PhRvL.113l1102A}, which clearly show a flattening of the
positron flux, a real {\it positron excess}. The spectra of
electrons and positrons obtained by AMS-02
\citep{2014PhRvL.113l1102A} are shown in Fig.~\ref{fig:amselpos}.

\subsection{Qualitative Evaluation of the PWN Contribution}\label{sec:qualitative}

An intriguing suggestion immediately made after this discovery was
that these extra positrons could come from dark matter annihilation
in the Galaxy (see e.g.
\cite{2012APh....39....2S,2016JCAP...05..031D} for a review). While
a dark matter related origin of the rise of $\chi$ is difficult to
rule out completely, astrophysical explanations seem favored at the
current time and the most natural place where to look for a source
of extra positrons is the magnetosphere of a pulsar, where a large
number of electron-positron pairs are created by every electron that
leaves the star surface.

These $e^--e^+$ pairs end up in the PW and are then
accelerated at its termination shock. The efficiency of conversion
of the pulsar spin-down power into particle acceleration is usually
thought to be very high, and measured to reach about 30\% in the
case of the Crab nebula, the prototype PWN. In addition, we know
from observations of many such objects that the particle spectrum is
usually a broken power law: from radio observations we infer a
particle spectral index $\gamma_R$ in the range $1<\gamma_R<2$,
while the X-rays come from particles with spectral index
$\gamma_X>2$. The boundary between the two regimes is inferred to
occur in the several hundreds GeV range. It is apparent then that
these are extremely promising candidates to explain the excess at
least from the spectral point of view: the excess is observed in the
energy region between few and 300 GeV (no data are available at
higher energies) and in this interval PWNe could well enter the
scene with their flat positron spectrum.

Of course as long as the particles are part of the PWN they are
confined within the remnant of the parent supernova explosion. Their
Larmor radii in the nebular magnetic field (of typical strength
around a few $\times$100 $\mu G$) are very small and diffusive
escape is not effective. The situation is different when the pulsar
breaks out of the remnant and finds itself in the ISM. The time at
which this happens and the energy that is left in the system at that
time were derived in \S~\ref{sec:intro}, where we found, for typical
values of the pulsar and ISM parameters, $T_{\rm esc}\approx few
\times 10^4$ yr and $E_{\rm res}\approx 10^{47}$ erg.

After time $T_{\rm esc}$ the supersonic motion of the pulsar through
the ISM transforms its PWN in a Bow Shock PWN. As we extensively
discussed in earlier sections, the plasma in front of the pulsar is confined
by the bow shock, but particles flowing through the tail are free to leave
the system and become part of the cosmic ray pool.

These particles are the interesting ones in view of explaining the
{\it positron excess}. Their spectrum, at the energies of interest,
is shown to be flat, similar to what
described above for their younger counterparts. In particular, radio
emission again shows a flat particle spectrum in the up to hundreds
of GeV. In addition, numerical simulations support the idea that
these low energy particles can easily escape from the system along
the tail with very little energy losses \citep{2005A&A...434..189B}.

These particles are easily shown to be numerous enough to explain
the excess. This can be done through the following
back-of-the-envelope calculation.

Let us go back to the rough description of the electron spectrum in
Eq.~(\ref{eq:elspec}), and let us assume that these particles are
accelerated in SNRs with the same spectrum as protons: given the
protons' spectrum measured at Earth (Eq.~\ref{eq:primspec})
$\gamma_{\rm inj,p}+\delta\approx 2.7$, one can promptly derive the
spectrum that must be injected by the sources as soon as the energy
dependence of the diffusion coefficient is known. For the latter we
assume a slope $\delta=1/3$, in agreement with a Kolmogorov's model
for the turbulence in the Galaxy (which is preferred to steeper
energy dependencies by considerations concerning the CR anisotropy
\citep{2012JCAP...01..011B}). We then obtain that electrons must be
injected with $\gamma_{\rm inj,el}\approx2.4$. The fraction of SN
explosion energy that is usually assumed to be converted into CR
acceleration is $\xi_{\rm CR}=0.1$ and most of it is carried by
protons. The ratio between accelerated electrons and protons that is
needed to interpret the CR measurements at Earth is usually $f_{\rm
e,p}\approx 0.01$. We can use this information to find the number of
electrons that SNRs inject in the Galaxy per unit time and unit
energy interval: \be Q_{e^-}(E)\approx (\gamma_{e^-}-2) {\cal R}\
\frac{\xi_{\rm CR} f_{e,p} E_{\rm
SN}}{E_0^2}\left(\frac{E}{E_0}\right)^{-\gamma_{el}}\ ,
\label{eq:snelinj} \ee with $\cal R$ the SN rate in the Galaxy and
$E_0=1$ GeV corresponding to the particles that carry most of the
energy in the distribution. If we now follow a similar line of
reasoning for the positrons, assuming that they are injected by
BSPWNe, which convert some fraction $\eta$ of $E_{\rm res}$ in a
distribution of electrons and positrons characterized by a flat
spectrum $E^{-\gamma_{pos}}$ with $\gamma_{pos}<2$, extending up to
$E_{\rm max}\approx 500$ GeV, we can write: \be Q_{e^+}(E)\approx
(2-\gamma_{pos}) {f_{II} \cal R}\ \frac{\eta E_{\rm res}}{E_{\rm
max}^2}\left(\frac{E}{E_{\rm max}}\right)^{-\gamma_{pos}}\ ,
\label{eq:snposinj} \ee where $E_{\rm res}$ is given in
Eq.~(\ref{eq:eres}) and $f_{\rm II}$ takes into account the fact
that only $\approx$ 80\% of the SN explosions give birth to a
pulsar. Since propagation affects electrons and positrons in the
same way, the ratio $\chi$ will simply scale as the injection, and
for energies large enough that the secondary positrons can be
neglected, will read \be \chi\hspace{-0.5mm}\approx\hspace{-0.5mm}
\frac{Q_{e^+}}{Q_{e^-}}\hspace{-0.5mm}=\hspace{-0.5mm}\frac{2-\gamma_{pos}}{\gamma_{el}-2}\
\frac{f_{II}\eta E_{\rm res}}{\xi_{\rm CR} f_{e,p} E_{\rm SN}}
\frac{E^{(\gamma_{el}-\gamma_{pos})}}{E_0^{(\gamma_{el}-2)}-E_{\rm
max}^{(\gamma_{pos}-2)}}\approx 0.035
\left(\hspace{-0.5mm}\frac{E}{30 {\rm
GeV}}\hspace{-0.5mm}\right)^{1/2} \label{eq:chiboe} \ee where the
last approximate equality holds for the typical values of the
parameters mentioned above. At 30 GeV this value of $\chi$ is within
a factor of 2 of the AMS-02 measurement reported in
Fig.~\ref{fig:amsfrac}, which strongly suggest that the contribution
of PWNe must definitely be taken into account when wondering about
the positron excess. In addition the energy dependence in
Eq.~(\ref{eq:chiboe}) also approximately agrees with the data.

\subsection{Positrons from BSPWNe}

A quantitative calculation of the flux of electrons and positrons
expected from BSPWNe was carried out by \cite{2011ASSP...21..624B},
who showed that BSPWNe are likely to contribute a large fraction of
the needed extra positrons, if not all of them. That work was
adopting some simplifying assumptions about the spatial distribution
of the SN explosions in the Galaxy (the presence of arms was not
considered) and was associating to every explosion a universal
pulsar with typical values of the parameters, rather than accounting
for the spread of parameter across the pulsar population. In the
following we show the results of a more refined version of the same
exercise, including the above mentioned missing ingredients  and the
most recent information available from AMS-02 on the spectrum of
primary protons \citep{2015PhRvL.114q1103A} and on the diffusion
coefficient \citep{2016PhRvL.117w1102A}. We first provide the
analytical expression for the contribution at Earth of a single
source and then sum up the contribution of all the potentially
relevant sources in the Galaxy, generated at random from an
appropriate distribution. Some other studies along the same
line were performed by \cite{2014JCAP...04..006D}, who considered
the contribution of putative nebulae associated with pulsars from
the ATNF
 catalogue, rather than a generic pulsar population.

The equation describing the transport of particles in the presence
of spatial diffusion, energy losses and catastrophic losses reads:
\be \frac{\partial f(E,\vec r, t)}{\partial t} - D(E)\nabla^2 f(E,
\vec r, t)+ \frac{\partial}{\partial E}\left(B(E) f(E,\vec r,
t)\right)+\frac{f(E, \vec r, t)}{T}=Q(E,\vec r, t)
\label{eq:transp1} \ee where $f$ is the number of particles of
energy $E$ per unit volume and energy interval at position $\vec r$
and time $t$, $D(E)$ is the particle diffusion coefficient,
$B(E)=dE/dt$ represents the particles' energy losses, $T$ is the
timescale for catastrophic losses (particle destruction) and
$Q(E,\vec r,t)$ is a source term (number of particles with energy
$E$ injected in the system per unit time and energy interval at
position $\vec r$ and time $t$ ). The solution of
Eq.~(\ref{eq:transp1}) can be found through the Green function method
\citep{1959SvA.....3...22S}, and allows one to compute the particle distribution
function at any location once $D(E)$, $B(E)$, $T$ and $Q(E,\vec
r,t)$ are known. In the absence of catastrophic losses, the Green
function reads: \be G(\vec r, t, E; \vec r_0, t_0,
E_0)=\frac{\exp\left[-\frac{\left(\vec r-\vec r_0\right)^2}{4
\lambda^2(E,E_0)}\right]}{\left(4 \pi
\lambda^2(E,E_0)\right)^{3/2}}\frac{\delta(t-t_0-\tau(E,E_0))}{\left|B(E)\right|}\
, \label{eq:green} \ee where $\delta$ is the Dirac delta function,
$\tau$ is the timescale of energy losses, and $\lambda$ is the
particle diffusion length. In formulae we have: \be
\tau(E,E_0)=\int_{E_0}^E \frac{dy}{B(y)} \, \, \, \, \, {\rm and} \,
\, \, \, \, \lambda^2(E,E_0)=\int_{E_0}^E \frac{D(y)}{B(y)}dy\ .
\label{eq:taulambda} \ee

The full solution of Eq.~(\ref{eq:transp1}) can be found through
convolution of the Green function in Eq.~(\ref{eq:green}) with the
appropriate source term.  Sources such as SNRs and BSPWNe are
localised in space and time. The injection by one such source,
located at position $\vec r_s$ at time $t_s$, can be described as
\be Q(E,\vec r, t)=N_{\rm inj}(E) \delta(\vec r - \vec r_s)
\delta(t-t_s)\ . \label{eq:sources} \ee Convolution of the Green
function in Eq.~(\ref{eq:green}) with the source term in
Eq.~(\ref{eq:sources}) provides an expression for the particle density
per unit volume and unit energy interval contributed at $\vec
r=(x,y,z)$ by a single source of space-time coordinates
$(x_s,y_s,z_s;t_s)$: \be
\begin{aligned}
f(E,\vec r,t)= & N_s(E_0)
\frac{\left|B(E_0)\right|}{\left|B(E)\right|}\
\frac{\exp\left[\frac{-(x-x_s)^2+(y-y_s)^2}{4
\lambda^2(E,E_0)}\right]}{\left(4 \pi
\lambda^2(E,E_0)\right)^\frac{3}{2}} \times \\
& \times \sum_{n=-\infty}^{\infty}
(-1)^n \exp \left[\frac{-(z-z_n')^2}{4 \lambda^2(E,E_0)}\right]
\end{aligned}
\label{eq:result}
\ee where \be E_0=\frac{E}{1-B(E)(t-t_s)/E}\ .
\label{eq:e0}
\ee In Eq.~(\ref{eq:result}) $z_n'=(-1)^n z_s+2nH$ and
the sum on the {\it rhs} is introduced to take into account the
boundary condition of particle escape from $\pm H$ (see e.g.
\citet{2012JCAP...01..010B} for a discussion).

Eq.~(\ref{eq:result}) can be used with any distribution of sources. In
particular, when a spatially uniform distribution of sources is
considered, it is easy to see that its integration over the source
volume leads to a particle spectrum as described by
Eq.~(\ref{eq:elspec}), which we used for our qualitative discussion.

\begin{figure}
\includegraphics[width=\textwidth]{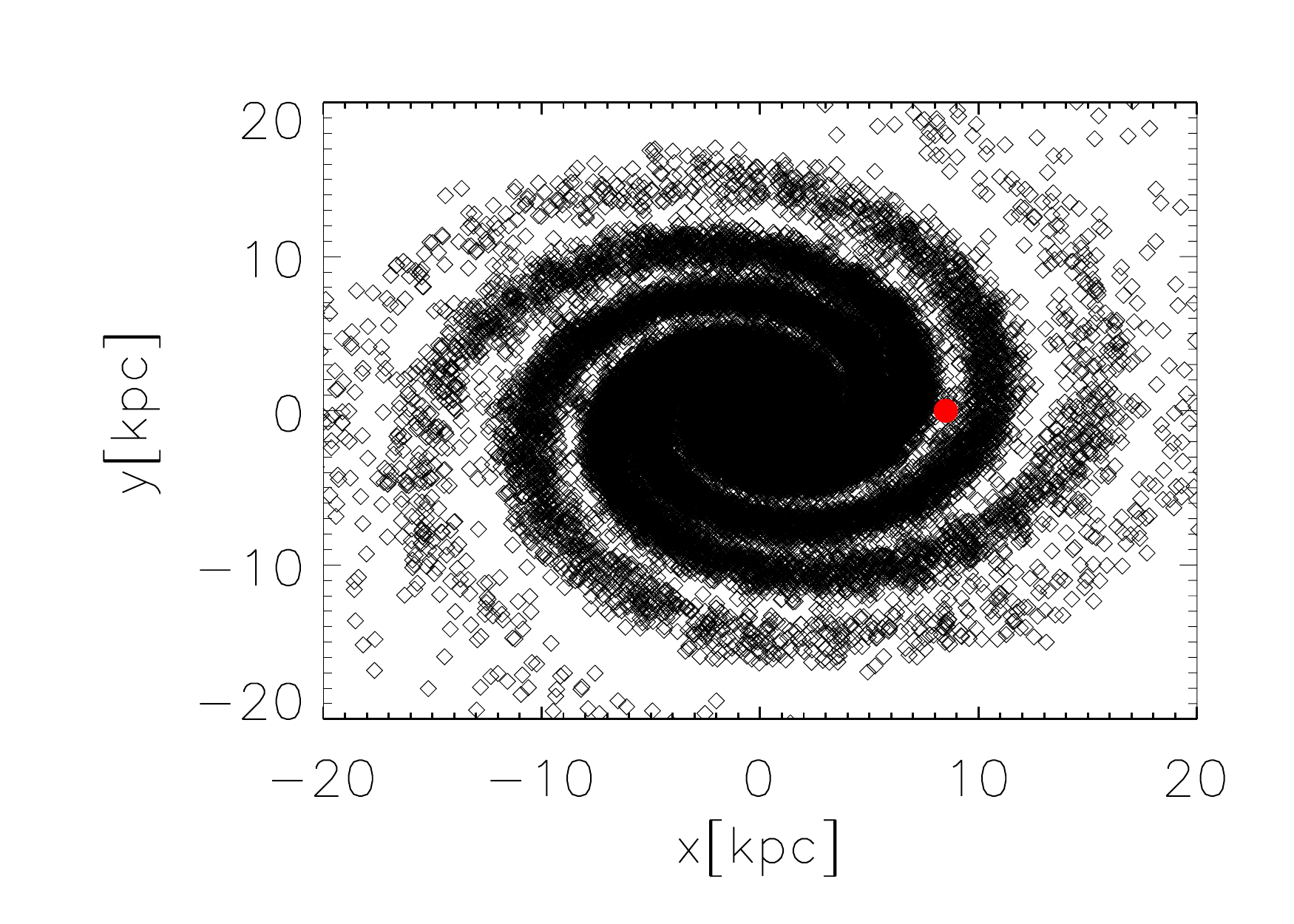} 
\caption{The projected positions in the Galactic plane of about
30000 SN explosions in one of the considered realisations of the
Galaxy. The red dot represents the position of the Earth.}
\label{fig:snlocation}
\end{figure}

However for a quantitative evaluation of the electron and positron
spectrum contributed by SNRs and PWNe the procedure to follow is
different. SN explosions must be generated at random places in the
Galaxy and at random times. In order to be sure to correctly
reproduce the spectrum of low energy CRs, one needs to include all
the explosions occurring in a time span corresponding to the
escape-time of these particles. To do this, we assume (as we always
will, hereafter) a Kolmogorov diffusion coefficient,  in agreement
with the the estimate by the AMS-02 team based on the B/C ratio
\citep{2016PhRvL.117w1102A}. The normalisation is taken according to
the most recent GALPROP analysis \citep{2011ApJ...729..106T},
resulting into: \be D(E)=1.33 \times 10^{28} {\rm cm}^2 {\rm
s}^{-1}\ \frac{H_0}{\rm kpc} \left(\frac{E}{3
GeV}\right)^\frac{1}{3}, \label{eq:diff} \ee where $H_0=3$ kpc is
the halo size. The escape time of 10 GeV particles is then
$\tau_{\rm esc}(10 GeV)=H_0^2/D({\rm 10 GeV})$=45 Myr. With an
explosion rate as described by \citet{2006ApJ...643..332F}, namely a
gaussian distribution around an average rate of $2.8$
explosions/century and a standard deviation of $0.1$ events/century,
the number of explosions that have to be taken into account is
$\approx 1.2 \times 10^6$. The explosions are assumed to occur at
random places in the Galaxy, spatially distributed as: \be n_{\rm
SN}(\vec \rho_s,z_s)\propto
\left(\frac{\rho_s}{R_\odot}\right)^\alpha\ \exp \left[-\frac{\beta
\left(\rho_s-R_\odot\right)}{R_\odot}\right]\ \exp
\left[-\frac{z_s}{h_{\rm SN}}\right] \label{eq:spatdistr} \ee where
a cylindrical coordinate system has been adopted, with $\rho_s$ the
radial coordinate of the explosion and $z_s$ its height above or
below the Galactic plane. The parameters appearing in
Eq.~(\ref{eq:spatdistr}) have the following values: the sun radial
distance from the Galactic centre is $R_\odot=8.5$ kpc, $h_{\rm
SN}=50$ pc (form \cite{2006ApJ...643..332F}), $\alpha=1.09$ and
$\beta=3.87$ from \cite{2015MNRAS.454.1517G}. The presence of
Galactic arms has been included on top of this profile, according to
the description given by \cite{2006ApJ...643..332F}. The resulting
distribution of SN explosions can be seen in
Fig.~\ref{fig:snlocation}, where the coordinates of about 30000
events are reported in a Galactocentric frame.

All these events are considered as actual contributors to the
primary electron flux at Earth, while only 80\% of them, randomly
selected, are assumed to be associated to type II explosions giving
origin to a pulsar and hence included in the count of positron
contributors. The flux of electrons and positrons is obtained as the
sum over all relevant sources of the single source contribution
described in Eq.~(\ref{eq:result}), where $N_s(E)=Q_s(E)/{\cal R}$
with $Q_s(E)$ given by Eqs.~\ref{eq:snelinj} and \ref{eq:snposinj}
for electrons and positrons respectively. The electrons' injection
spectral index is assumed to be the same as that of protons, and the
spectrum of the latter is taken equal to the fit to the newest data
provided by the AMS-02 collaboration \citep{2015PhRvL.114q1103A}.
The SN explosion energy that enters the normalisation is assumed to
be $E_{\rm SN}=10^{51}$ erg for all events. As far as the positrons
are concerned, observations of PWNe suggest that $\gamma_{pos}$ (and
accompanying $\gamma_{el}$) is different from source to source but
without an obvious relation to any other observable. The results
that are shown in the following refer to the choice
$\gamma_{pos}=1.2$ up to a maximal energy $E_{\rm max}=500$ GeV. A
steeper power law with index $2.3$ is assumed at higher energies.

$E_{\rm res}$ will also vary from source to source
depending both on the pulsar properties and on the environment in
which the explosion occurs. In the absence of a good and easy recipe
for how to vary the latter, previous work can be improved in realism
by taking into account variations of the former. The surface
magnetic field is assumed to be the same for all pulsars and
corresponding to the average value proposed by \citet{2006ApJ...643..332F},
$B_*=4.5\times 10^{12}$, but the pulsar's spin period at birth and
the proper motion vary among different objects according to the
distributions provided by the same authors.

\begin{figure}
\includegraphics[width=\textwidth]{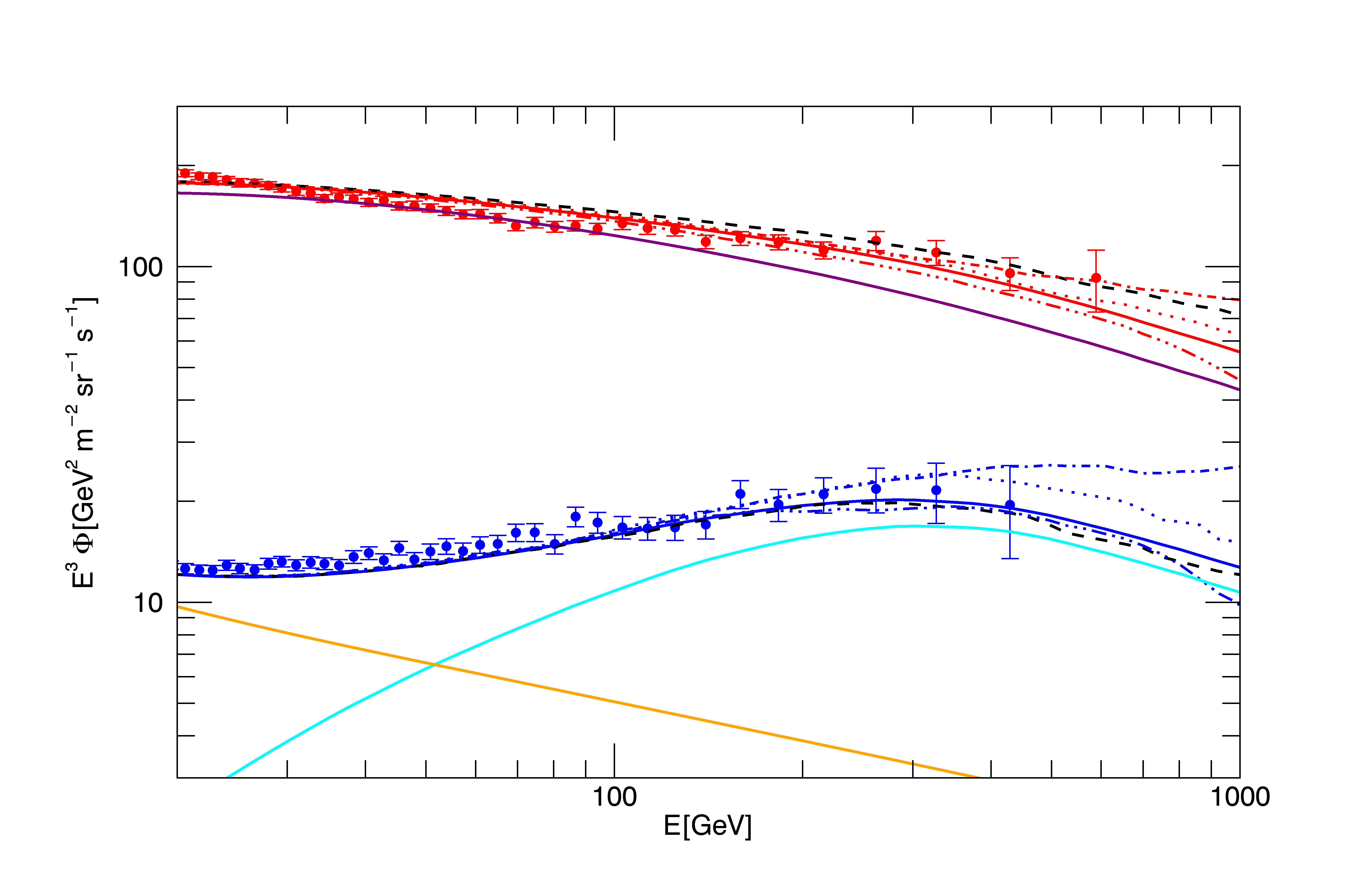}
\caption{The electrons (red) and positrons (blue) spectrum at Earth
as computed within the described model is compared with AMS-02
measurements. The legend for the different curves is as follows. The
solid thick curves refer to ensemble averages over the 30 different
realisations of the Galaxy, while the other curves correspond to
individual realisations. Averages are shown for the total electron
(red) and positron (blue) spectra, and also for the different
contributions: electrons and positrons of secondary origin (orange);
electrons and positrons from Bow Shock PWNe (light blue); primary
electrons from SNRs (purple). The shed black curve refers to a
particular realisation that seems to fit especially well the
positron fraction data.} \label{fig:elposspec}
\end{figure}

In addition to primary positrons from BSPWNe and primary electrons
from the same sources plus SNRs, we also account for secondary
electrons, generated from CR interactions with the ISM. The density
of the latter is taken to correspond to 1 particle/cm$^3$ and the
incident proton spectrum is taken again from AMS-02
\citep{2015PhRvL.114q1103A}.

In Fig.~\ref{fig:elposspec} the spectrum of electrons and positrons
resulting from this calculation is shown together with the AMS-02
data \citep{2014PhRvL.113l1102A}. Some of the individual curves,
corresponding to different realisations, are also shown in order to
give an idea of how different the electrons' and positrons' spectra
can be at high energy depending on the source configuration. A
conversion efficiency of $\eta=$22\% of the pulsar spin-down energy
into acceleration of pairs has been adopted.

\begin{figure}
\includegraphics[width=0.5\textwidth]{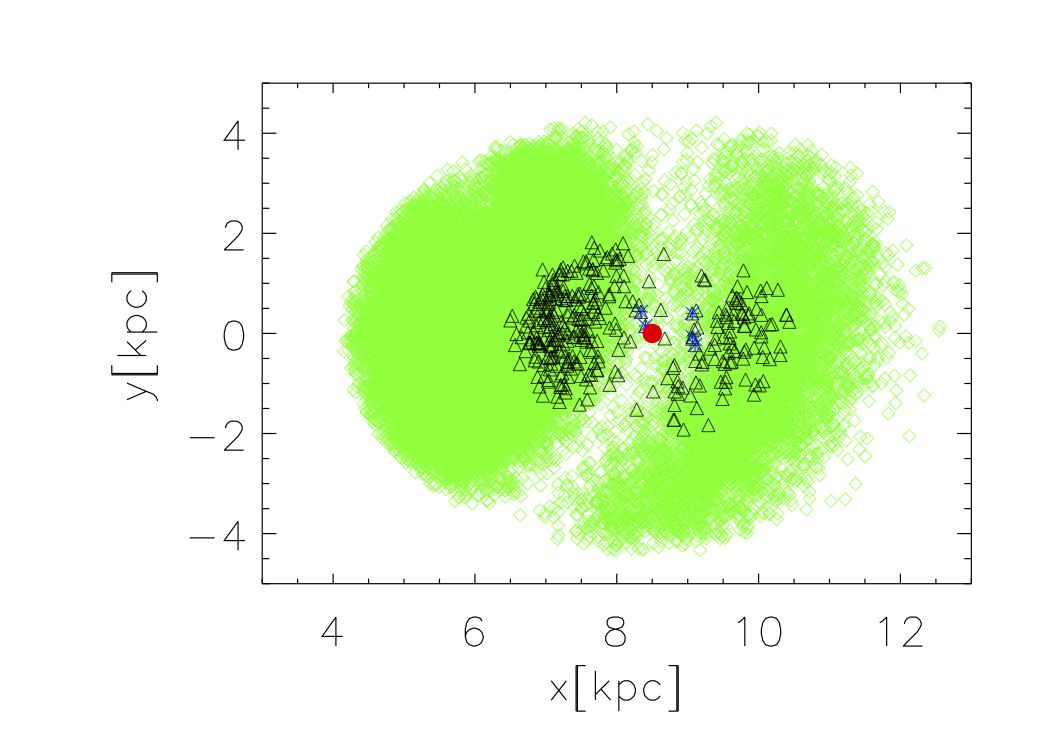}
\includegraphics[width=0.5\textwidth]{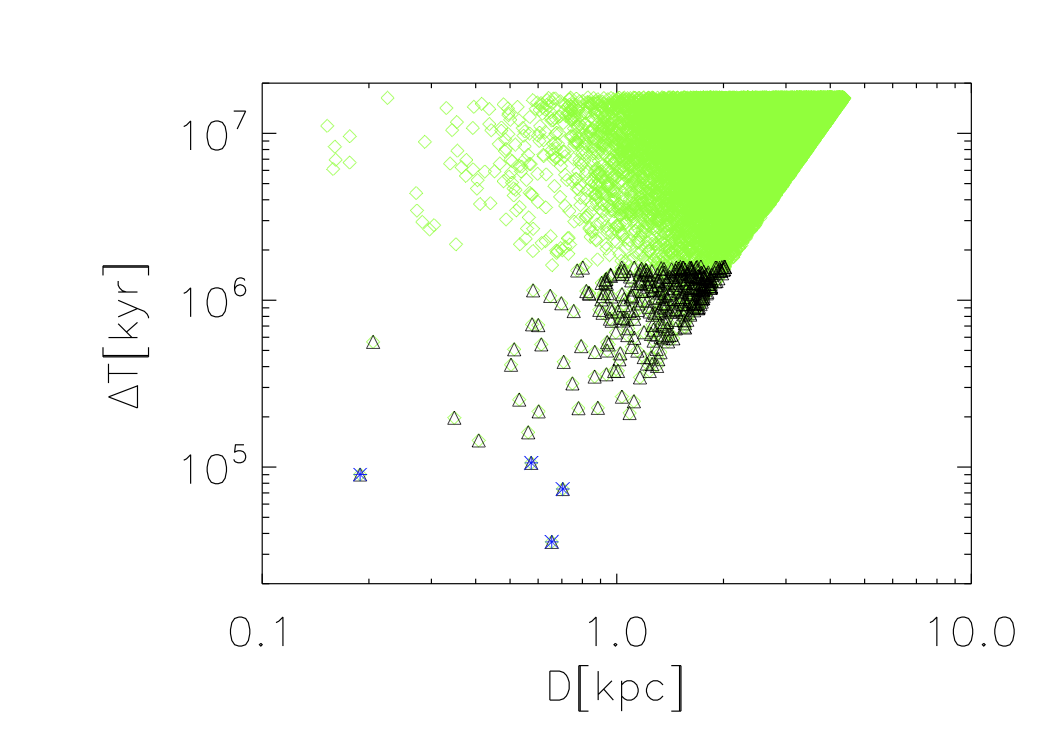}
\caption{In the left panel we show the projection onto the Galactic
plane of the positions of all sources that contribute positrons to
Earth at three different energies: 10 GeV (green diamonds), 100 GeV
(black triangles), 1 TeV (blue asterisks). The Earth position is
shown as a red circle. In the right panel, the distance in time and
space of the contributing sources is shown with the same notation as
on the left.}
\label{fig:srcnumber}
\end{figure}

The difference between the different realisations at high energies
depends on the strong decrease with increasing electron energy of
the number of contributing sources. Fig.~\ref{fig:srcnumber} serves
the purpose of highlighting this effect. The two panels of the
figure show the sources that in this calculation actually make a
contribution to the flux of positrons (and electrons) of different
energies at Earth. It is apparent that while at 10 GeV the number of
relevant sources (green diamonds) is large (more than 2000 in the
particular realisation of the Galaxy to which the plots refer), it
strongly decreases with increasing energy: at 100 GeV about 300
sources are relevant (black triangles in the plots), and only a handful at 1
TeV (blue asterisks). The panel on the left shows the projected
position in the Galactic plane. The fact that the Earth is in an
inter-arm region, poorly populated with sources is evident from the
plot and plays an important role: this is likely the main reason for
the need of a larger value of $\eta$ with respect to the calculation
by \cite{2011ASSP...21..624B}.

From the left panel of Fig.~\ref{fig:srcnumber} one sees instead
what a small range of distances in space and time can contribute to
positrons at the highest energies: already at 100 GeV all the
leptons that reach Earth come from sources closer than 2 kpc and
younger than 2 Myr.

\begin{figure}
\includegraphics[width=0.75\textwidth]{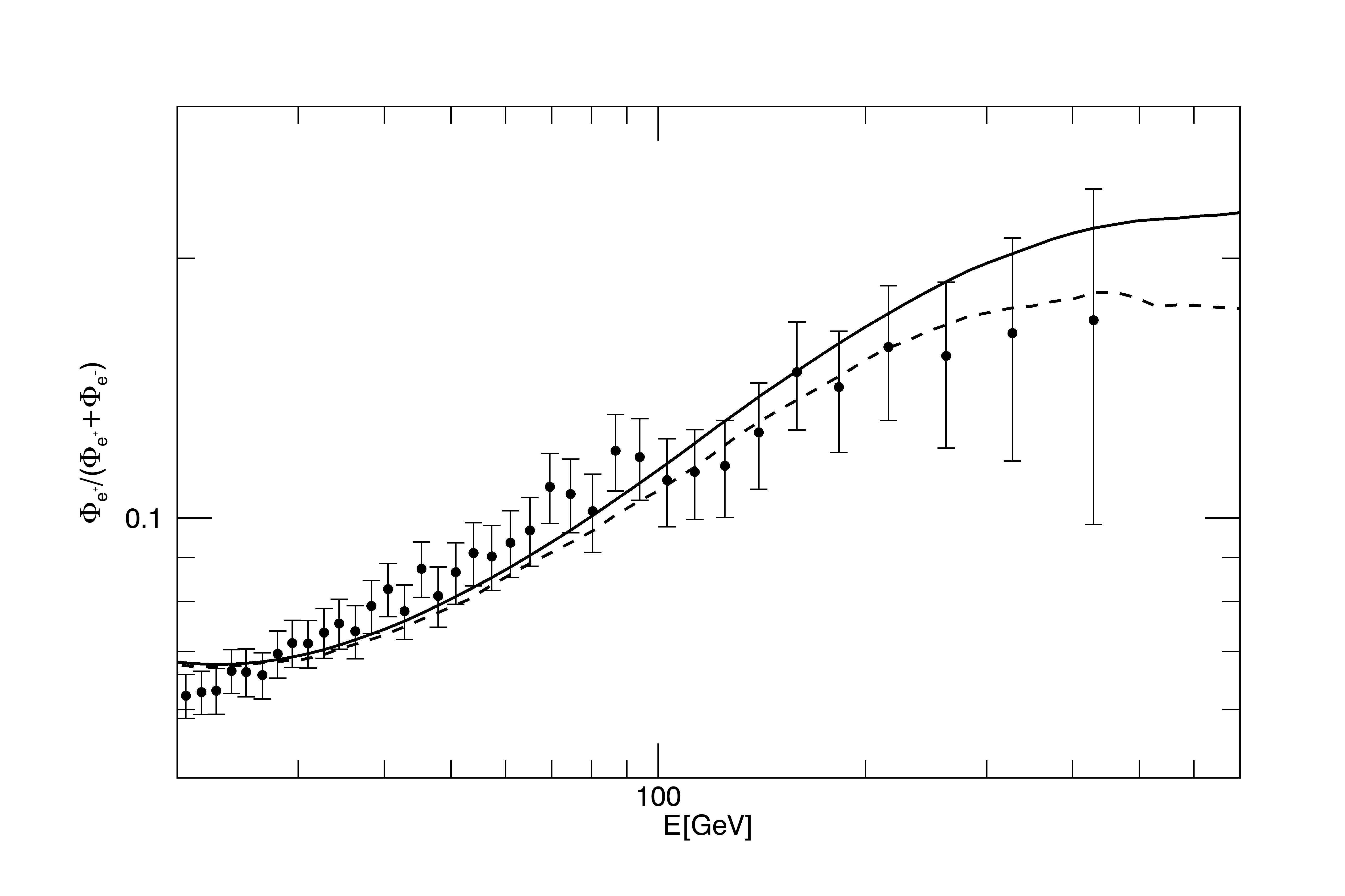} 
\caption{The positron fraction as a function of energy compared with
AMS02 data. The thick solid curve corresponds to the average spectra
shown as the solid (blue and red) curves of
Fig.~\ref{fig:elposspec}.  The dashed curve refers to the same
configuration shown by the corresponding curves of
Fig.~\ref{fig:elposspec}.} \label{fig:posfrac}
\end{figure}

In Fig.~\ref{fig:posfrac}, finally, the rise of the positron
fraction with energy that this model provides is compared with
AMS-02 data. The different curves are drawn following the same
notation as in Fig.~\ref{fig:elposspec}. A generally good agreement
is seen between the model and the data.

Further refinement of this kind of calculation are possible and
desirable, possibly with inclusion of some model for the density
distribution of the ISM. However the main point this section aims at
making is that bow shock PWNe are expected to play an important role
as contributors of cosmic ray positrons in the Galaxy. That their
contribution to the positron excess be non-negligible seems
unavoidable and this must be properly assessed and taken into
account before using positrons as a diagnostics of other phenomena.
In this sense, achieving a better understanding of the physics of
these sources is a task with potential far reaching implications for
cosmic ray and dark matter physics.


\begin{acknowledgements}
The authors thank the referee for constructive comments.
A.M.~Bykov and E.~Amato thank the staff of ISSI for their generous
hospitality and assistance. A.M.~Bykov and A.E.~Petrov were
supported by the RSF grant 16-12-10225. A.B. thanks A.V.~Artemyev,
S.M.~Osipov, and G.G.~Pavlov for discussions. Some of the modeling
was performed at the ``Tornado'' subsystem of the St.~Petersburg
Polytechnic University supercomputing center.
\end{acknowledgements}


\end{document}